\documentclass[12pt]{article}
\usepackage{hyperref,natbib,amsmath,amssymb, amsthm}
\usepackage{import, color,graphicx}

\newcommand{\blu}[1]{\textcolor{black}{#1}} 

\usepackage{epsf}
\usepackage{epsfig}
\usepackage{amsfonts}
\usepackage{setspace}
\usepackage[boxed]{algorithm}
\usepackage{algpseudocode}
\usepackage{algorithmicx}

\makeatother

\usepackage{caption}
\usepackage{subcaption}
\usepackage[normalem]{ulem}

\DeclareMathOperator{\Diag}{Diag}

\newcommand{\blind}{0}

\newcommand{\bm}[1]{\mbox{\boldmath $#1$}}

\def\spacingset#1{\renewcommand{\baselinestretch}%
{#1}\small\normalsize} \spacingset{1}

\begin{document}

\if0\blind
{
\title{\vspace{-1cm} On-site surrogates for \blu{large-scale} calibration}
\author{
	Jiangeng Huang\thanks{Corresponding author: Department of Statistics, Virginia Tech, \href{mailto:huangj@vt.edu}{\tt huangj@vt.edu}}\
\and 
Robert B.~Gramacy\footnotemark[1]
\and 
	Micka\"{e}l Binois\thanks{Mathematics and Computer Science Division, Argonne National Laboratory}, 
\and
\and
Mirko \blu{Libraschi}\thanks{Baker Hughes, a GE Company}
}
\date{}
\maketitle
}\fi

\if1\blind
{
  \bigskip
  \bigskip
  \bigskip
  \begin{center}
    {\LARGE\bf Bayesian calibration for a large computer experiment using local emulation}
\end{center}
  \medskip
  \bigskip
} \fi

\begin{abstract}
Motivated by a computer model calibration problem from the oil and
gas industry, involving the design of a honeycomb seal, we develop a
new Bayesian methodology to cope with limitations in the canonical
apparatus stemming from several factors. We propose a new strategy of on-site
design and surrogate modeling for a computer simulator acting on a
high-dimensional input space that, although relatively speedy, is prone to
numerical instabilities, missing data, and nonstationary dynamics. Our aim is
to strike a balance between data-faithful modeling and computational
tractability in a calibration framework---tailoring the computer model to a
limited field experiment.  Situating our {\em on-site surrogates} within the
canonical calibration apparatus requires updates to that framework.  We
describe a novel yet intuitive Bayesian setup that carefully decomposes
otherwise prohibitively large matrices by exploiting the sparse blockwise
structure.  Empirical illustrations demonstrate that this approach
performs well on toy data and our motivating honeycomb example.

\end{abstract}

\if0\blind
{
\bigskip
\noindent {\bf Keywords:}
Bayesian calibration, big data, computer experiment, local Gaussian process,
hierarchical model, uncertainty quantification }\fi

\spacingset{1} 

\section{Introduction}
\label{sec:intro}

With remarkable advances in computing power, today's complex physical systems
can be simulated comparatively cheaply and to high accuracy by using mature
libraries.  The ability to simulate has dramatically driven down the cost of
scientific inquiry in engineering settings, at least at initial
proof-of-concept stages.  Even so, computer models often idealize the
system---they are biased---or require the setting of tuning parameters:
inputs unknown or uncontrollable in actual physical processes in
the field.

An excellent example is the simulation of a free-falling object, which is a
potentially involved if well-understood enterprise from a modeling
perspective. Acceleration due to gravity might be known, but possibly not
precisely. Coefficients of drag may be completely unknown.  A model
incorporating both factors but not others such as ambient air disturbance or
rotational velocity  could be biased in consistent but unpredictable ways.

Researchers are interested in calibrating such models to experimental data.
With a flexible yet sturdy apparatus, a limited number of field observations
from physical experiments can provide valuable information to fine tune,
improve fidelity, understand uncertainty, and correct bias between simulations
and physical phenomena they model. When done right, tuned and
bias-corrected simulations are more realistic, forecasts more reliable, and
these can inform simulation redevelopment, if necessary.  

Here we are motivated by a calibration and uncertainty quantification goal in
the development of a so-called \textit{honeycomb seal}, a component in
high-pressure centrifugal compressors, with collaborators at Baker Hughes, a
General Electric company (BHGE).  Several studies in the literature treat
similar components from a mechanical engineering perspective
\citep[e.g.,][]{D'Souza:Childs:2002}.  To our knowledge, however, no one has
yet coupled mathematical models and field experimentation in this setting.
Using a commercial simulator and a limited field experiment, our BHGE
colleagues performed a nonlinear least-squares (NLS) calibration as a proof of
concept. The results left much to be desired.

Although we were initially optimistic that we could readily improve on this
methodology, a careful exploratory analysis on computer model and field data
revealed challenges hidden just below the surface.  These included data size,
dimensionality, computer simulation reliability, and the nonstationary nature
of the dynamics under study. Taken separately, each stretches the limits of
the canonical computer model calibration setup, especially in our favored
Bayesian setting. Taken all at once, these challenges demanded a fresh
perspective.

Contributions by \citet[][KOH]{Kennedy:O'Hagan:2001} and \citet{Higdon:2004}
lay the foundation for flexible Bayesian calibration of computer experiments,
tailored to situations where simulations are computationally expensive and
cheap, respectively.  Our situation is somewhere in between, as we describe in
more detail in Section \ref{sec:honeycomb}.  To set the stage
and establish some notation, we offer the following brief introduction.
Denote by $\mathbf{x} \in \mathbb{R}^{p_x}$ the controllable inputs in a physical
experiment and by $\mathbf{u} \in \mathbb{R}^{p_u}$ any additional (tuning) parameters
to the computer model that are unobservable or uncontrollable (or even
meaningless, such as mesh size) in the field. In the KOH framework, the
physical field observations $y^F(\mathbf{x})$ are connected with computer model
simulations $y^M(\mathbf{x}, \mathbf{u}^*)$ through a discrepancy term, or bias correction
$b(\mathbf{x})$, between simulation and field as follows:
\begin{equation}
y^F(\mathbf{x}) = y^M(\mathbf{x}, \mathbf{u}^*) + b(\mathbf{x}) +\epsilon.
 \label{eq:koh}
\end{equation}
Here, $\mathbf{u}^*$ is the unknown ``true" or ``best" setting for the calibration
input parameters, and $\epsilon\overset{\mathrm{iid}}{\sim} \mathcal N(0,
\sigma^2_\epsilon)$ represents random noise in the field measurements.

The main distinguishing feature between KOH and the work of
\citeauthor{Higdon:2004}~is the treatment of $y^M(\cdot, \cdot)$.  If
simulation is fast, then \citeauthor{Higdon:2004}~describe how evaluations may
be collected on-demand within the inferential procedure, for each choice of
$\mathbf{u}$ entertained, with bias $b(\cdot)$ trained directly on residuals
$y^F(\mathbf{X}^F) - y^M(\mathbf{X}^F,\mathbf{u})$ observed at a small number
of $N_F$ field data input sites, $\mathbf{X}^F$.  When simulations are slow on
not readily available for on-demand evaluation, then KOH prescribe surrogate
modeling to obtain a fitted $\hat{y}^M(\cdot,\cdot)$ from $N_M$ training
evaluations $[(\mathbf{X}^M, \mathbf{U}^M), \mathbf{Y}^M]$, with inference
being joint for the bias $b(\cdot)$ and tuning parameter settings,
$\mathbf{u}$, via a Bayesian posterior. If Gaussian processes (GPs) are used
both for the surrogate model and bias, a canonical choice in the computer
experiments literature \citep{Sacks1989,Santner2003}, then that posterior
enjoys a large degree of analytical tractability.  Numerical methods such as
Markov chain Monte Carlo (MCMC) facilitate learning in $\mathbf{u}$-space,
potentially averaging over any GP hyperparameters, such as characteristic
lengthscale or nugget.  For GP details, see \citet{Rasmussen2006}.

The KOH framework has been successfully implemented in many applications and
has demonstrated empirically superior predictive power for new untried
physical observations. The method is at the same time highly flexible and well
regularized.  Its main ingredients, coupled GPs ($\hat{y}^M$ and $\hat{b}$)
and a latent input space ($\mathbf{u}$), have separately been proposed as
tactics for adding fidelity to fitted GP surfaces, in particular as a thrifty
means of relaxing stringent stationarity assumptions
\citep[][]{ba:joseph:2012,bornn:shaddick:zidek:2012}. However, KOH is not
without its drawbacks.  One is identifiability, which is not a primary focus
of this paper \citep[see, e.g.,][]{plumlee2017bayesian,tuo2015}. Of more
pressing here are computational demands, especially in the face of the rapidly
growing size of modern computer experiments, both in the number of runs $N_M$
and in the input dimension $p_x$ or, to a lesser extent, $p_u$. GPs require
calculations cubic in $N_M$ to decompose large $N_M \times N_M$ covariance
matrices, limiting experiment sizes to the small thousands in practice. KOH
compounds the issue with $(N_M + N_F) \times (N_M + N_F)$ matrices. Bayesian
analysis in input high dimension ($p_x + p_u$), coupled with the large $N_M$
to adequately cover such a big computer simulation space, is all but
impossible without modification.
 
Inroads have recently been made in order to effectively and tractably
calibrate in settings where the computer experiment is orders of magnitude
larger than typical.  For example, \citet{gra:etal:2015} simplified KOH with
three modern ideas: modularization \citep{Liu:2009} to simplify joint
inference, local GP approximation \citep{gramacy:apley:2015} for fast
nonstationary modeling, and derivative-free optimization
\citep{AuCo04a,Le09b} for point estimation.  While effective, Bayesian
posterior uncertainty quantification (``the baby'') was all but thrown out
(``with the bath water'').

Here we propose a setup that borrows some of these themes, while at the same
time backing off on others. We develop a flavor of local GP approximation that
we call an {\em on-site surrogate, or OSS} that does not require
modularization in order to fit within the KOH framework.  As a result, we are
able to stay within a Bayesian joint inferential setup, although we find it
effective to perform a preanalysis via optimization, in part to prime the
MCMC.  We show how our OSSs accommodate a degree of nonstationarity while
imposing a convenient sparsity structure on otherwise huge $(N_M + N_F) \times
(N_M + N_F)$ coupled-GP covariance matrices, leading to fast decomposition
under partitioned inverse identities.  The result is a tractable calibration
framework that is both more accurate out-of-sample and more descriptive about
uncertainties than BHGE's NLS.

The remainder of the paper is outlined as follows. Section \ref{sec:honeycomb}
describes the honeycomb seal application, challenges stemming from its
simulation, and subsequent attempts to calibrate via a limited field data.
Section \ref{sec:localemu} introduces our novel OSS strategy for emulation
within a calibration framework and application within an
optimization/point-estimate setting. Section \ref{sec:fullBayes} expands this
setup in Bayesian KOH-style. Returning to our motivating example, Section
\ref{sec:results} demonstrates calibration results from both optimization and
fully Bayesian approaches, including comparison with the simpler NLS strategy
at BHGE. Section \ref{sec:discussion} concludes this paper with a brief
discussion.

\section{Honeycomb seal}
\label{sec:honeycomb}

The honeycomb seal is an important component widely used in BHGE's
high-pressure centrifugal compressors to enhance rotor stability in  oil and
gas applications or to control leakage in aircraft gas turbines. The seal(s)
and applications at BHGE are described by $p_x = 13$ design variables
$\mathbf{x}$ characterizing geometry and flow dynamics: rotational speed, cell
depth, seal diameter and length, inlet swirl, gas viscosity, gas temperature,
compressibility factor, specific heat, inlet/outlet pressure, and clearance.
The field experiment, from BHGE's component-level honeycomb seal test
campaign, comprises $N_F = 292$ runs varying a subset of those conditions,
$\mathbf{X}^F$, believed to have greatest variability during turbomachinery
operation: clearance, swirl, cell depth, seal length, and seal diameter.
Measured outputs include direct/cross stiffness and damping, at multiple
frequencies. Here our focus is on the direct stiffness output $y \equiv
k_{\mathrm{dir}}$ at 28 Hz.

A few hundred runs in thirteen input dimensions is hardly sufficient to
understand honeycomb seal dynamics to any reasonable degree in this highly
nonlinear setting.  Fortunately, the rotordynamics of seals like the honeycomb
are relatively well understood, at least from a mathematical and computational
modeling standpoint.  Although input dimension is somewhat high by
computer model calibration standards, library-based numerical routines provide
ready access to calculations for direct/cross stiffness and damping for inputs
like those listed above.  In what follows, we provide some insight into one
such solver and the advantages as well as challenges to using it (along with
the field data) to better understand and predict the dynamics of our honeycomb
seal.

\subsection{{\tt ISOTSEAL} simulator}
\label{sec:isotseal}

A simulator called {\tt ISOTSEAL}, developed at Texas A\&M University
\citep{isotseal}, offers a relatively speedy evaluation (about one second) of
the response(s) of interest for the honeycomb seal under study at BHGE. {\tt
ISOTSEAL} is built on bulk-flow theory, calculating gas seal force
coefficients based on seal flow physics.  Our BHGE colleagues have developed
an {\sf R} interface mapping seventeen scalar inputs for the honeycomb seal
experiment into the format required for {\tt ISOTSEAL}.  Thirteen of those
inputs match up with the columns of $\mathbf{X}^F$ (i.e., they are
$\mathbf{x}$'s); four are tuning parameters $\mathbf{u}$, which could not be
controlled in the field.  These comprise statoric and rotoric friction
coefficients $n_s, n_r$ and exponents $m_s, m_r$.  They are the {\em friction
factors} of the honeycomb seal. In the turbulent-lubrication model from
bulk-flow theory, the shear stress $f$ is a function of the friction
coefficient $n$ and exponent $m$ through the Blasius model $f =n
\mathrm{Re}^m$, where $\mathrm{Re}$ is the Reynolds number \citep{Hirs:1973}.
Applied separately for the stator ($s$) and rotor ($r$), friction factors $n$
and $m$ must be determined empirically from experimental data. To protect
BHGE's intellectual property, but also for practical considerations, we work
with friction factors coded to the unit cube.
\[
(n_s, m_s, n_r, m_r)^\top \rightarrow (u_1, u_2, u_3, u_4)^\top \in [0,1]^4
\]
These are treated as calibration parameters $\mathbf{u}$, with the goal of learning
their setting via field data and {\tt ISOTSEAL} simulations. 

Although {\tt ISOTSEAL} is fast and has a reputation for delivering outputs
faithful to the underlying physics, we identified several drawbacks in our
application.  For some input settings it fails to terminate, especially with
friction factors ($\mathbf{u}$) near the boundary of their physically
meaningful ranges.\footnote{At least for the commercial version of the
simulator in use at BHGE, paired with their input--mapping front-end.  The
{\sf R} wrapper aborts the simulation and returns {\tt NA} after seven seconds
of execution.} For others, where a response is provided, numerical
instabilities and diverging approximation are evident. Although evaluations
are operationally deterministic, in that providing the same input
yields the same output, the behavior can seem otherwise unpredictable. Even
subtle numerical ``jitters'' of this sort can thwart conventional GP
interpolation \citep{gra:lee:2012}. As we show below, {\tt ISOTSEAL}'s jitters
are sometimes extreme.  Others have commented on similar drawbacks
\citep{vann:2011}; modern applications of {\tt ISOTSEAL} may be pushing the
boundaries of its engineering.

\begin{figure}[ht!]
  \centering
\includegraphics[width=0.32\linewidth, trim=0 15 15 50, clip]{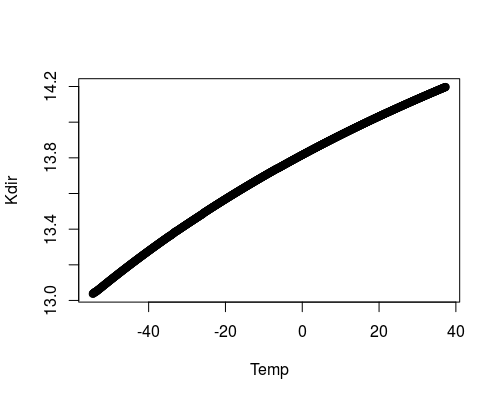}
\includegraphics[width=0.32\linewidth, trim=0 15 15 50, clip]{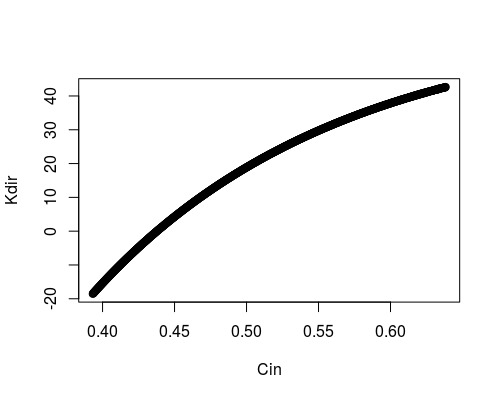}
\includegraphics[width=0.32\linewidth, trim=0 15 15 50, clip]{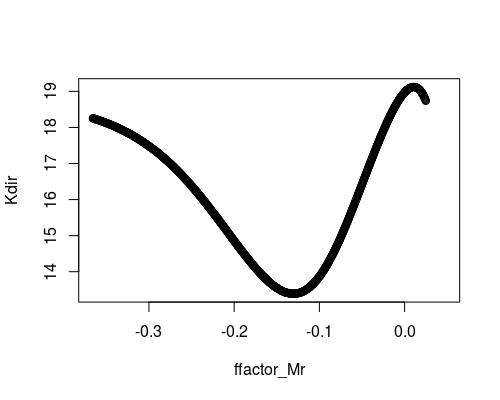}
\includegraphics[width=0.32\linewidth, trim=0 15 15 50, clip]{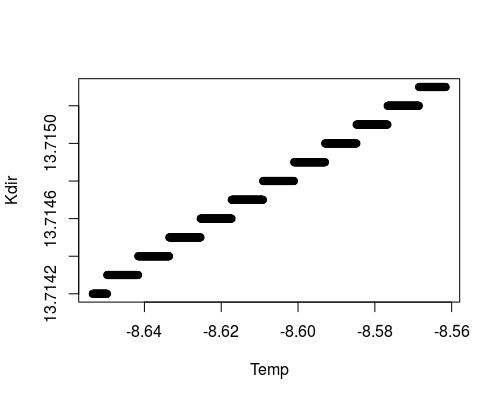}
\includegraphics[width=0.32\linewidth, trim=0 15 15 50, clip]{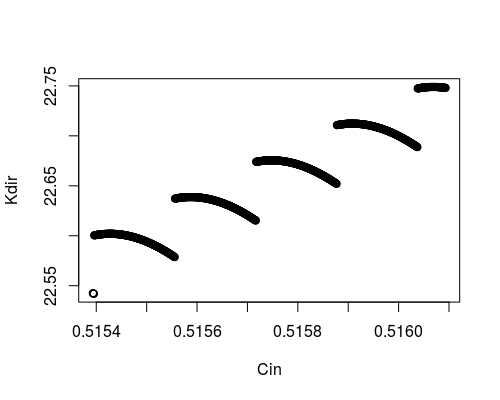}
\includegraphics[width=0.32\linewidth, trim=0 15 15 50, clip]{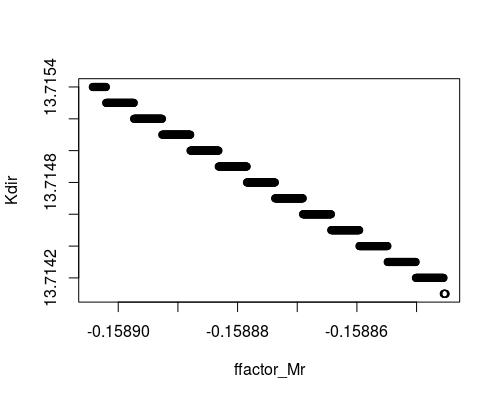}
\includegraphics[width=0.32\linewidth, trim=0 15 15 50, clip]{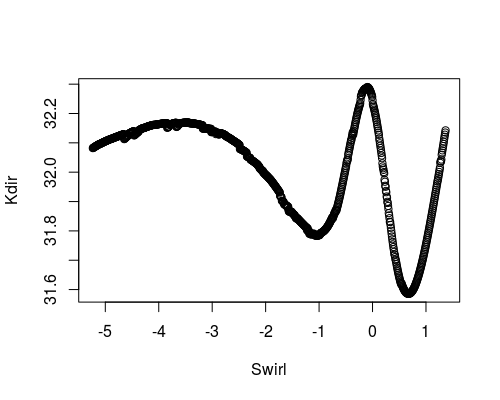}
\includegraphics[width=0.32\linewidth, trim=0 15 15 50, clip]{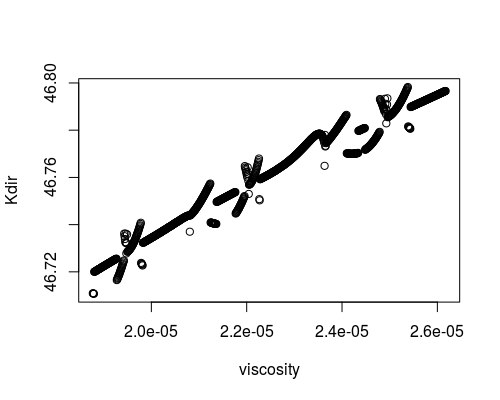}
\includegraphics[width=0.32\linewidth, trim=0 15 15 50, clip]{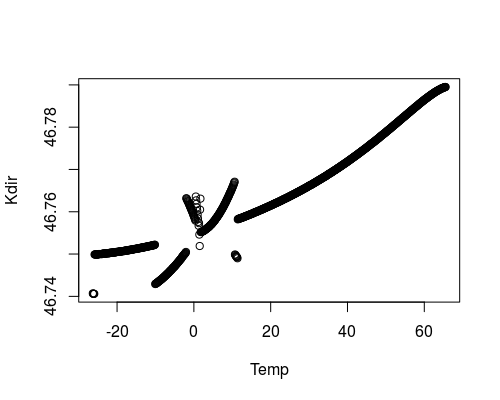}
\includegraphics[width=0.32\linewidth, trim=0 15 15 50, clip]{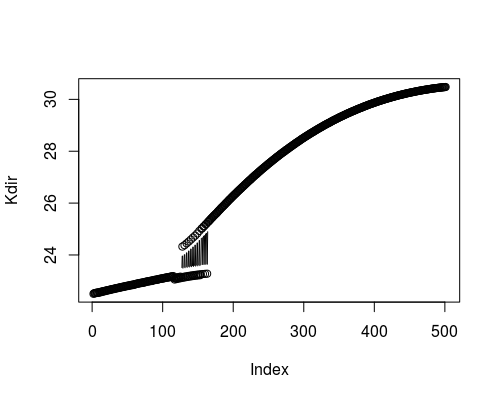}
\includegraphics[width=0.32\linewidth, trim=0 15 15 50, clip]{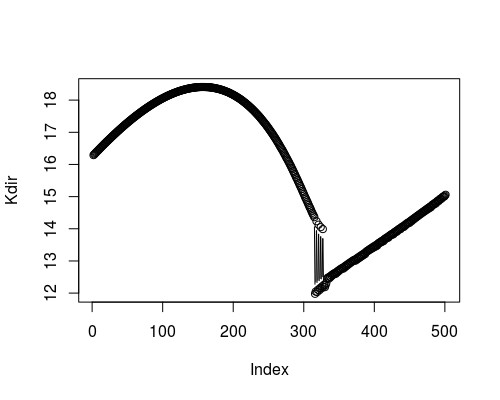}
\includegraphics[width=0.32\linewidth, trim=0 15 15 50, clip]{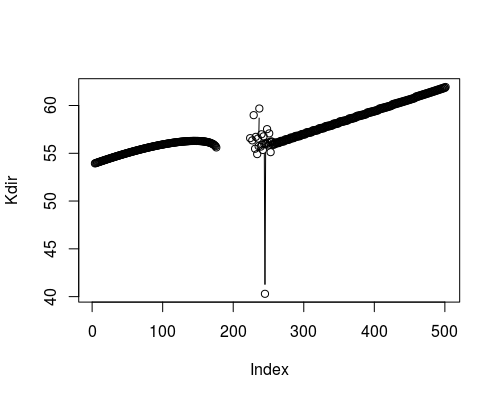}
\caption{Local plots of ISOTSEAL response surface for direct
stiffness (Kdir). Row 1: change of one input in grid in wide input ranges. Row 2:
zoomed-in versions of row 1, changing one input in a much denser grid. Row 3:
inexact simulations, changing one input in a grid in input space. Row 4:
input trajectory between two arbitrary points from the input space, varying
all inputs in grids. 
}
\label{fig:loc1}
\end{figure}

Figure \ref{fig:loc1} shows example outputs $\mathbf{y}^M$ obtained by varying
one input at a time in a narrow range, while fixing the others at sensible
values (first three rows); and varying all inputs in grids between two
arbitrary points (fourth row).  The first row shows ideal settings: the
response is a smooth function of the input over the range(s) entertained.  The
second row, however, zooming in on the same input--response scenarios,
reveals a ``staircase/striation'' effect at small
scales. The third row shows more concerning macro-level behavior over both
narrow and wide input ranges. According to BHGE's rotordynamics experts, these
``staircase'' and discontinuity features could be related to tolerances
imposed on first-order equilibrium and flow equations implemented in {\tt
ISOTSEAL}.

The last row illustrates unpredictable regime-changing behavior and gaps due
non-terminating simulation. Particular challenges exhibited by the bottom row
notwithstanding, dynamics are clearly nonstationary from a global perspective.
A great example of this is in the first column of the third row, where the
response is at first slowly changing and then more rapidly oscillating.  In
that example, the regime change is smooth. In other cases, however, as in the
middle column of the  bottom row, a ``noisy'' discontinuity separates a
hill-like feature from a steadier slope. An ordinary GP model, even with a
nugget deployed to smooth over noiselike features by treating them as genuine
noise \citep{gra:lee:2012}, could not accommodate such regime changes, smooth
or otherwise.

Consequently, initial attempts to emulate {\tt ISOTSEAL}-generated response
surfaces via the canonical GP in the full (17-dimensional) input space were
not successful.  Even with space-filling designs sized in the several
thousands, pushing the limits of the $\mathcal{O}(N^3)$ bottleneck of large
matrix decompositions, we were unable to adequately capture the distinct
features we saw in smaller, more localized experiments.  Modest reductions in
the input dimension---holding some inputs fixed---and, similarly, reductions
in the width of the input domain for the remaining coordinates led to
unremarkable improvement in terms of accuracy in out-of-sample predictions.
Global nonstationarity, local features, numerical artifacts, and high input
dimension proved to be a perfect storm.  Section \ref{sec:localemu} uses those
unsuccessful proof-of-concept fits as a benchmark, showing how our proposed
on-site surrogate offers a far more accurate alternative, at least from a
purely out-of-sample emulation perspective.

\subsection{Nonlinear least-squares calibration}
\label{sec:nls}

To obtain a crude calibration to the small amount of field data
they had, our BHGE colleagues performed a nonlinear least-squares
analysis. Starting in a stable part of the input space, from the perspective
of {\tt ISOTSEAL} behavior, they used a numerical optimizer---a Nash variant
of Marquardt NLS via QR linear solver, {\tt nlfb} \citep{nlmrt}---to tune
$\mathbf{u}$-values, that is, the four friction factors, based on a quadratic
loss between simulated $y_i^M(\mathbf{x}_i, \mathbf{u})$ and observed
output $y_i^F(\mathbf{x}_i)$ at the input training data sites $\mathbf{X}^F$.
\begin{align}
\hat{\mathbf{u}} = \mathrm{arg}\min_\mathbf{u} \left\{ \dfrac{1}{N_F} 
\sum_{i=1}^{N_F} \left[ y^F_{i}(\mathbf{x}_i)-y^M_i(\mathbf{x}_i, \mathbf{u}) \right]^2 \right\},
\label{eq:nls}
\end{align}

In search for $\hat{\mathbf{u}}$, each new $\mathbf{u}$-value tried by the
{\tt nlfb} optimizer triggered $N_F$ calls to {\tt ISOTSEAL}, one for each row
of the design parameters $\mathbf{X}^F$, much in the style of
\citet{Higdon:2004} but without estimating a bias correction. To cope with
failed {\tt ISOTSEAL} runs, {\tt nlfb} monitors the rate of missing values in
evaluations. When the missingness rate is below a threshold (e.g., 10\%), a
predetermined large residual value (100 on the original scale) is imputed for the
missed residual to discourage convergence toward solutions nearby. Once above
the threshold, {\tt nlfb} reports an error message and is started afresh.

\begin{figure}[ht!]
  \centering
\includegraphics[width=0.49\linewidth, trim=0 0 0 0, clip]{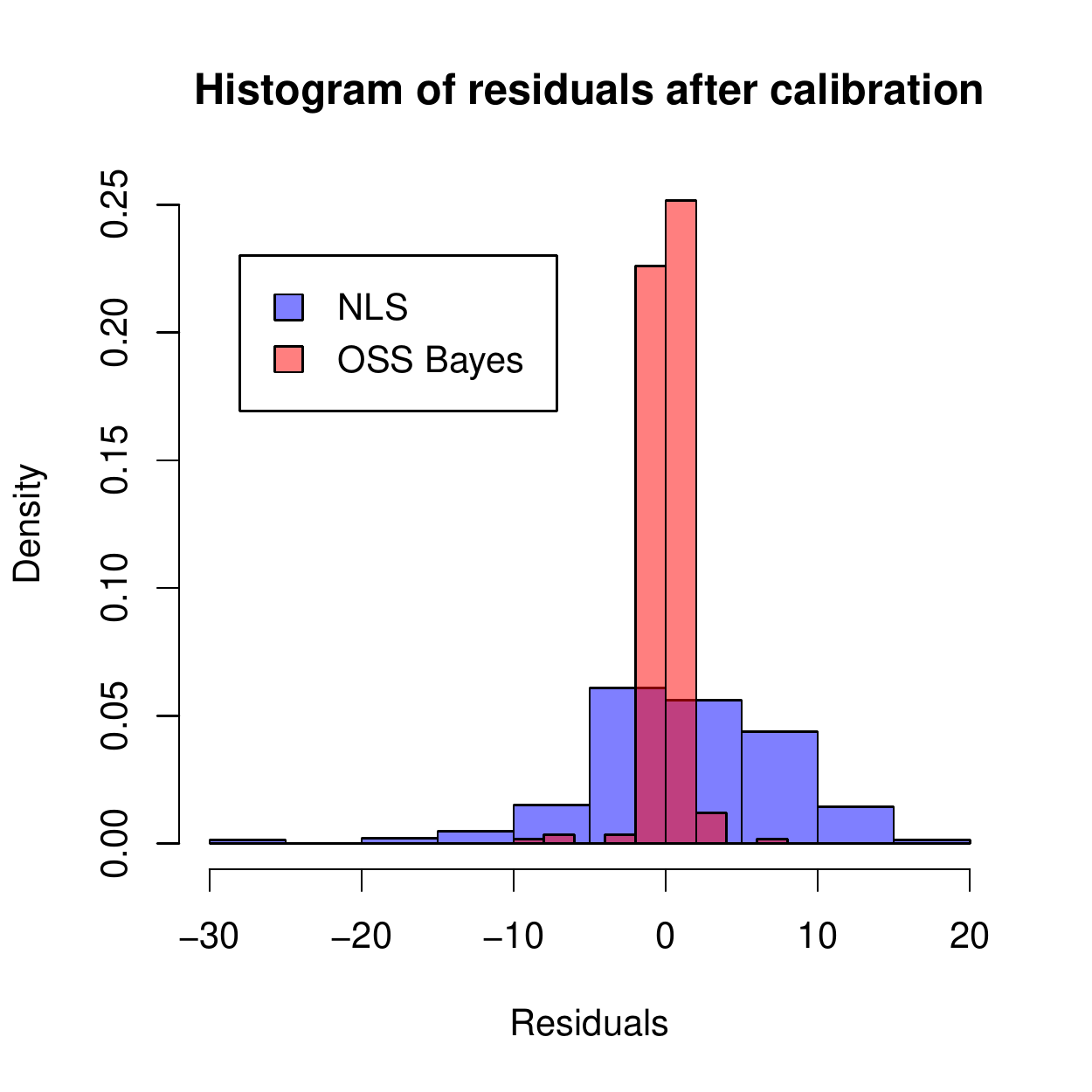}
\includegraphics[width=0.49\linewidth, trim=0 0 0 0, clip]{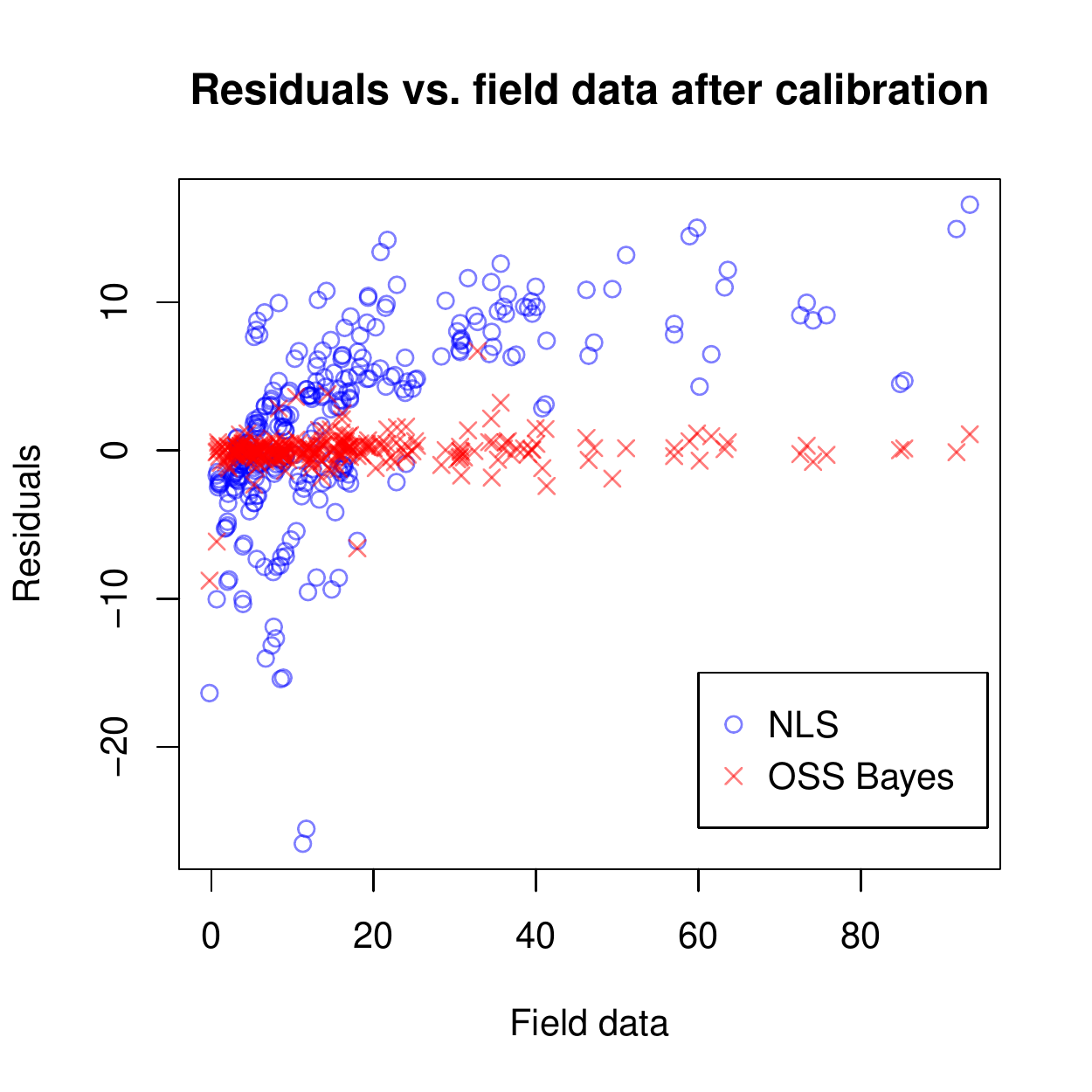}
\caption{Observed in-sample residuals between NLS  calibrated {\tt ISOTSEAL} and OSS Bayes 
from field data.  The left panel shows histograms of the residuals; the right
panel shows them versus the true response. 
The NLS has in-sample RMSE $= 6.605$.
The OSS Bayes has in-sample RMSE $=1.125$,
which is further discussed in Section \ref{sec:oosresults}. }
 \label{fig:nls}
\end{figure}

We repeated this experiment, starting instead from 100 random space-filling
$\mathbf{u}$ values in hopes of improving on our BHGE colleagues' results
with a best value at RMSE $=8.567$ and having a strong \blu{benchmark} for later
comparison. Because this NLS setup does not model a discrepancy between
$\mathbf{y}^M$ and $\mathbf{y}^F$, converged solutions have large quadratic
loss, even in-sample. Among 100 restarts, two failed; and the other losses,
mapped to the scale of $\mathbf{y}^F$ by taking the square root, had the
following distribution.
\begin{center}
\begin{tabular}{rrrrrr}
min & 25\% & med & mean & 75\% & max \\
\hline
6.605 & 8.161 & 8.401 & 10.117 & 9.099 & 25.787
\end{tabular}
\end{center}
The blue/circle marks in Figure \ref{fig:nls} show  observed residuals between
field data and NLS calibrated {\tt ISOTSEAL} with the best solution we
obtained, $\hat{u}=(0.000, 0.000, 0.821, 0.996)^\top$.  Notice that three
of four friction factors are set at their limit values. This restart
benefited from a serendipitous initialization, having initial RMSE of 9.219
converging to 6.605. However, Figure \ref{fig:nls} shows that many large
residuals still remain (blue/circles).
The red/crosses comparator is based on our proposed methodology and is
described in subsequent sections.  For comparison, and to whet the reader's
appetite, we note that the in-sample RMSE we obtained was 1.125.  Out-of-sample results are
provided in Section \ref{sec:oosresults}.  We attribute NLS's relatively poor
performance to two features.  One is its inability to compensate for biases in
{\tt ISOTSEAL} runs, relative to the outcome of field experiments.  Another is
that the solutions found were highly localized to the
neighborhood of the starting configuration. A post mortem analysis revealed
that this was due primarily to large missingness rates.

Although we were confident that we could improve on this methodology and
obtain more accurate predictions by correcting for bias between field and
simulation in a Bayesian framework, it quickly became apparent that a
standard, KOH-style analysis would be fraught with difficulty.  In a test run,
we used a space-filling design $\mathbf{X}^M$ and fit a global GP emulator in
the 17-dimensional space of {\tt ISOTSEAL} runs $\mathbf{y}^M$ thus obtained.
That surrogate offered nice-looking predictive surfaces and provided posterior
surfaces for calibrated friction factors substantially different from those
obtained from NLS (e.g., away from the boundary), but unfortunately the
surrogates were highly inaccurate out of sample, as illustrated below.

\section{Local design and emulation for calibration}
\label{sec:localemu}

Failed attempts at surrogate modeling {\tt ISOTSEAL}, either generally or for
the specific purpose of calibration to field data [see Section
\ref{sec:isotseal}], motivate our search for a new perspective.  Local
emulation has been proposed in the recent literature \citep{gra:etal:2015} as
a means of circumventing large-data GP surrogate modeling for calibration,
leveraging the important insight that surrogate evaluation is required only at
field data locations $\mathbf{X}^F$, of which we have relatively few ($N_F =
292$).  But in that context the input dimension was small, and here we are
faced with the added challenges of numerical instability, nonstationary
dynamics, and missing data. In this section we port that idea to our setting
of on-site surrogates, leveraging relatively cheap
{\tt ISOTSEAL} simulation, while mitigating problems of big $N_M$, big $p_x +
p_u$, and challenging simulator dynamics.

\subsection{On-site surrogates}
\label{sec:oss}

{\em On-site surrogates} (OSSs) reduce a $p = p_x + p_u =17$-dimensional
problem into a $p_u =4$-dimensional problem by building as many surrogates as
there are field data observations, $N_F = 292$.  Let $\mathbf{x}$ denote a generic
design variable setting and $\mathbf{u}$ a generic tuning vector (e.g., friction
factor in {\tt ISOTSEAL}). Then the mapping from one big surrogate to many
smaller ones may be conceptualized by the following chart:
\begin{align}
\hat{y}^M(\mathbf{x}, \mathbf{u}) \longrightarrow  \hat{y}^M(\mathbf{x}_i, \mathbf{u}) 
\longrightarrow  \hat{y}_i^M(\mathbf{u}), \quad \mbox{for } \; i = 1, 2, \dots, N_F.
\end{align} 
That is, rather than building one big emulator for the entire $p$-dimensional
input space $\hat{y}^M(\mathbf{x}, \mathbf{u})$, we instead train separate
emulators $\hat{y}_i^M(\mathbf{u})$ focused on each site $\mathbf{x}_i$ where
field data has been collected. In this way, OSSs are a divide-and-conquer
scheme that swap joint modeling in a large $(\mathbf{x},\mathbf{u})$-space,
where design coverage and modeling fidelity could at best be thin, for many
smaller models in which, separately, ample coverage is attainable with
modestly sized design in $\mathbf{u}$-space only. Fitting and simulation can
be performed in parallel, since the calculations for each field data site
$\mathbf{x}_i$, $i=1,\dots, N_F$ are both operationally and statistically
independent. Nonstationary modeling is implicit, since each surrogate focuses
on a different part of the input space.  If simulations are erratic for some
$(\mathbf{x}_i, \mathbf{u})$, say, the OSS indexed by $i$ can compensate by
smoothing over with nonzero nuggets.  If dynamics are well behaved for other
sites $j$, OSSs can interpolate after the typical fashion.

In some ways, OSSs are akin to an {\em in situ} emulator \citep{gul:2018}.
Whereas the in situ emulator is tailored to uncertainty quantification around
nominal inputs, OSSs are applied in multitude for each element of
$\mathbf{X}^F$ in the calibration setting. Another distinction is the role of
design in building OSSs.  Here we propose separate designs at each
$\mathbf{x}_i$ to learn each $\hat{\mathbf{y}}_i^M(\mathbf{u})$, rather than
working with design subsets. A maximin Latin hypercube sample (LHS) is
preferred for their space-filling and uniform margin properties \citep[see,
e.g.,][]{morris:1995}.  We use {\tt maximinLHS} in the {\tt lhs} \citep{lhs}
for {\sf R}.

Specifically, at each of the $N_F = 292$ field data sites, we create novel
1000-run maximin LHS designs for friction factors in $p_u = 4$-dimensional
$\mathbf{u}$-space. In this way, we separately design a total of $N_M=292,000$ {\tt
ISOTSEAL} simulation runs. With about one second for evaluation (for
successfully terminating runs and about seven seconds waiting 
to terminate a failed run), this is a manageable workload requiring about
81 core-hours, or about one day on a modern hyperthreaded multicore
workstation.

Let $\mathbf{y}_i^M = y^M(\mathbf{U}_i)$ be a vector holding the $n_i$
converged {\tt ISOTSEAL} runs (out of the 1,000) at the $i^\mathrm{th}$ site, for
$i=1,\dots,N_F$. $\mathbf{U}_i$ is the corresponding $n_i \times p_u$ on-site
design matrix. In our {\tt ISOTSEAL} experiment, where $N_F=292$, a total of
$N_M=
\sum_{i=1}^{N_F}n_i= 286,282$ runs terminated successfully. Most sites (241)
had a complete set of $n_i = 1000$ successful runs. Of the 51 with missing
responses of varying multitudes, the smallest was $n_{238} = 574$.

Each OSS comprises a fitted GP regression between successful
on-site {\tt ISOTSEAL} run outputs $\mathbf{y}_i^M$ and $\mathbf{U}_i$.
Specifically, $\hat{y}_i^M (\mathbf{U}_i)$ is built by fitting a stationary
zero-mean GP using a scaled and nugget-augmented separable Gaussian power
exponential kernel
\begin{align}
V_i(\mathbf{u}, \mathbf{u}') = \tau_i^2\exp\left\{ - \sum_{k=1}^{p_u} 
\frac{||\mathbf{u}_{ik} - \mathbf{u}'_{ik}||^2}{\theta_{ik}}  + \delta_{u,u'} \eta_i\right\},
 \label{eq:kernel}
\end{align}
where $\tau_i^2$ is a site-specific scale parameter, $\bm{\theta}_i=(\theta_{i1},
\theta_{i2},  \dots, \theta_{ip_u} )^\top$ is vector of site-specific
lengthscales, $\eta_i$ is a nugget parameter,\footnote{Note that the nugget
$\eta_i$ augmentation is applied only when $\mathbf{u}'$ and $\mathbf{u}$ are
identically indexed, i.e., on the diagonal of a symmetric covariance matrix;
not simply when their values happen to coincide.} 
and $\delta_{u,u'}$ is the Kronecker delta.  Denote the set of
hyperparameters of the $i^\mathrm{th}$ OSS as $\bm{\phi}_i=\{\tau^2_i,
\bm{\theta}_i, \eta_i\}$, for $i=1, 2, \dots, N_F$. Although nuggets $\eta_i$
are usually fit to smooth over noise, here we are including them to smooth
over any deterministic numerical ``jitters.'' Other mean and covariance
structures may be reasonable, so in what follows let $\bm{\phi}_i$ stand in
generically for the estimable quantities of each OSS. Although numerous
options for inference exist, we prefer plug-in maximum likelihood estimates (MLEs)
$\hat{\bm{\phi}}_i$, calculated in parallel for each $i=1,\dots,N_F = 292$ via {\tt
L-BFGS-B} \citep{byrd:etal:1995} using analytic derivatives via {\tt mleGPsep}
in the {\tt laGP} package \citep{laGP,gramacy:jss:2016} for {\sf R}. As we
illustrate momentarily, this simple OSS strategy provides far more accurate
emulation out-of-sample than does the best global alternative we could muster with
a commensurate computational effort.

\subsection{Merits of on-site surrogates}
\label{sec:merits}

To build a suitable global GP competitor, we created an $N_M = 8000$-run
maximin LHS in $p=17$ input dimensions, fit a zero-mean GP based on a
separable covariance structure (\ref{eq:kernel}), and estimated the
19-dimensional hyperparameters $\hat{\bm{\phi}}_g=\{\tau^2_g, \bm{\theta}_g,
\eta_g\}$ via MLE.  We chose 8,000 runs because that demanded a comparable
computational effort to the OSS setup described in Section \ref{sec:oss}.
Although the {\tt ISOTSEAL} simulation effort for 8,000 runs is far less than
the 292K for the OSSs, the hyperparameter inference effort and subsequent
prediction for an $N_M = 8000$-sized design is commensurate with that required
for our 292 size $n_i \approx 1000$ OSS calculations. Repeated matrix
decompositions in likelihood and derivative calculations in search of the MLE,
requiring $\mathcal{O}(N_M^3)$ flops for the global surrogate, represented a
heavy burden even when parallelized by multi-threaded linear algebra libraries
such as the Intel Math Kernel Library. Similarly threaded calculations of
$O(n_i^3)$ flops were faster even in 292 copies, in part because fewer
evaluations were needed to learn hyperparameters $\hat{\bm{\phi}}_i$ in the
lower-dimensional $\mathbf{u}$-space.\footnote{The OSSs learn $|\bm{\phi}_i| =
6$ compared to $|\bm{\phi}_g| = 19$ for the global analog.  The latter thus
demands more expensive gradient calculations.  Moreover, the former generally
converges to the same local optima when reinitialized, whereas the latter have
many local minima due to nonstationary and locally ``jittery'' responses.
Multiple restarts are required to mitigate the chance of finding vastly
inferior local optima.}

\begin{figure}[ht!]
  \centering
 \begin{minipage}{6cm}
\includegraphics[scale=0.5,trim=10 40 10 50,clip]{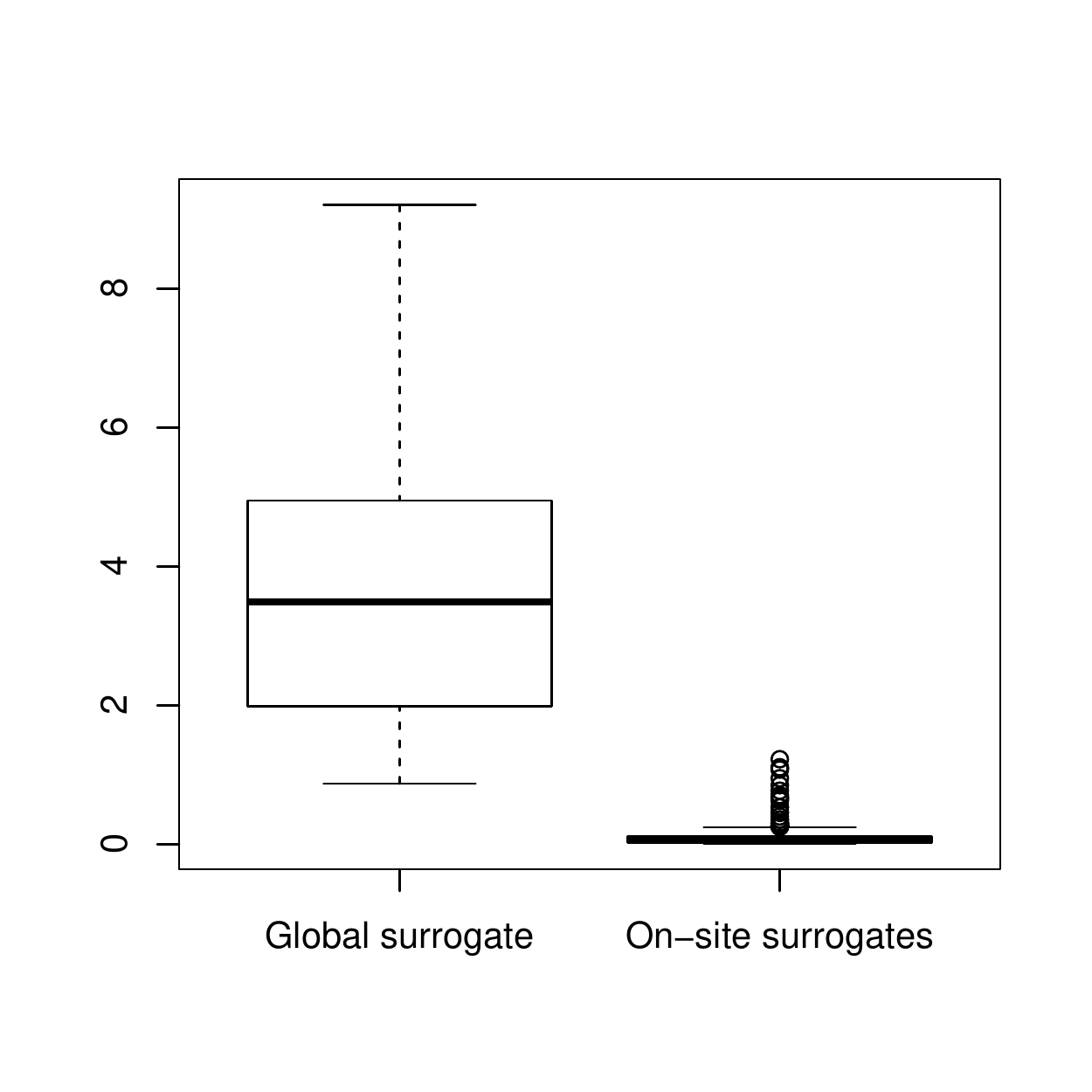}
\end{minipage} \hspace{1cm}
\begin{tabular}{r|rr}
& global & OSS \\
\hline
min & 0.871 & 0.008 \\
25\% & 1.991 & 0.023 \\
med & 3.492 & 0.050 \\
mean & 3.619 &  0.120 \\
75\% & 4.928 & 0.112 \\
max & 9.207 &  1.223
\end{tabular} 
  \caption{Boxplots of 292 out-of-sample RMSEs, where each RMSE is computed by using novel  
  $n_i' \leq 1,000$ on-site data from both global surrogate and OSSs.}
  \label{fig:rmse}
\end{figure}

Since the OSSs were trained on a much larger corpus of simulations, it is
perhaps not surprising that they provide more accurate predictions out of
sample. To demonstrate that empirically, Figure \ref{fig:rmse} summarizes the
results of emulation accuracy from both global surrogate and OSSs. For our
calibration goal, we need accurate emulation only at locations where we
have field data $\mathbf{X}^F$. Therefore we entertain out-of-sample
prediction accuracy only for those $\mathbf{X}^F$ sites. At each of the 292 field input
sites $\mathbf{x}_i$, we design $\mathbf{U}_i'$ with 1,000 runs each, the same
amount as the training set, via maximin LHS. In total we collected
$N_M'=286,224$ testing {\tt ISOTSEAL} runs, which is fewer than we ran since
some came back missing. A pair of RMSEs, based on the OSSs and global
surrogates, were calculated at each site $i=1, 2, \dots, N_F = 292$ based on
the $n_i'
\approx 1,000$ testing runs located there. The distribution of these values
is summarized in Figure \ref{fig:rmse}. From those boxplots, one can easily
see that the OSSs yield far more accurate predictions.

\begin{figure}[ht!]
  \centering
\includegraphics[width=0.32\linewidth, trim=0 10 20 5, clip=TRUE]{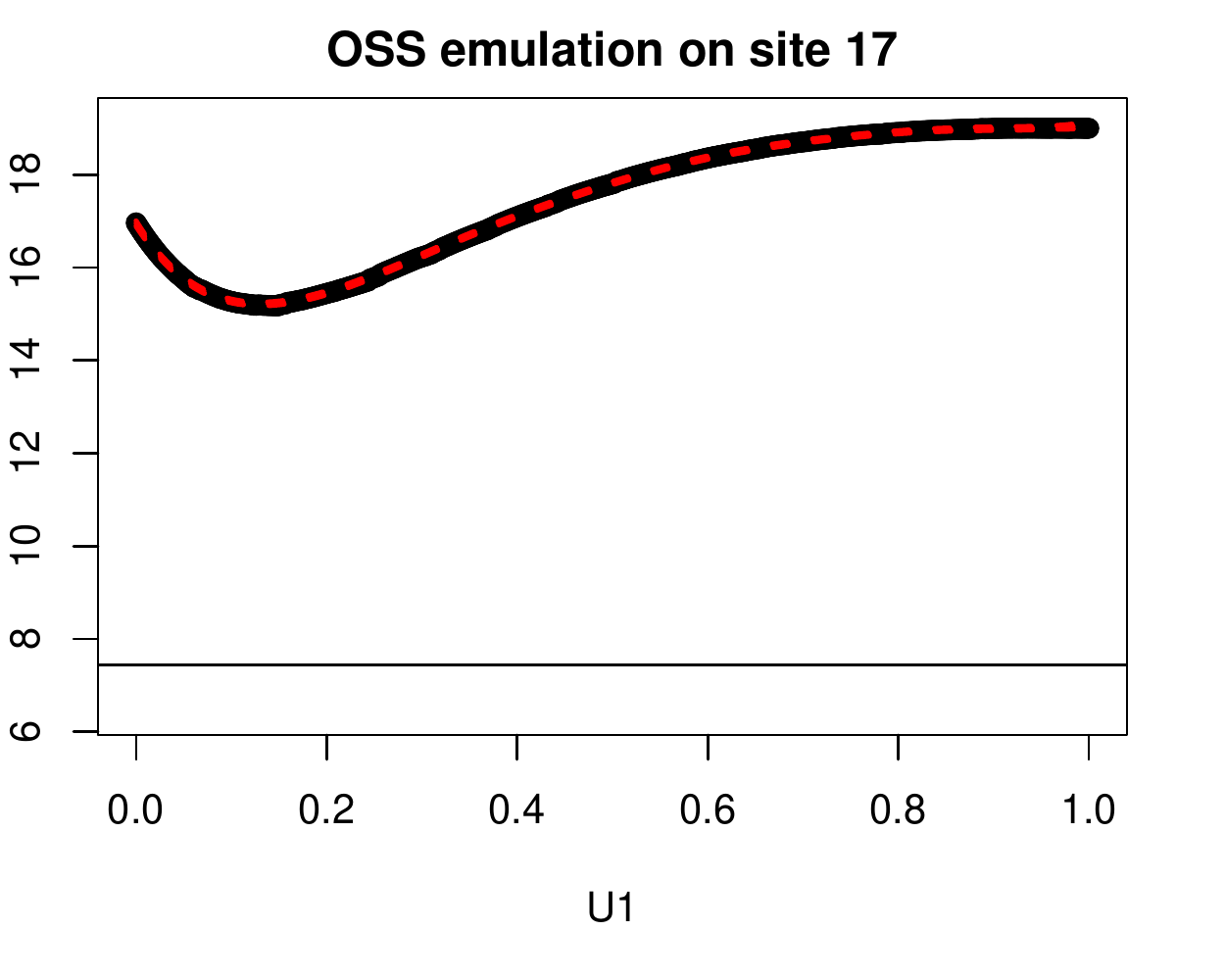}
\includegraphics[width=0.32\linewidth, trim=0 10 20 5, clip=TRUE]{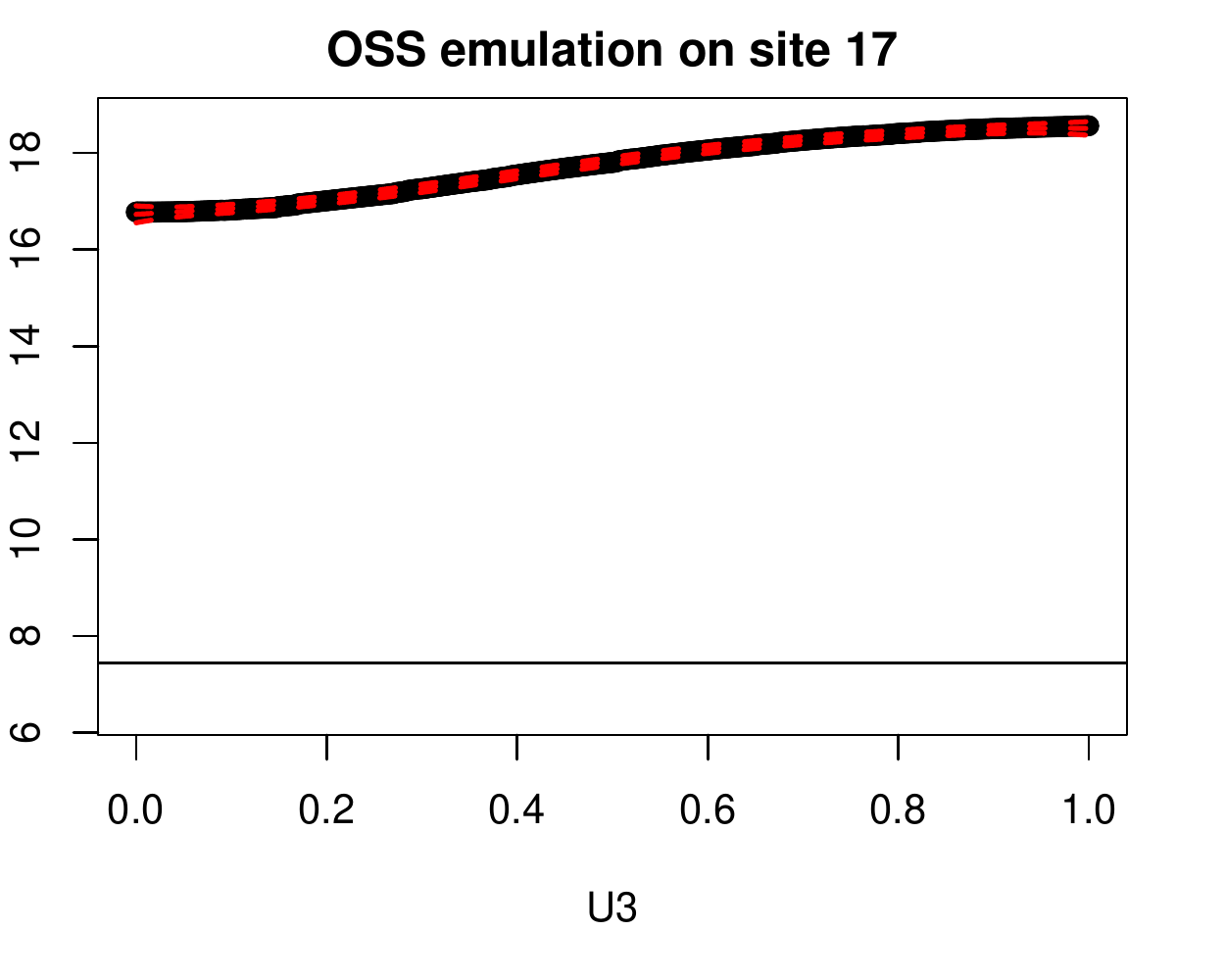}
\includegraphics[width=0.32\linewidth, trim=0 10 20 5, clip=TRUE]{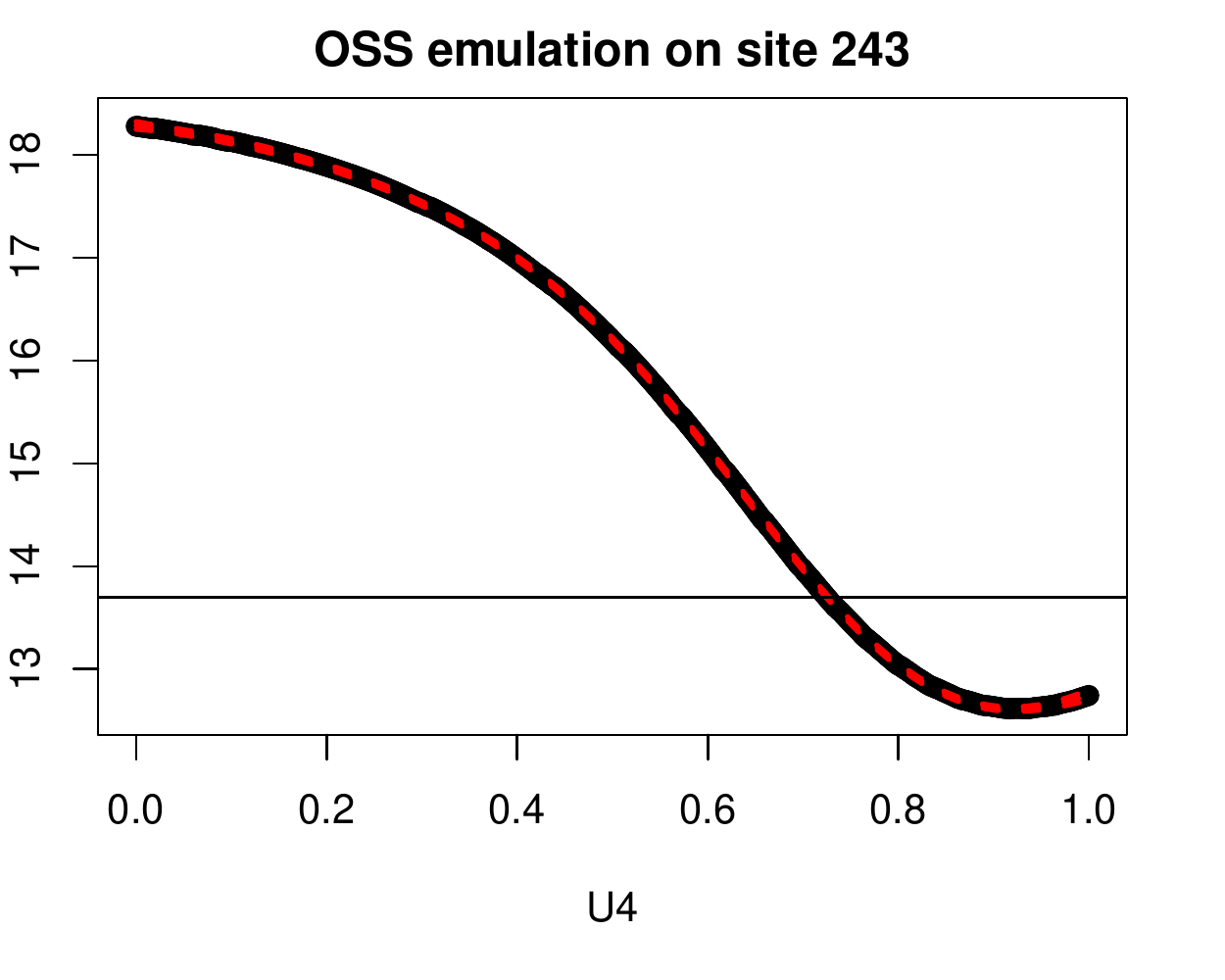}\\
\includegraphics[width=0.32\linewidth, trim=0 10 20 5, clip=TRUE]{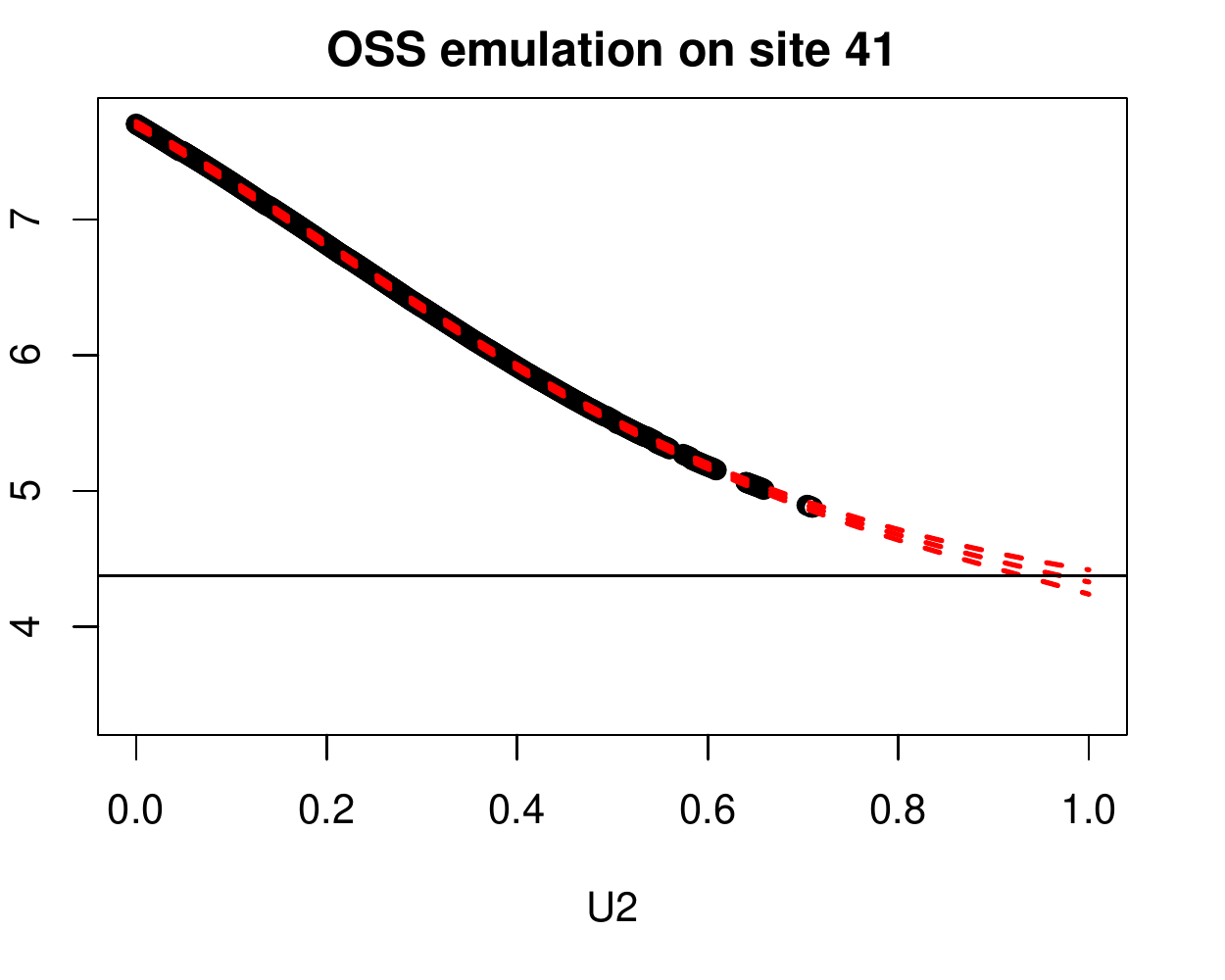}
\includegraphics[width=0.32\linewidth, trim=0 10 20 5, clip=TRUE]{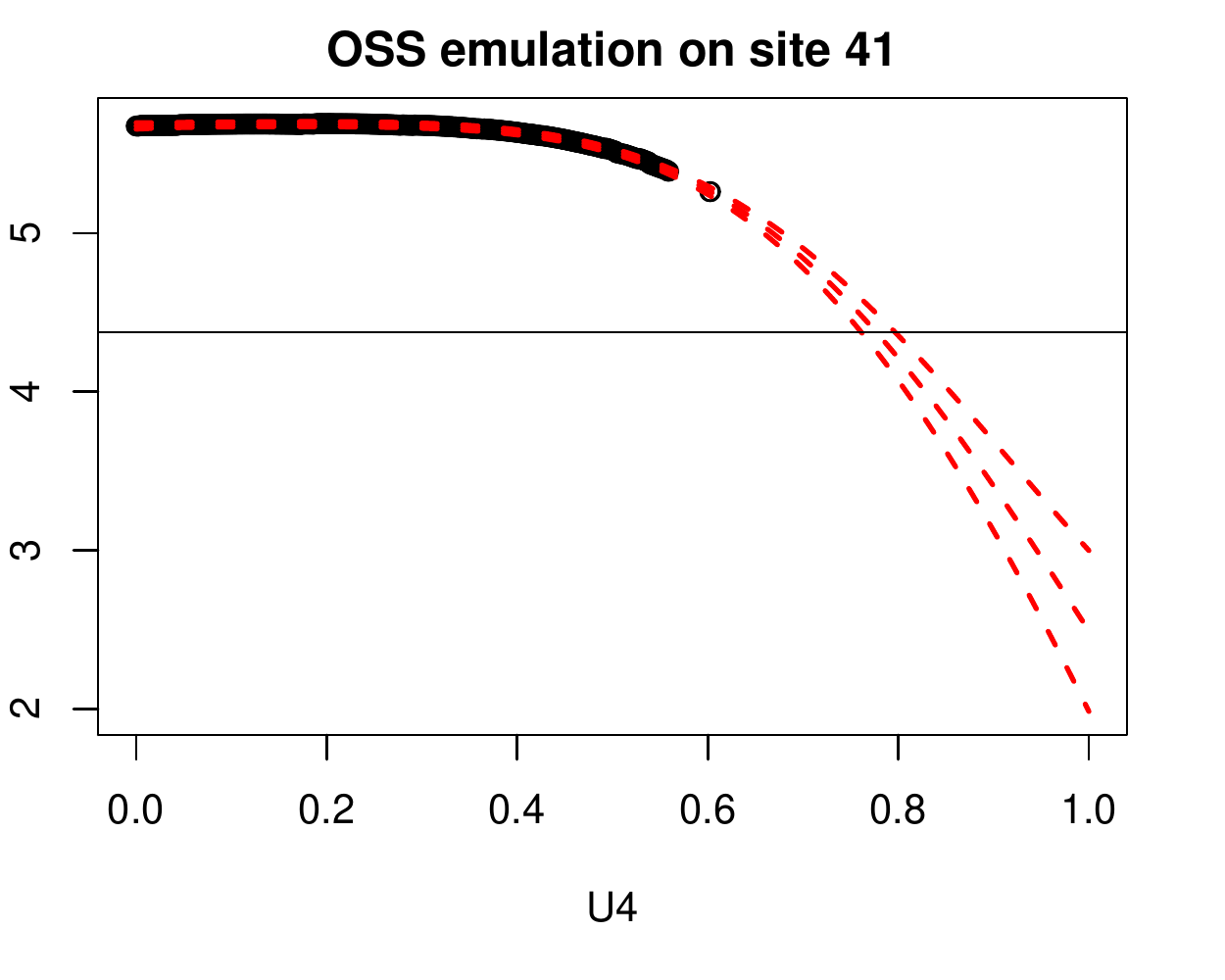}
\includegraphics[width=0.32\linewidth, trim=0 10 20 5, clip=TRUE]{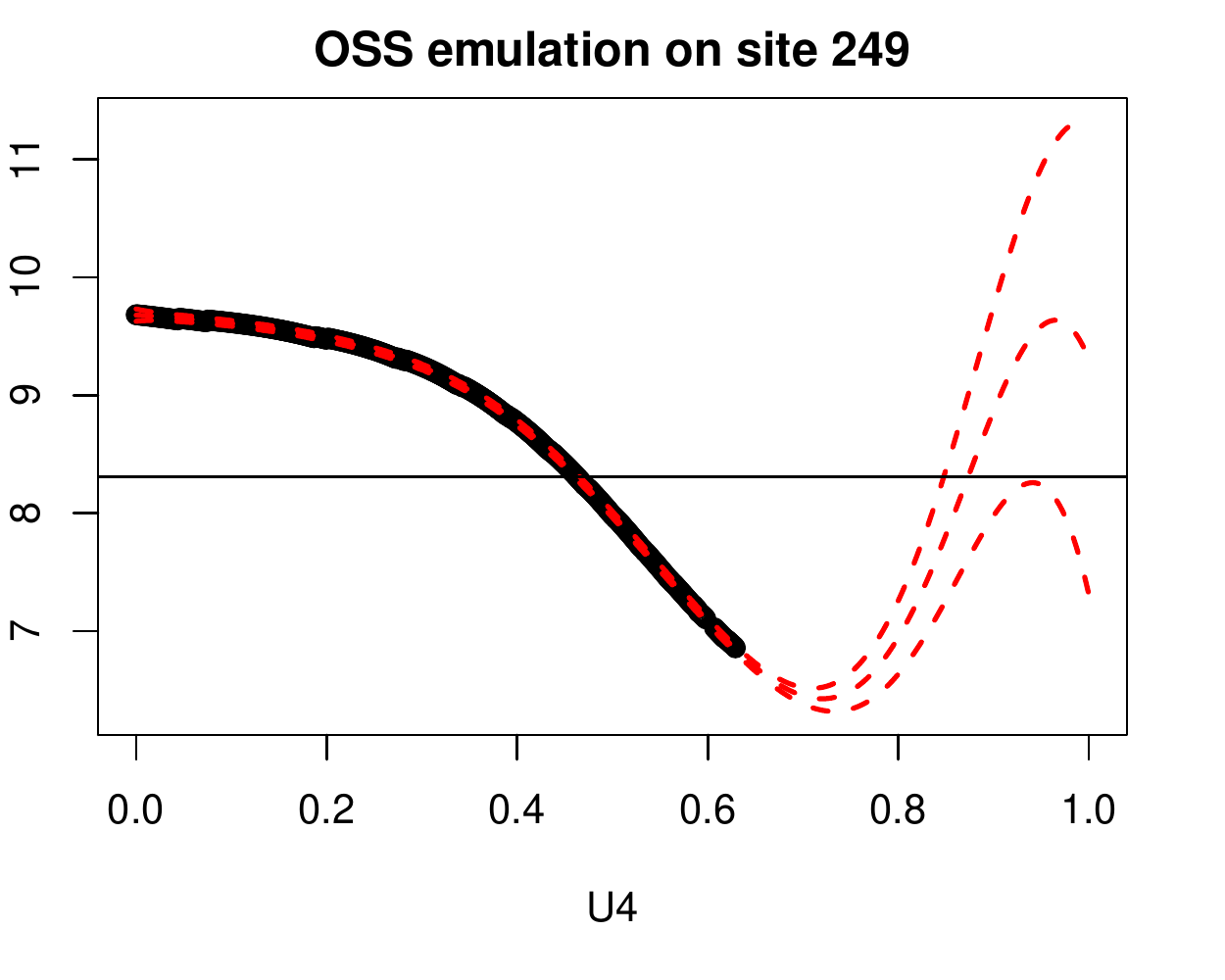}\\
\includegraphics[width=0.32\linewidth, trim=0 10 20 5, clip=TRUE]{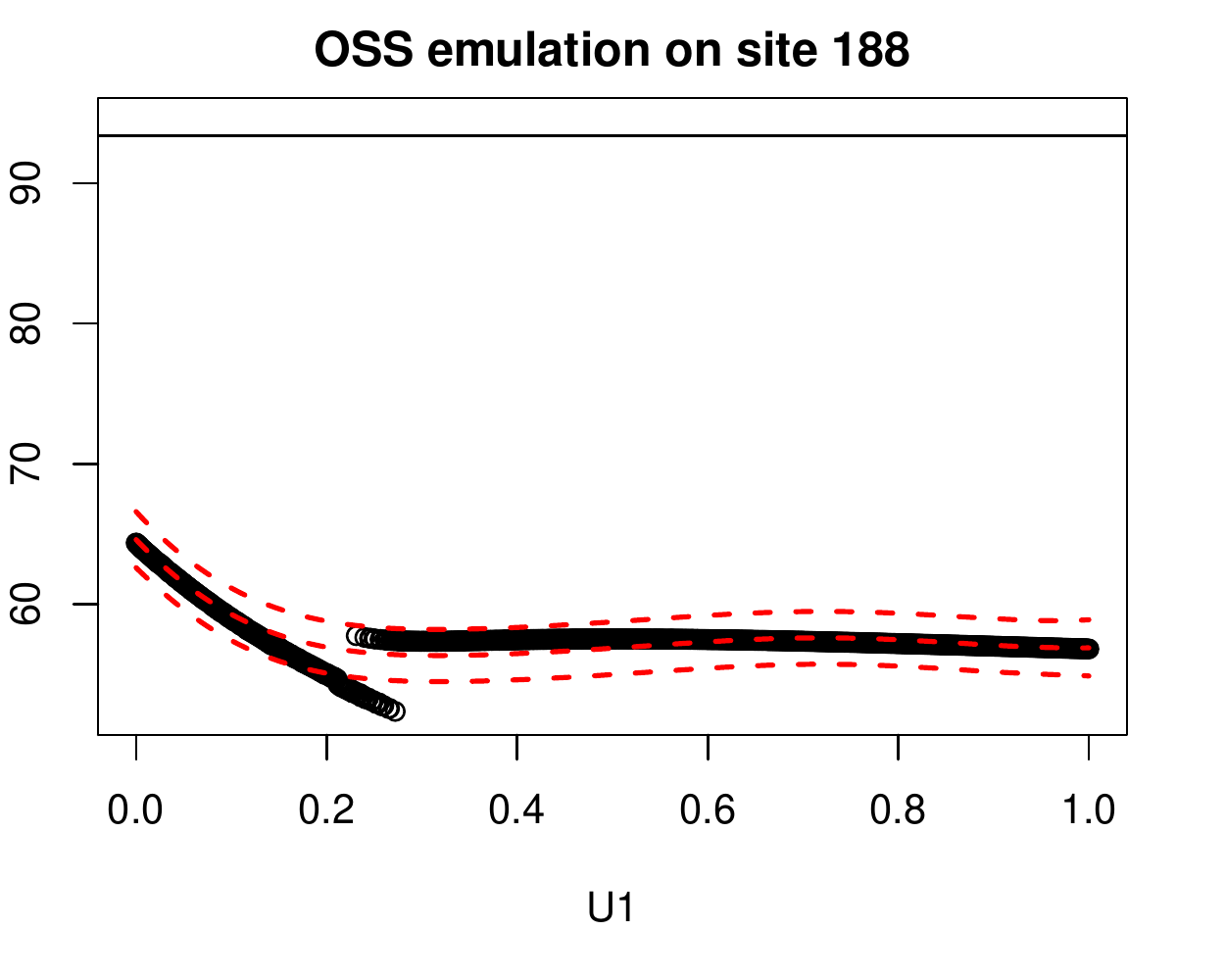}
\includegraphics[width=0.32\linewidth, trim=0 10 20 5, clip=TRUE]{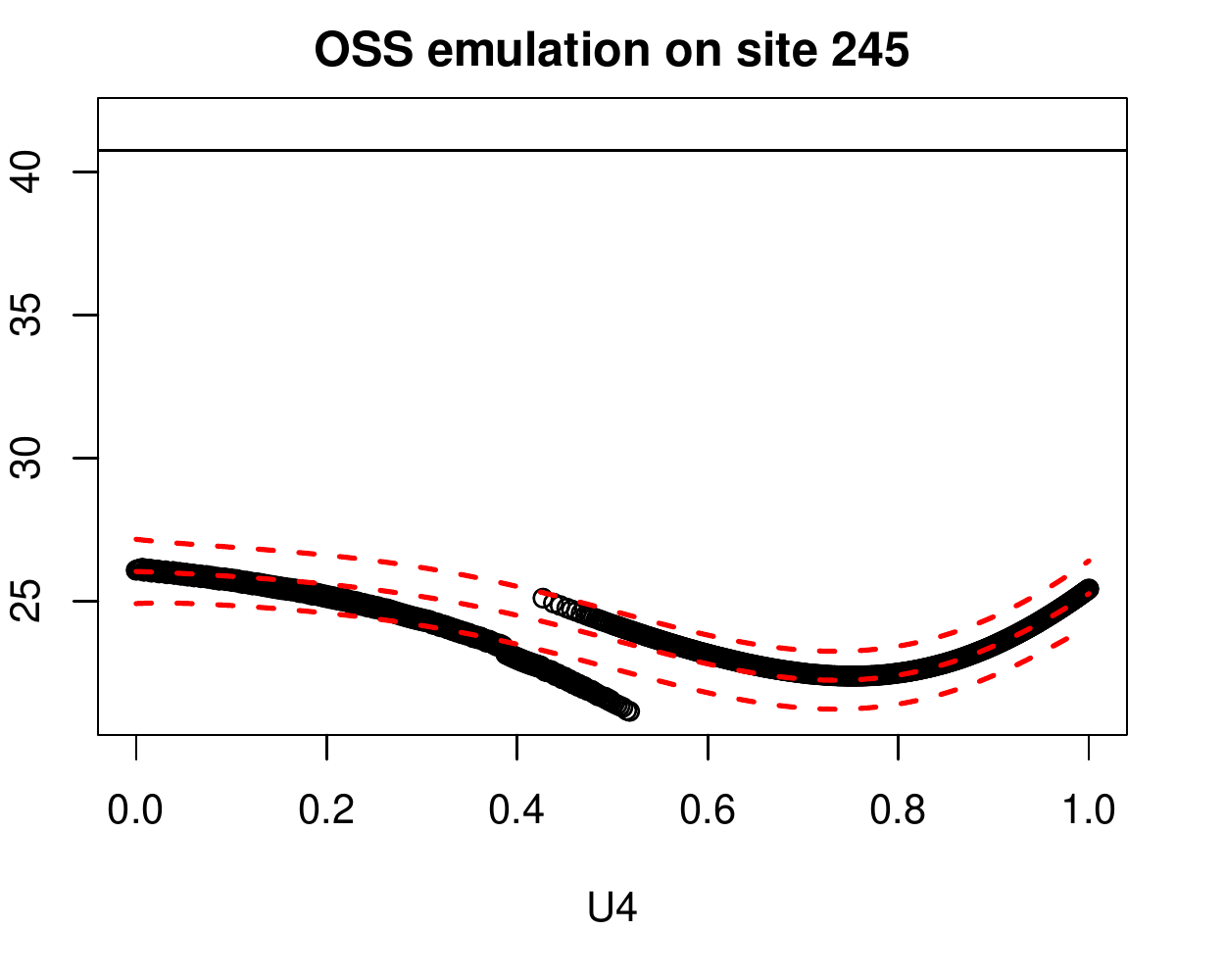}
\includegraphics[width=0.32\linewidth, trim=0 10 20 5, clip=TRUE]{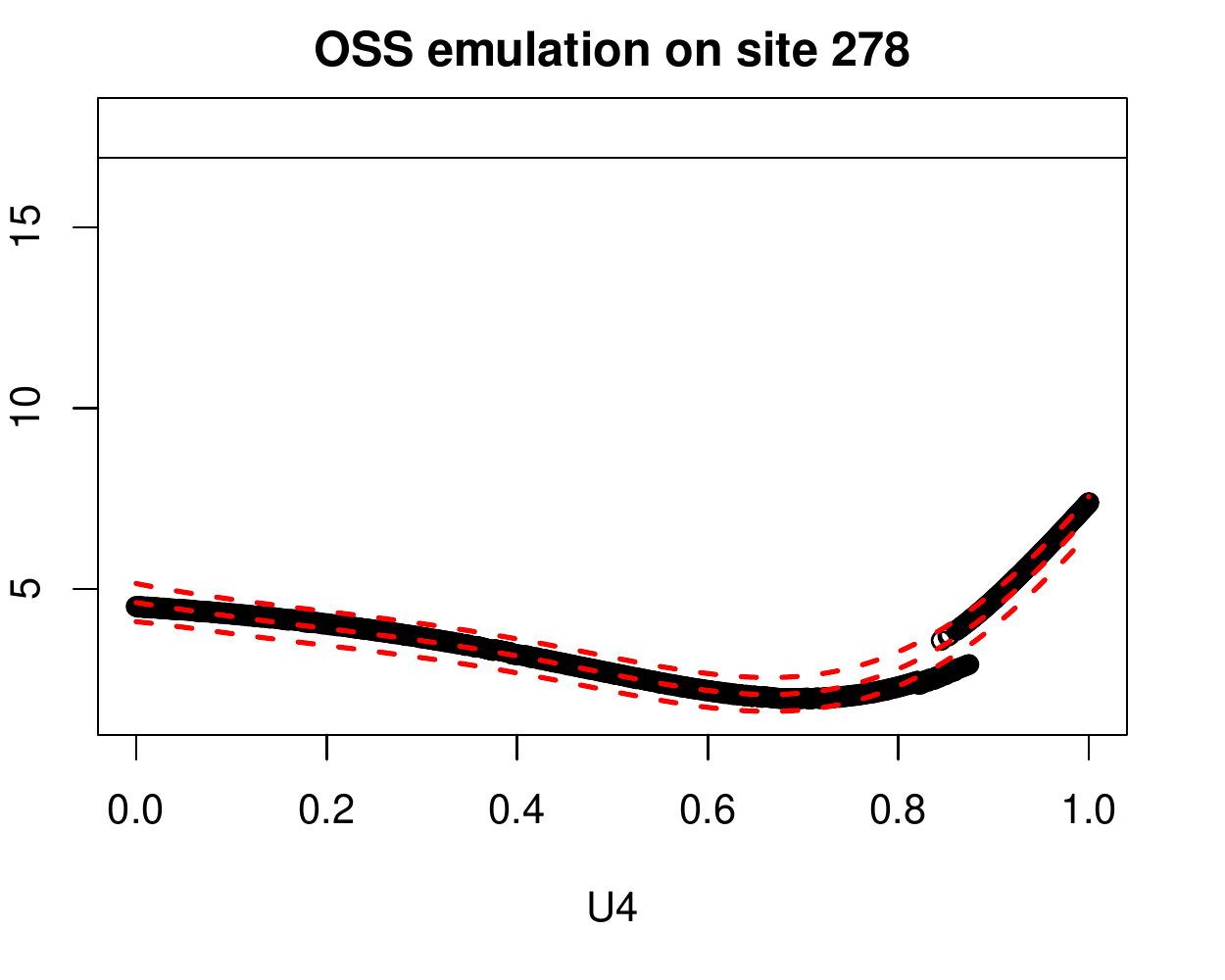}
  \caption{Profile plots of OSSs via predictive means and 95\% predictive
  intervals (dashed-red).  First row shows three well-behaved
  cases; middle row illustrates extrapolations to partially missing
  regimes; last row shows three cases where smoothing is required in order to
  cope with discontinuities. Red lines are the predicted mean 
  (solid) and 95\% predictive intervals (dashed). 
  Black horizontal lines show the field response
  $y_i^F$ at that location, $\mathbf{x}_i$, with $i$ provided in the main title.}
  \label{fig:loc2}
\end{figure}

Figure \ref{fig:loc2} supplements those results with a window into the
behavior of the OSSs, in three glimpses. The first row shows three relatively
well-behaved input settings by varying two $\mathbf{u}$-coordinates at
$\mathbf{x}_{17}^F$, and one at $\mathbf{x}_{243}^F$.  In all
three cases, the three dashed-red lines describing the predictive distribution
(via mean, and 95\% interval) completely cover the {\tt ISOTSEAL} simulations
in that space.  Both flat (middle panel) and wavier dynamics (outer
panels) are exhibited, demonstrating a degree of nonstationary flexibility.
The horizontal line indicates the field data $y_i^F$ value, and in two of
those cases there is a substantial discrepancy between $y^M(\mathbf{x}_i,
\mathbf{u})$, and $y_i^F$ for the range of $\mathbf{u}$-values on display. The
middle row in the figure shows what happens when {\tt ISOTSEAL} runs fail to
converge, again via two $\mathbf{u}$-coordinates for one OSS, at
$\mathbf{x}_{41}^F$, and one for another $\mathbf{x}_{249}^F$. Notice that
failures happen more often toward the edges of $\mathbf{u}$-space, but not
exclusively. In all three cases the extrapolations are sensible and reflect
diversity in waviness (first two flatter, third one wavier) that could not
be accommodated by a globally stationary model.  All three have the
corresponding $y_i^F$-value within range, but only in the extrapolated regime.
The last row of the figure shows how a nugget is used to smooth over
bifurcating regime changes in the output from {\tt ISOTSEAL},  offering a
sensible compromise and commensurately inflated uncertainty in order to cope
with both regimes.  All three cases map to outlying RMSE values (open
circles beyond the whiskers OSS boxplot) from Figure $\ref{fig:rmse}$.
Although they are among the hardest to predict out of sample, the overall
magnitude of the error is small.  Since the corresponding
$y_i^F$-values (horizontal lines) are far from $y^M(\mathbf{x}_i,
\mathbf{u})$, and $\hat{y}_i^M(\mathbf{u})$ in the $\mathbf{u}$-range under
study, a substantial degree of bias correction is needed to effectively
calibrate in this part of the input space.

\subsection{Calibration as optimization with on-site surrogates}
\label{sec:optim}

Even with accurate OSSs at all field data locations, Bayesian calibration can
still be computationally challenging in large-scale computer experiments. In
the KOH framework (\ref{eq:koh}), both $\mathbf{u}^*$ and a bias correcting GP
$b(\mathbf{x})$, via hyperparameters $\bm{\phi}_b$, are unknown and must jointly be
estimated.  The size of that parameter space, using a separable Gaussian
kernel (\ref{eq:kernel}) for $b(\cdot)$, is large (19d) in our
motivating honeycomb seal application.  MCMC in such a high-dimensional space
is fraught with computational challenges.

As an alternative to the fully Bayesian method, presented shortly in Section
\ref{sec:fullBayes} taking advantage of a sparse matrix structure, and to
serve as a smart initialization of the resulting MCMC scheme, we propose here
an adaptation of \citet{gra:etal:2015}'s modularized \citep{Liu:2009}
calibration as optimization.  Instead of sampling a full posterior distribution,
$\hat{b}(\cdot)$ and $\hat{\mathbf{u}}$ are calculated as
\begin{equation}
\hat{\mathbf{u}} = \mathrm{arg}\max_\mathbf{u} \left\{ p(\mathbf{u}) \left[ \max_{\bm{\phi}_b} 
p_b(\bm{\phi}_b \mid \mathbf{D}^{B}_{N_F}(\mathbf{u}))\right] \right\},
\label{eq:opt}
\end{equation}
which explores different values of $\hat{\mathbf{u}}$ via the resulting
posterior probability of discrepancy hyperparameters $p_b(\phi_b \mid
\mathbf{D}^{B}_{N_F}(\mathbf{u}))$ applied to a data set of residuals
$\mathbf{D}^B_{N_F}(\mathbf{u})$. Specifically,
$\mathbf{D}^{B}_{N_F}(\mathbf{u})=(\mathbf{X}_{N_F}^F,
\hat{\mathbf{y}}_{N_F}^{B|u})$ is the observed field inputs
$\mathbf{X}_{N_F}^F$ and discrepancies
$\hat{\mathbf{y}}_{N_F}^{B|\mathbf{u}}=\mathbf{y}_{N_F}^F-\hat{\mathbf{y}}_{N_F}^{M|\mathbf{u}}$
given a particular $\mathbf{u}$. The probability $p_b(\cdot \mid \cdot) $
refers to the marginal likelihood of the GP with parameters
$\hat{\bm{\phi}}_b$ fit to those residuals via their own ``inner''
derivative-based optimization routine. The object in Eq.~(\ref{eq:opt})
basically encodes the idea that $\mathbf{u}$-settings leading to better-fitting GP bias
corrections are preferred. A uniform prior $p(\mathbf{u})$ is a sensible
default; however, we prefer independent $u_j \sim \mathrm{Beta}(2,2)$
in each coordinate as a means of regularizing the search by mildly penalizing
boundary solutions, in part because we know that frictions factors at the
boundaries of $\mathbf{u}$-space lean heavily on the surrogate as runs of {\tt
ISOTSEAL} fail to converge there. Of course, any genuine prior information on
$\mathbf{u}$ could be used here to further guide the calibration.

Actually, this approach is not unlike the NLS one described in Section
\ref{sec:nls}, augmented with OSSs (rather than raw {\tt ISOTSEAL} runs) and
with bias correction. Instead of optimizing a least-squares criterion,  our GP
marginal likelihood-based loss is akin to a spatial Mahalanobis criterion
\citep{bastos:ohagan:2009}. In practice, the log of the criteria in
Eq.~(\ref{eq:opt}) can be optimized numerically with library methods
such as ``L-BFGS-B",  via {\tt optim} \citep{lbfgsb}, or  {\tt nloptr}
\citep{nloptr}. Since the optimizations are fast but local and since the surface
being optimized can have many local optima, we entertain a large set of
random restarts---in parallel---in search for the best (most global)
$\hat{\mathbf{u}}$ and $\hat{b}(\cdot)$.

To economize on space, we summarize here the outcome of this approach on the
honeycomb seal, alongside its fully Bayesian KOH analog. A more detailed
discussion of fully Bayesian calibration is provided in Section
\ref{sec:fullBayes} with results in Section \ref{sec:results}. As mentioned
above, our main use of this procedure is to prime the fully Bayesian KOH MCMC.
Foreshadowing somewhat, we can see from Figure \ref{fig:res} that the point
estimates $\hat{\mathbf{u}}$ so-obtained are not much different from the
maximum {\em a posteriori} (MAP) found via KOH, yet at a fraction of the
computational cost.  Since MCMC is inherently serial and our
randomly initialized optimizations may proceed in parallel, we can get a good
$\hat{\mathbf{u}}$ in about an hour, whereas getting a good (effective) sample
size from the posterior takes about a day.

\section{Fully Bayesian calibration via on-site surrogates}
\label{sec:fullBayes}

The approach in Section \ref{sec:optim} is Bayesian in the sense that 
marginal likelihoods are used to estimate hyperparameters to the GP-based
OSSs and discrepancy $b(\cdot)$, and priors are entertained for
the friction factors $\mathbf{u}$. However, the modularized approach to joint
modeling, via residuals from (posterior) predictive quantities paired with
optimization-based point \blu{estimation}, makes the setup a poor man's Bayes at
best.  In the face of big data---large $N_M$, $N_F$ and $p_u$---such a setup
may represent the only computationally tractable alternative. However, in our
setting with moderate $N_F$ and $N_M = \sum_{i=1}^{N_F} n_i$ composed of
independently modeled computer experiments of moderate size ($n_i \leq 1000$),
fully Bayesian KOH-style calibration is within reach.  As we show below, a
careful application of partition inverse identities allows the implicit
decomposition of a huge matrix via its sparse structure.

\subsection{KOH setup using OSS}

Our OSSs from Section \ref{sec:oss} are trained via $p_u$-dimensional on-site
designs $\mathbf{U}_1, \mathbf{U}_2, \dots, \mathbf{U}_{N_F}$.  Their row
dimension, $n_i \leq 1000$, depends on the proportion of {\tt ISOTSEAL} runs
that successfully completed.  Collect these $N_M = \sum_{i=1}^{N_F} n_i$
outputs of those simulations, each tacitly paired with inputs $\mathbf{x}_i$,
as $\mathbf{y}^M=(\mathbf{y}_1,
\mathbf{y}_2, \dots, \mathbf{y}_{N_F})^\top$. The KOH framework compensates
for surrogate biased computer model predictions under an unknown setting
$\mathbf{u}$ by estimating a discrepancy $b(\cdot)$ via $N_F$ field data runs
$\mathbf{y}^F$ observed at $N_F \times p_x$ inputs $\mathbf{X}^F$: 
\[
\mathbf{y}^F=y^M(\mathbf{U}) + b(\mathbf{X}^F), \quad \mbox{where} \quad
\mathbf{U}=[\mathbf{u}^\top; \cdots; \mathbf{u}^\top]^\top
\]
stacks $N_F$ identical $p_u$-dimensional row vectors $\mathbf{u}^\top$.  Under joint
GP priors, for each of $N_F$ OSSs and $b(\cdot)$, the sampling model can be
characterized by the following multivariate normal (MVN) distribution.
\begin{align}
\begin{bmatrix} \mathbf{y}^M \\ \mathbf{y}^F \end{bmatrix}
=  \begin{bmatrix} \mathbf{y}_1 \\ \mathbf{y}_2\\ \vdots \\\mathbf{y}_{N_F} \\ \mathbf{y}^F 
\end{bmatrix}
= \begin{bmatrix} 
y_1(\mathbf{U}_1)\\ y_2(\mathbf{U}_2)\\ \vdots \\ y_{N_F}(\mathbf{U}_{N_F}) \\ y^M(\mathbf{U}) + b(\mathbf{X}^F) 
\end{bmatrix} \sim \mathcal{N}_{N_M + N_F}(\mathbf{0}, \mathbb{V}(\mathbf{u}))
\label{eq:d}
\end{align}
Generally speaking, $\mathbb{V}(\mathbf{u})$ would be derived by
hyperparameterized pairwise inverse distances between inputs on
$(\mathbf{x},\mathbf{u})$-space. In our OSS setup, however, it has a special
structure owing to the independent surrogates fit at each $\mathbf{x}_i$,
for $i=1,\dots,N_F$.

Let $\mathbf{V}_i \equiv V_i(\mathbf{U}_i, \mathbf{U}_i)$ denote the $n_i
\times n_i$ covariance  matrix for the $i^\mathrm{th}$ OSS, for example, following
Eq.~(\ref{eq:kernel}).  This notation deliberately suppresses dependence on
hyperparameters $\bm{\phi}_i$, which is a topic we table momentarily to
streamline the discussion here. Similarly, $V_b \equiv
V_b(\mathbf{X}^F)$.  Let $V_i(\mathbf{U}) \equiv V_i(\mathbf{U}, \mathbf{U}_i)$
be the $n_i \times N_F$ matrix of the $i^\mathrm{th}$ OSS's cross-covariances
between field data locations, paired with $\mathbf{u}$-values, and
$(\mathbf{x}_i, \mathbf{U}_i)$ design locations.  Since the $i^\mathrm{th}$
OSS is tailored to $\mathbf{x}_i$ only, independent of the other
$\mathbf{X}^F$, this matrix is zero except in the $i^\mathrm{th}$ row.
Let $\mathbf{v} \mathbb{I}_{N_F}$ be a $N_F \times N_F$ diagonal
matrix holding $V_i(\mathbf{u}', \mathbf{u}')$ values.  Although expressed as
a function of $\mathbf{u}'$ it is not actually a function of $\mathbf{u}'$
because the distance between $\mathbf{u}'$ and itself is zero.  Using
Eq.~(\ref{eq:kernel}) would yield $\mathbf{v} \mathbb{I}_{N_F} =
\mathrm{Diag}[\tau_i^2 (1 + \eta_i)]$.  With those definitions, we have the following:
\begin{equation}  
\mathbb{V}(\mathbf{u}) = \begin{bmatrix} \mathbf{V}_1 & \mathbf{0} & \mathbf{0} & \mathbf{0}& V_1(\mathbf{U})^\top \\ 
\mathbf{0} & \mathbf{V}_2 & \mathbf{0} & \mathbf{0} & V_2(\mathbf{U})^\top \\ 
\mathbf{0} & \mathbf{0} & \ddots & \mathbf{0} & \vdots \\ \mathbf{0} & \mathbf{0} & \mathbf{0} &  \mathbf{V}_{N_F} &  V_{N_F}(\mathbf{U})^\top \\
V_1(\mathbf{U}) &  V_2(\mathbf{U}) & \dots & V_{N_F}(\mathbf{U}) & \mathbf{v} \mathbb{I}_{N_F}  + V_b(\mathbf{X}^F)  \end{bmatrix} 
\equiv \begin{bmatrix} \mathbb{V}_{o} & \mathbb{V}^\top_{ob}(\mathbf{u})  \\  \mathbb{V}_{ob}(\mathbf{u}) & \mathbb{V}_{b}  \end{bmatrix}.  
\label{eq:Vd} 
\end{equation}

Although $\mathbb{V}(\mathbf{u})$ is huge, being $(N_M + N_F) \times (N_M +
N_F)$ or roughly $292292 \times 292292 > 85$ billion entries in our honeycomb
setup, it is sparse, having several orders of magnitude fewer nonzero
entries---about 292 million in our setup.  That is still too big, even for
sparse matrix manipulation.  Fortunately, the block diagonal structure makes
it possible to work with, via more conventional libraries. Toward that end,
denote by $\mathbb{V}_{o}=\Diag[\mathbf{V}_i(\mathbf{U}_i,
\mathbf{U}_i)]$ the huge $N_F \cdot (n_i \times n_i)$ upper-left block diagonal
submatrix from the OSSs.  Let $\mathbb{V}_{b}=\mathbf{v}\mathbb{I}_{N_F}  +
V_b(\mathbf{X}^F)$ represent the remaining (dense) lower-right block, corresponding to the bias.
Abstract by $\mathbb{V}_{ob}(\mathbf{u})$ and $V_{ob}^\top(\mathbf{u})$
the remaining, symmetric, rows and columns on the edges.  Recall that the
$V_i(\mathbf{U})$ therein are themselves sparse, comprising a single
row of nonzero entries.

Before detailing in Section \ref{sec:kohoss} how we use these blocks, first
focus on the specific operations required.  A fully Bayesian approach to
inference for $\mathbf{u}$ via posterior $p(\mathbf{u} \mid \mathbf{y}^M,
\mathbf{y}^F)
\propto p( \mathbf{y}^M, \mathbf{y}^F \mid \mathbf{u}) \cdot p(\mathbf{u})$ involves
evaluating an MVN likelihood
\begin{equation}
p( \mathbf{y}^M, \mathbf{y}^F \mid \mathbf{u}) \propto
|\mathbb{V}(\mathbf{u}) |^{-\frac{1}{2}} \times \exp \left\{-\dfrac{1}{2} 
\begin{bmatrix}\mathbf{y}^M \\ \mathbf{y}^F\end{bmatrix}^\top  
\mathbb{V}^{-1}(\mathbf{u}) \begin{bmatrix}\mathbf{y}^M \\ \mathbf{y}^F\end{bmatrix} \right\}.
\label{eq:lik}
\end{equation}
The main computational challenges are manifest in the inverse
$\mathbb{V}^{-1}(\mathbf{u})$ and determinant $|\mathbb{V}(\mathbf{u})|$
calculations, both involving $\mathcal{O}((N_M+N_F)^3)$ flops in addition to
$\mathcal{O}((N_M+N_F)^2)$ storage, assuming a dense representation.  However,
substantial savings comes from the sparse structure (\ref{eq:Vd}) of
$\mathbb{V}(\mathbf{u})$ and that only a portion---the edges---involves
$\mathbf{u}$.

\subsection{On-site surrogate decomposition}
\label{sec:kohoss}

Partition inverse and determinant equations \citep[e.g.,][]{Petersen:2008}
provide convenient forms for the requisite decompositions of
$\mathbb{V}(\mathbf{u})$:
\begin{align}
\mathbb{V}^{-1}(\mathbf{u}) &=\begin{bmatrix} \mathbb{V}_{o} & \mathbb{V}_{ob}^\top(\mathbf{u})  \\  
\mathbb{V}_{ob}(\mathbf{u}) & \mathbb{V}_{b}  \end{bmatrix} ^{-1} \nonumber 
\!\!\!\!\!\!
 = \begin{bmatrix} \mathbb{V}_{o}^{-1} + \mathbb{V}_{o}^{-1} \mathbb{V}_{ob}^\top(\mathbf{u})  
\mathbb{C}^{-1}(\mathbf{u}) \mathbb{V}_{ob}(\mathbf{u}) \mathbb{V}_{o}^{-1} 
& - \mathbb{V}_{o}^{-1}\mathbb{V}_{ob}^\top(\mathbf{u})  \mathbb{C}^{-1}(\mathbf{u}) \\ 
-\mathbb{C}^{-1}(\mathbf{u}) \mathbb{V}_{ob}(\mathbf{u}) \mathbb{V}_{o}^{-1} 
& \mathbb{C}^{-1}(\mathbf{u})\end{bmatrix} \\
|\mathbb{V}(\mathbf{u})| &=\det\begin{bmatrix} \mathbb{V}_{o} & \mathbb{V}_{ob}^\top(\mathbf{u}) \\  
\mathbb{V}_{ob}(\mathbf{u}) & \mathbb{V}_{b}  \end{bmatrix}= \det(\mathbb{V}_{o}) \times \det(\mathbb{C(\mathbf{u})}),
\label{eq:Vdet}
\end{align}
where $\mathbb{C}(\mathbf{u})= \mathbb{V}_{b} - \mathbb{V}_{ob}(\mathbf{u})
\mathbb{V}_{o}^{-1}\mathbb{V}_{ob}^\top(\mathbf{u})$.  Eq.~(\ref{eq:Vdet})
involves a potentially huge $N_M \times N_M$ component $\mathbb{V}_{o}$, with 
$N_M=286,282$ in the honeycomb example. Since it is block diagonal,
thanks to the OSS structure, we have
\begin{equation}
\mathbb{V}_{o}^{-1}
= \Diag[\mathbf{V}_i^{-1}]  
\quad \mbox{ and } \quad\det(\mathbb{V}_{o})= \prod_{i=1}^{N_F} \det[\mathbf{V}_i].
\label{eq:Vo}
\end{equation}
In this way, an otherwise $\mathcal{O}(N_M^3)$ operation may instead by calculated via
$N_F \times \mathcal{O}(n_i^3)$ calculations, potentially in parallel.  If some
$n_i$ are big, then the burden could still be substantial. However,
both are constant with respect to $\mathbf{u}$, so only one such
decomposition is required, even when entertaining thousands of potential
$\mathbf{u}$.  With $n_i
\leq 1000$ in our honeycomb application, these calculations require mere
seconds, even in serial.

Similar tricks extend to other quantities involved in Eq.~(\ref{eq:Vdet}).
Consider $\mathbb{V}_{o}^{-1}\mathbb{V}_{ob}^\top (\mathbf{u})$, which appears
multiple times in original and transposed forms.  We have
\begin{align}
\mathbb{V}_{o}^{-1}\mathbb{V}_{ob}^\top(\mathbf{u}) 
= \Diag [ \mathbf{V}_i^{-1} V_i(\mathbf{U})] =\Diag [\mathbf{h}_i(\mathbf{u})]
\quad \mbox{where} \quad \mathbf{h}_i(\mathbf{u}) = \mathbf{V}_i^{-1} V_i(\mathbf{u}),
\label{eq:VV}
\end{align}
and $V_i(\mathbf{u})$ is a vector holding the nonzero part of
$V_i(\mathbf{U})$.  In other words,
$\mathbb{V}_{o}^{-1}\mathbb{V}_{ob}^\top(\mathbf{u})$ is a $N_M \times N_F$
matrix comprising $N_F$ column vectors, whose $n_i$ nonzero entries
$\mathbf{h}_i$ follow a block structure for columns $i=1,\dots N_F$.
Each $\mathbf{h}_i(\mathbf{u})$ can be updated in parallel for new $\mathbf{u}$.

Next consider $\mathbb{C}(\mathbf{u}) = \mathbb{V}_{b} -\mathbb{V}_{ob}(\mathbf{u})
\mathbb{V}_{o}^{-1}\mathbb{V}_{ob}^\top(\mathbf{u})$, which appears in each block of
Eq.~(\ref{eq:Vdet}). $\mathbb{C}(\mathbf{u})$ is dense but is easy to compute
because it is just $N_F \times N_F$.  Recall from Eq~.\ref{eq:Vd} that
$\mathbb{V}_{b}=\mathbf{v}\mathbb{I}_{N_F}+{V} _b(\mathbf{X}^F)$, which
requires inversion only once because it is constant in $\mathbf{u}$. The next
part $\mathbb{V}_{ob}(\mathbf{u})
\mathbb{V}_{o}^{-1}\mathbb{V}_{ob}^\top(\mathbf{u})$ extends nicely from
$\mathbb{V}_{o}^{-1}\mathbb{V}_{ob}^\top(\mathbf{u}) =
\mathrm{Diag}[V_i(\mathbf{u})^\top \mathbf{h}_i(\mathbf{u})]$ following Eq.~(\ref{eq:VV}), an $N_F
\times N_F$ diagonal matrix whose entries can be calculated alongside the
$\mathbf{h}_i(\mathbf{u})$, similarly parallelized over $i=1,\dots,N_F$. 

Combining $\mathbb{V}_{o}^{-1}\mathbb{V}_{ob}^\top(\mathbf{u})$ and
$\mathbb{C}(\mathbf{u})$ results gives $\mathbb{V}_{o}^{-1}
\mathbb{V}_{ob}^\top(\mathbf{u}) \mathbb{C}^{-1}(\mathbf{u}) = \mathbf{H}(\mathbf{u})
\circ \mathbb{C}^{-1}(\mathbf{u})$, where ``$\circ$'' is the Hadamard product applied columnwise to $\mathbb{C}^{-1}(\mathbf{u})$ and where $\mathbf{H}(\mathbf{u}) = [\mathbf{h}_1(\mathbf{u}) ; \dots ;
\mathbf{h}_{N_F}(\mathbf{u})]$.  More concretely,
\[
\mathbb{V}_{o}^{-1}
\mathbb{V}_{ob}^\top(\mathbf{u}) \mathbb{C}^{-1}(\mathbf{u})
= \begin{bmatrix} 
c_{1,1}\mathbf{h}_1(\mathbf{u}) & c_{1,2}\mathbf{h}_1(\mathbf{u}) & \dots &  c_{1,N_F}\mathbf{h}_1(\mathbf{u}) \\ 
c_{2,1}\mathbf{h}_2(\mathbf{u}) & c_{2,2}\mathbf{h}_2(\mathbf{u}) & \dots & c_{2,N_F}\mathbf{h}_2(\mathbf{u}) \\ 
\vdots & \vdots & \ddots & \vdots  \\ 
c_{N_F,1}\mathbf{h}_{N_F}(\mathbf{u}) & c_{N_F,2}\mathbf{h}_{N_F}(\mathbf{u}) & \dots & c_{N_F,N_F}\mathbf{h}_{N_F}(\mathbf{u}) 
\end{bmatrix},
\]
where $c_{i,j}$ are scalar elements of $\mathbb{C}^{-1}(\mathbf{u})$.

Returning to Eq.~(\ref{eq:Vdet}), combining with Eq.~(\ref{eq:Vo}), establishes the determinant analog.
\begin{align}
|\mathbb{V}(\mathbf{u})|= \det(\mathbb{V}_{o}) \times \det(\mathbb{C}(\mathbf{u}))= 
\prod_{i=1}^{N_F} \det[\mathbf{V}_i] \times \det(\mathbb{C}(\mathbf{u}))
\end{align}
The first component, $\prod_{i=1}^{N_F} \det[\mathbf{V}_i(\mathbf{U}_i,
\mathbf{U}_i)]$, is composed of $\mathcal{O}(n_i^3)$ computations,
constant in $\mathbf{u}$.  Only the second component,
$\det(\mathbb{C}(\mathbf{u}))$ needs to be updated with new 
$\mathbf{u}$.

In summary, OSSs can be exploited to circumvent huge matrix
computations involved in likelihood evaluation (\ref{eq:lik}), yielding a
structure benefiting from a degree of precalculation, and from parallelization
if desired.  These features come on top of largely improved emulation accuracy
demonstrated in Section \ref{sec:merits}, compared with the global alternative.

\subsection{Priors and computation}
\label{sec:mcmc}

As briefly described in Section \ref{sec:optim}, we consider two priors on
$\mathbf{u}$, the friction factors in our honeycomb example.  The
first is independent uniform, $u_j \stackrel{\mathrm{iid}}{\sim}
\mathrm{Unif}(0,1)$. The second is $u_j
\stackrel{\mathrm{iid}}{\sim} \mathrm{Beta}(2,2)$ as a means of regularizing
posterior inference.  The marginal posterior for $\mathbf{u}$ is known to
sometimes concentrate on the boundaries $\mathbf{u}$-space, because of
identifiability challenges in the KOH framework \citep[see,
e.g.,][]{gra:etal:2015}.  Furthermore, we know that {\tt ISOTSEAL} is least
stable in that region.  $\mathrm{Beta}(2,2)$ slightly discourages that
boundary and commensurately elevates the posterior density of central values.
This choice has the added benefit of providing better mixing in the MCMC
described momentarily.

The coupled GPs involved in the KOH setup are hyperparameterized by scales,
lengthscales, and nuggets as in Eq.~(\ref{eq:kernel}).  A fully Bayesian
analysis would include these in the parameter space for posterior sampling,
augmenting the dimension by an order of magnitude in many cases.  In other
words, the posterior becomes $p(\mathbf{u}, \mathbf{\Phi} \mid \mathbf{y}^M,
\mathbf{y}^F)$, where $|\Phi| \in \mathcal{O}(p + p_x)$, $p=p_x+p_u$ 
for surrogate and $p_x$ for discrepancy, which would work out
to more than thirty parameters in our honeycomb example. Because of that high
dimensionality, a common simplifying tactic is to fix those $\mathbf{\Phi}$ at
their MLE or MAP setting $\hat{\mathbf{\Phi}}$, found via numerical optimization.  In
our OSS setup, with $N_F = 292$ independent surrogates, the burden of
hyperparameterization is exacerbated, with $|\Phi| \in
\mathcal{O}(N_F p_u + p_x)$ being several orders of magnitude higher in
dimension, over one thousand for honeycomb. This all but demands a setup where
point estimates are first obtained via maximization, as in Section
\ref{sec:optim}. That leaves only $\mathbf{u}$ for posterior sampling via
$p(\mathbf{u} \mid
\mathbf{y}^M, \mathbf{y}^F, \hat{\mathbf{\Phi}})$. Additionally, we
initialize our Monte Carlo search of the posterior with $\hat{\mathbf{u}}$
values found via Section \ref{sec:optim}.
 
Following KOH, we employ MCMC \citep{Hastings:1970, Gelfand:1990} to sample from
the posterior in a Metropolis-within-Gibbs fashion \cite[see,
e.g.,][]{hoff2009first}. Each Gibbs step utilizes a marginal random-walk
Gaussian proposal $u_j' = u_j + s_j$, $s_j
\stackrel{\mathrm{iid}}{\sim} \mathcal N(0, \sigma_j^2)$, $j=1,\dots, p_u$.
A pilot tuning stage was used to tune the $\sigma_j$, leading to
$\sigma=(0.02, 0.01, 0.2, 0.1)^\top$ in the honeycomb example.  Figures
\ref{fig:trace}--\ref{fig:res} in Section \ref{sec:calibresults} indicate
good mixing and adequate posterior exploration of the four-dimensional space
of friction factors.


\section{Empirical results}
\label{sec:results}

Before detailing the outcome of this setup on our motivating honeycomb
example, we illustrate the methodology in a more controlled setting.

\subsection{Illustrative example}

\begin{figure}[ht!]
\centering
\includegraphics[width=0.32\linewidth, trim=0 10 20 15, clip]{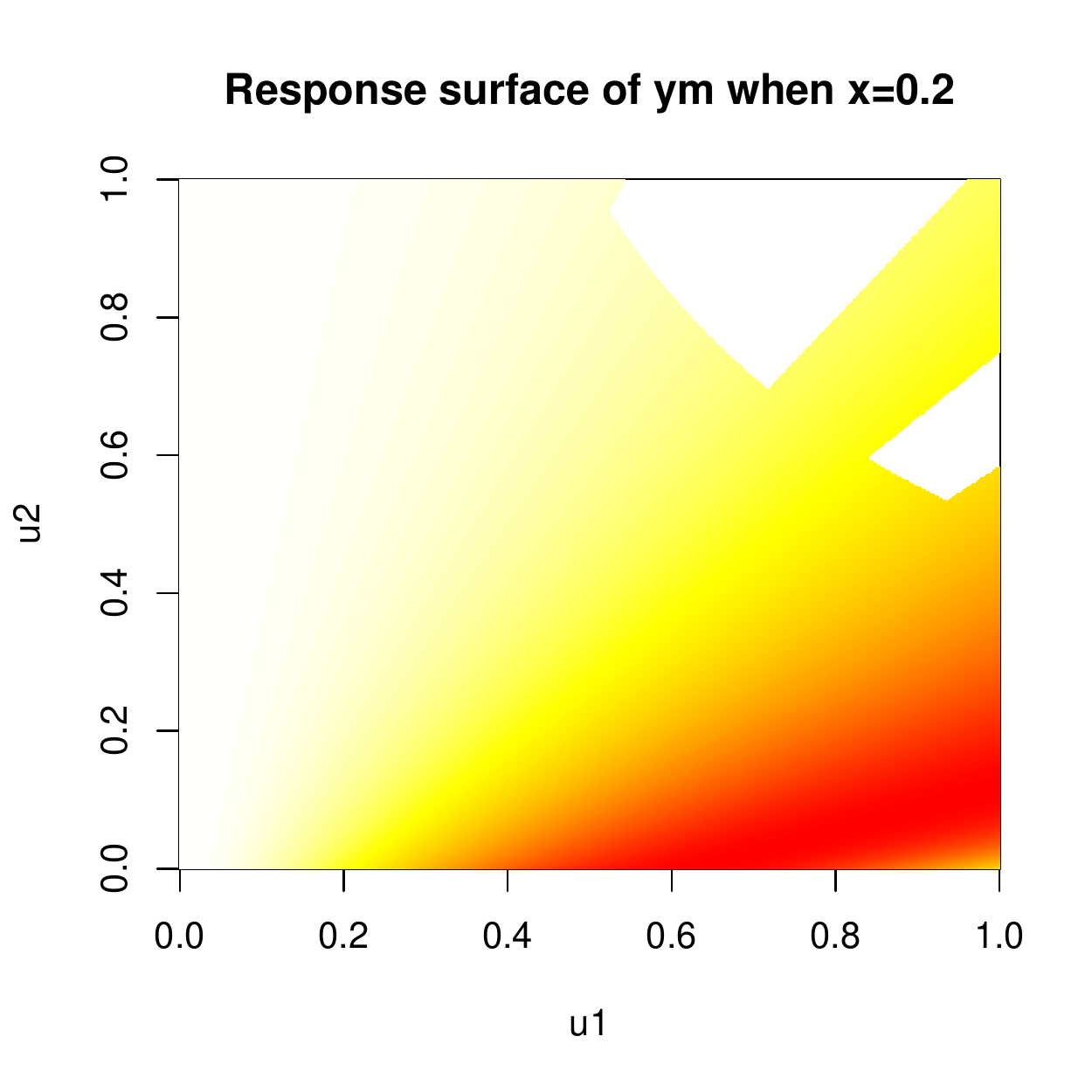}
\includegraphics[width=0.32\linewidth, trim=0 10 20 15, clip]{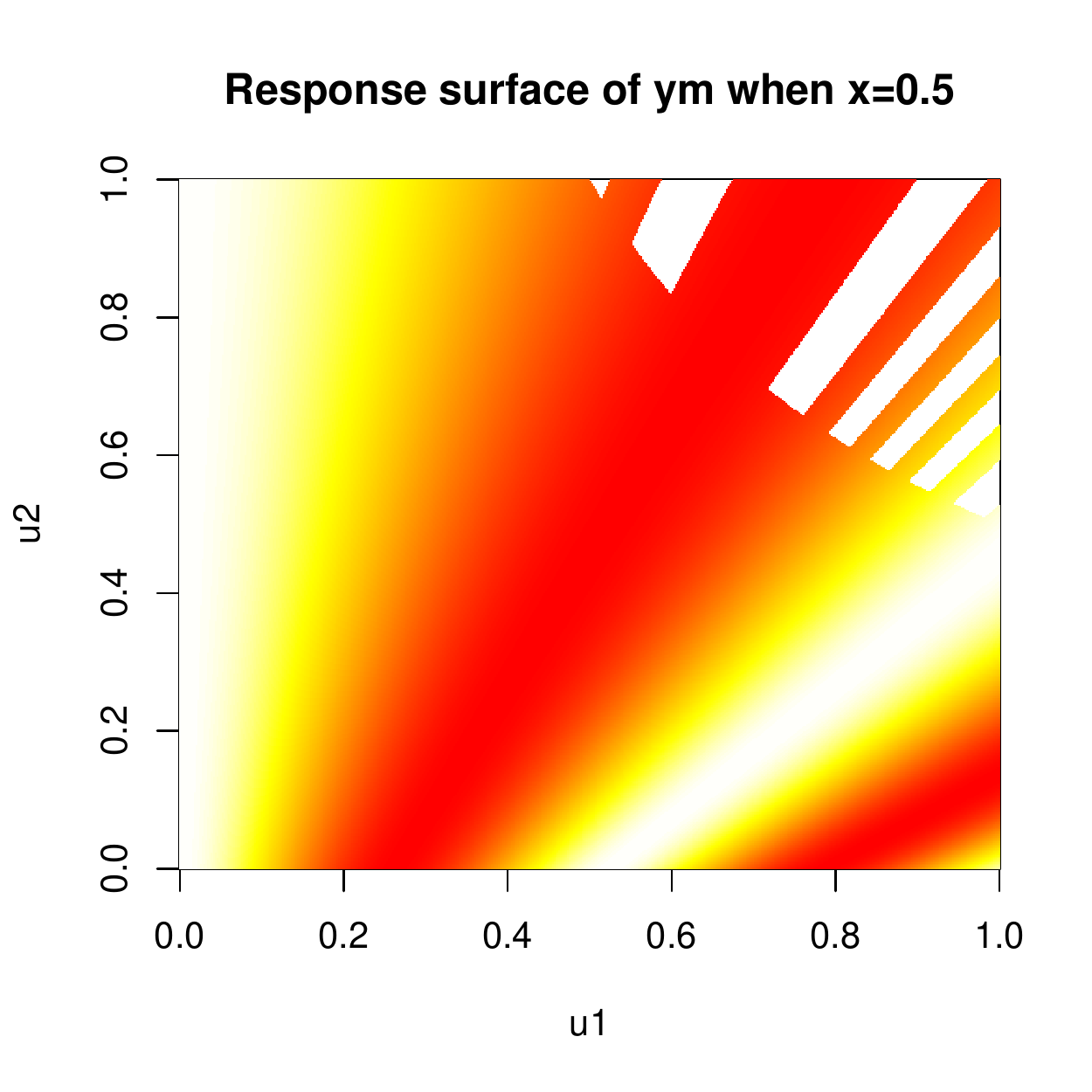}
\includegraphics[width=0.32\linewidth, trim=0 10 20 15, clip]{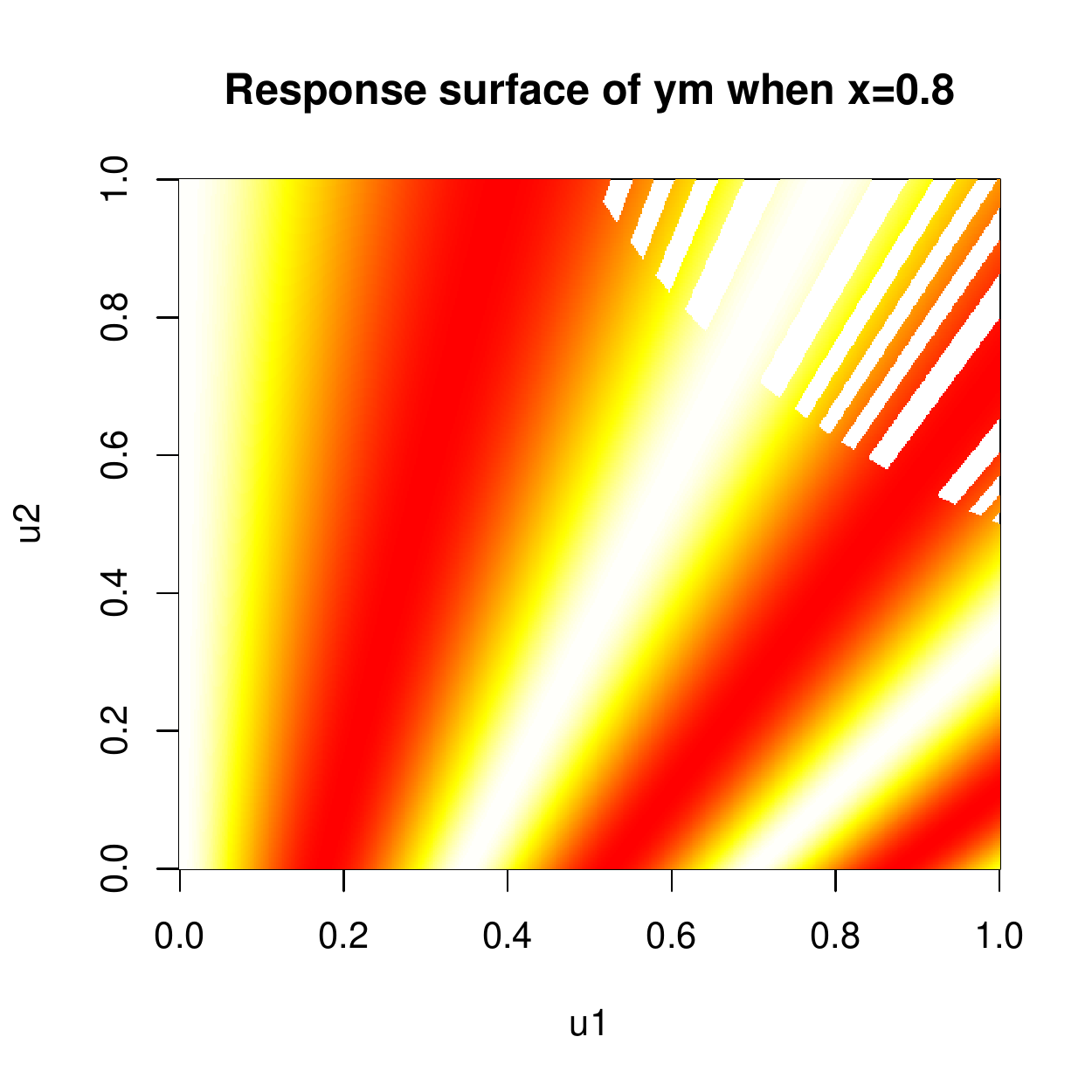}
\caption{Response surfaces illustrating computer model following (\ref{eq:toy2}) with
missing values under three settings of $x$.}
\label{fig:toy}
\end{figure}

Consider a mathematical model $y^{M^*}(\cdot)$ with three inputs $(x, u_1, u_2)$, following
\begin{equation}
y^{M^*}(x, u_1, u_2) = \cos \left( \dfrac{25 \sin(x)\times x \times u_1}{x + u_2} \right),
\label{eq:toy2}
\end{equation}
where $x
\in [0, 1]$ is a one-dimensional field input and $\mathbf{u}=(u_1, u_2) \in
[0, 1]^2$ are two-dimensional calibration parameters.  Suppose the real process follows
\[
y^R(x) = y^{M^*}(x, 0.8, 0.2) + b(x) \quad \mbox{where} \quad b(x) = \sin(4x).
\]
Mimicking the features of {\tt ISOTSEAL}, suppose the computer model $y^M$ is
unreliable in its evaluation of the mathematical model $y^{M^*}$, sometimes
returning {\tt NA} values.  Specifically, suppose the response is missing when
the $\mathbf{u}$ input is in its upper quartile, $u1 \times u2 > 0.5$, and 
$[5y^{M^*}] \mod 2 \equiv 0$, where $[\cdot]$ rounds to the nearest integer.
Figure \ref{fig:toy} provides an illustration.
Each panel in the figure shows the response as a function of $(u_1, u_2)$
for a different setting of $x$.  Observe the nonstationary dynamics manifest
in increasing waviness of the surface as $x$ increases.  Similarly, the
pattern of missingness becomes more complex for increasing $x$.  Therefore, a
global surrogate would struggle on two fronts: with stationarity as well as
with (nonmissing) coverage of the design in $\mathbf{u}$-space.

Now consider observing $N_F$ field realizations of $y^R(x) + \varepsilon$,
where $\varepsilon \sim \mathcal{N}(0, 0.02^2)$, under a maximin LHS in
$x$-space, and two variations on a computer experiment toward a calibrated
model. The first involves a global GP surrogate fit to $N_M = 500$ computer
model evaluations via a maximin LHS in $(x, \mathbf{u})$-space, where 33
(6.6\%) came back {\tt NA}. The second uses OSSs trained on $n_i = 200$
maximin LHSs in $\mathbf{u}$-space, paired with $x_i^F$ for $i=1,\dots,N_F$.
Of the $N_M = 2,000$ such simulations, 95 came back missing (4.75\%).  The
sizes of these computer experiment designs were chosen so that the computing
demands required for the global and OSS surrogates were commensurate.  Counting
flops, the global approach requires about $500^3 = 1.25 \times 10^8$, whereas
the OSSs need $10\times 200^3 = 8\times 10^7$, which can be 10-fold
parallelized if desired.

\begin{figure}[ht!]
  \centering
 \begin{minipage}{6cm}
\includegraphics[scale=0.5,trim=10 40 10 50,clip]{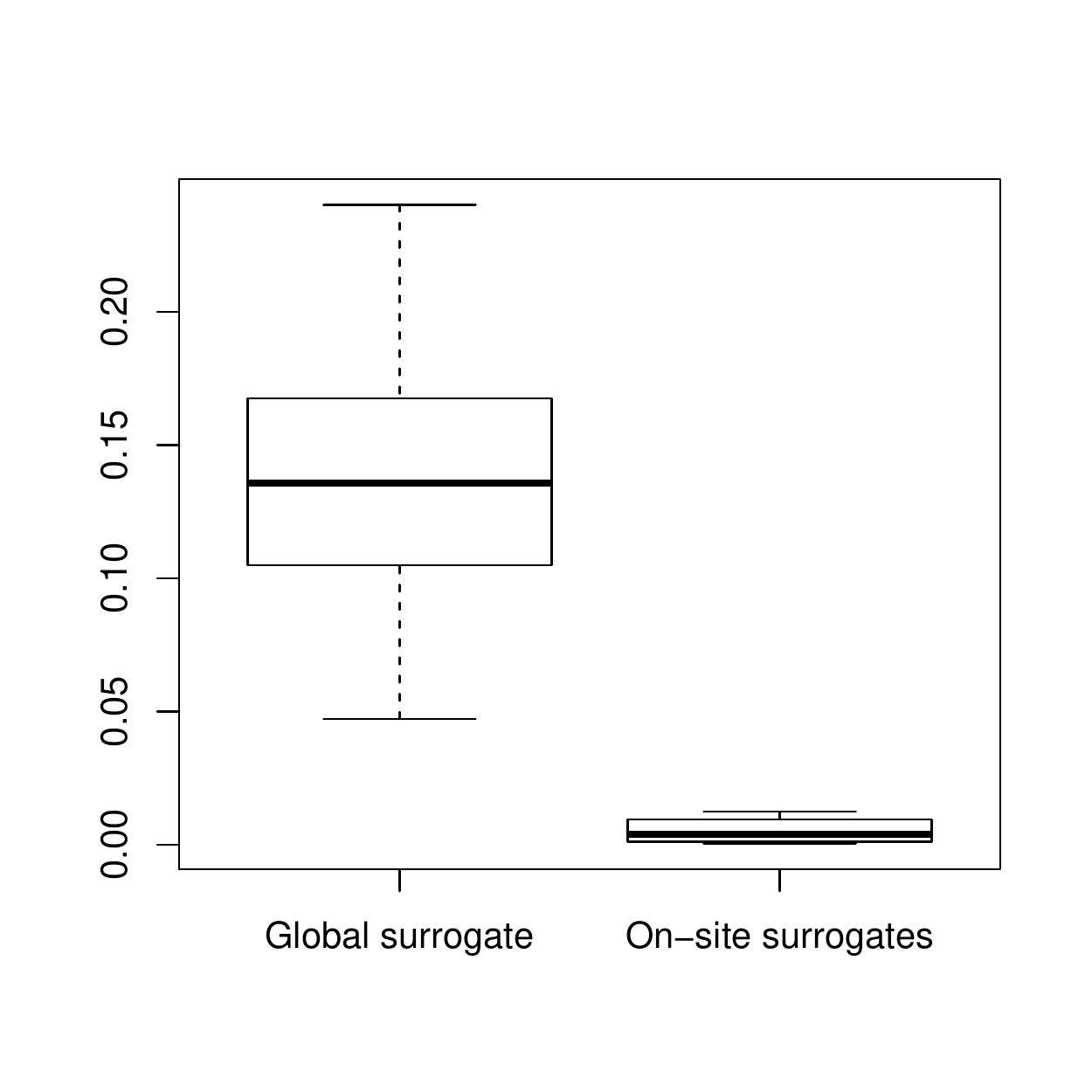}
\end{minipage} \hspace{1cm}
\begin{tabular}{r|rr}
& global & OSS \\
\hline
min & 0.0472 & 0.0004 \\
25\% & 0.1090 & 0.0012 \\
med & 0.1357 & 0.0039 \\
mean &  0.1398 &  0.0052 \\
75\% & 0.1673 & 0.0090 \\
max & 0.2403 &  0.0124
\end{tabular} 
\caption{Boxplots of 10 out-of-sample RMSEs, where each RMSE is computed by using novel  
$n_i' \leq 200$, for $i=1,\dots,N_F$.}
  \label{fig:toyrmse}
\end{figure}

Before turning to calibration, consider first the accuracy of the two
surrogates.  Mirroring Figure \ref{fig:rmse} for {\tt ISOTSEAL} in our
honeycomb example, Figure \ref{fig:toyrmse} shows the result of an
out-of-sample comparison of otherwise identical design.
The story here is similar to the one for {\tt ISOTSEAL}. Clearly, the
OSSs are more accurate.  They are better able to capture the nonstationarity
nature of computer model $y^M(\cdot, \cdot)$ nearby to the field sites.

\begin{figure}[ht!]
\centering
\includegraphics[width=0.425\linewidth, trim=0 10 20 15, clip]{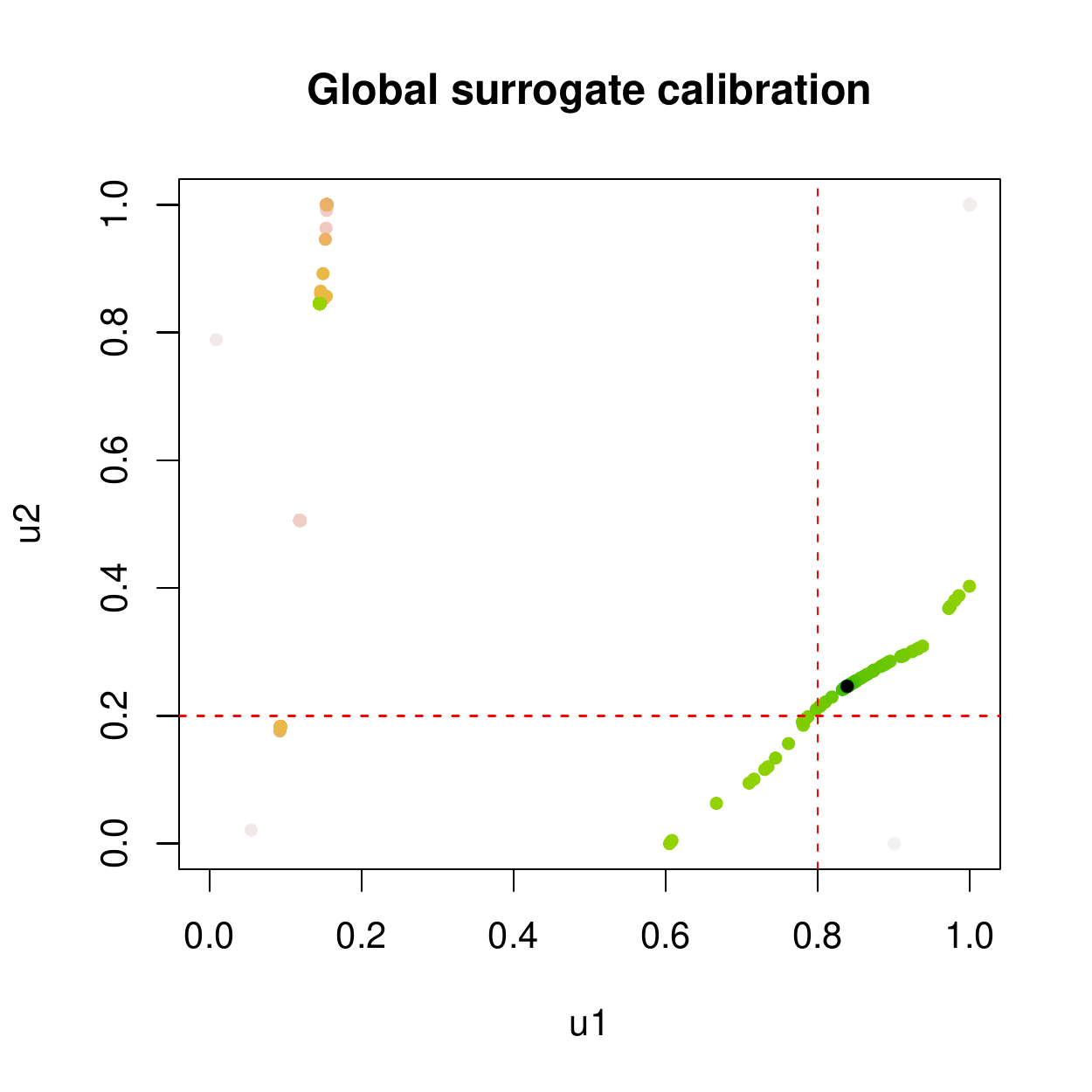}
\includegraphics[width=0.425\linewidth, trim=0 10 20 15, clip]{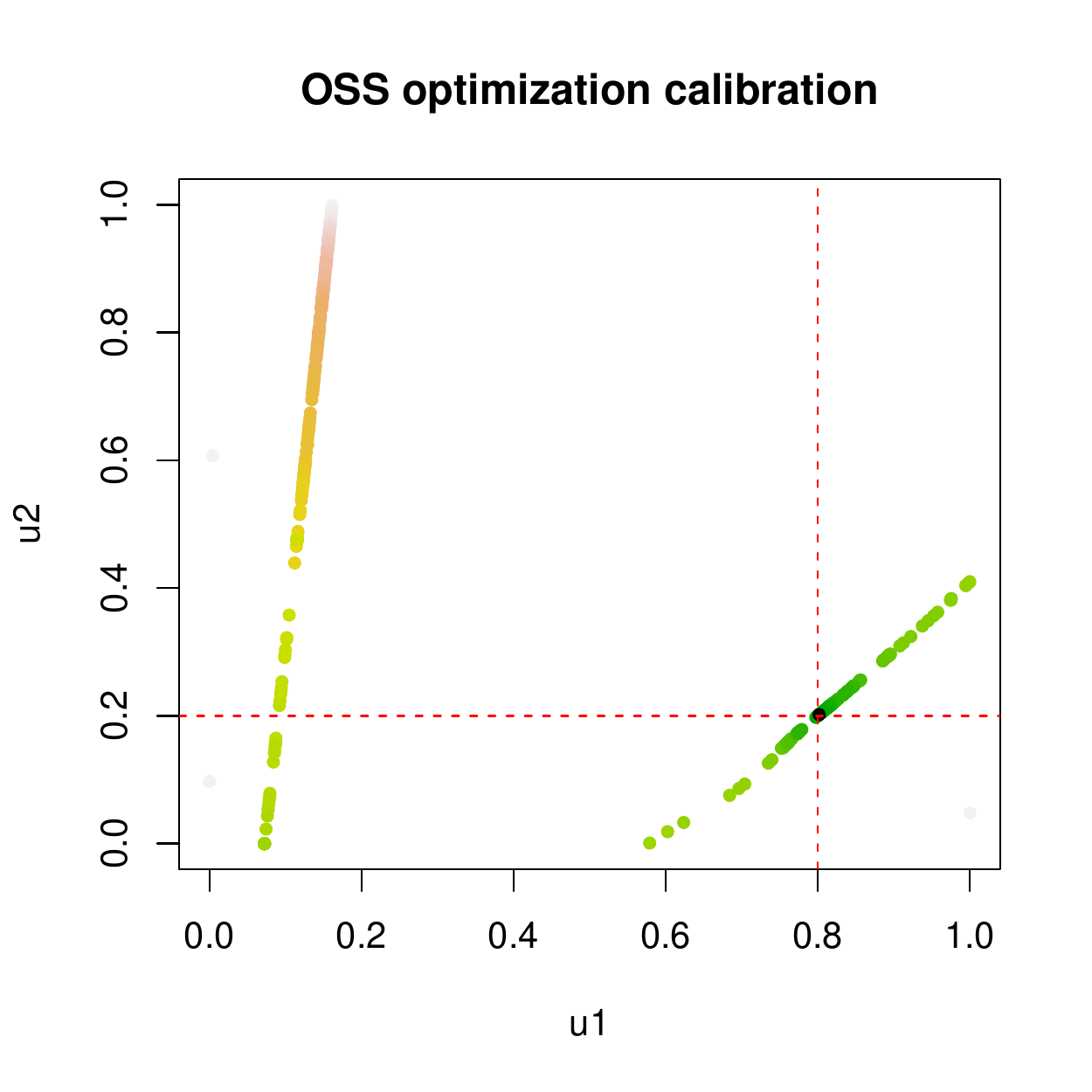} \\
\includegraphics[width=0.425\linewidth, trim=0 10 25 15, clip]{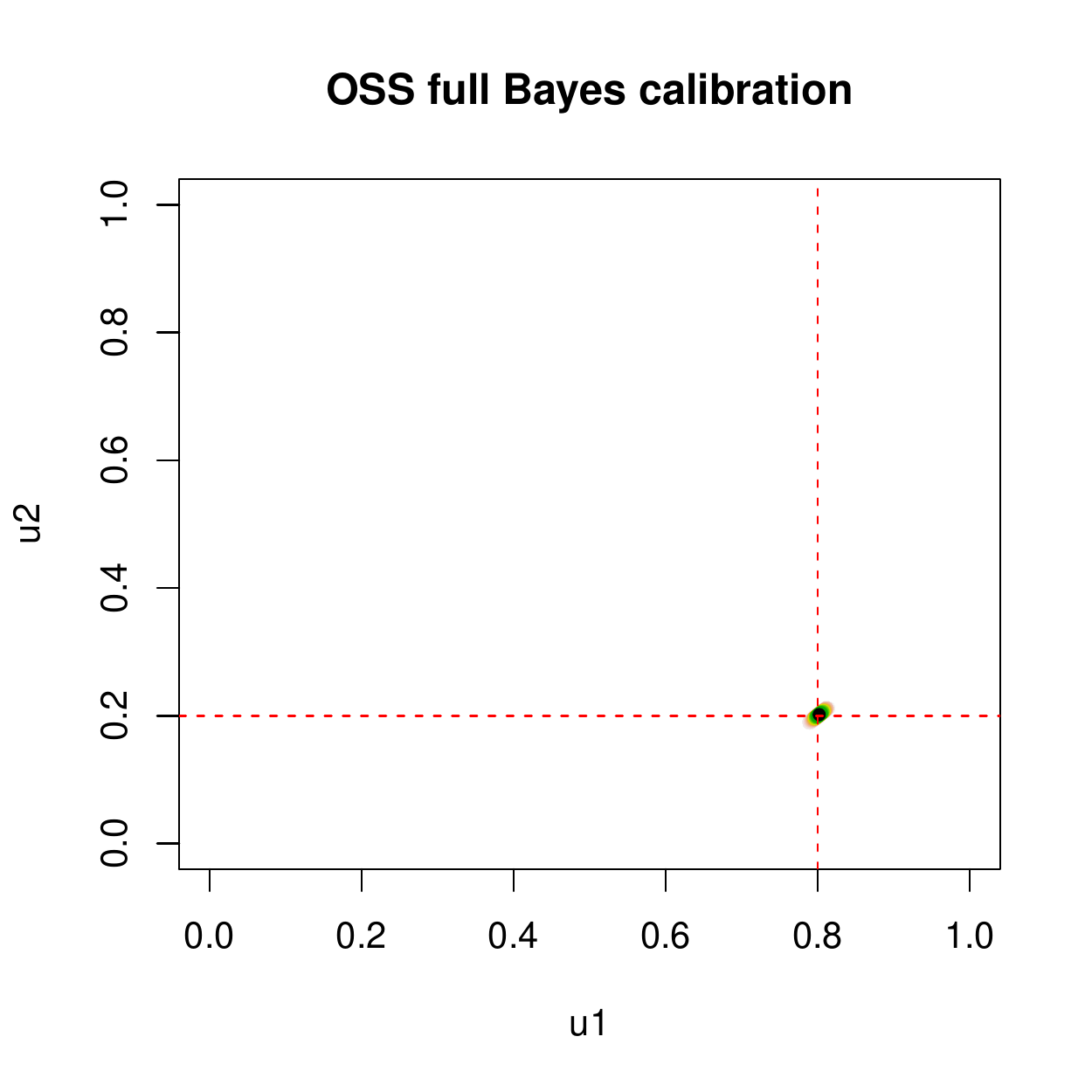}
\includegraphics[width=0.425\linewidth, trim=0 10 25 15, clip]{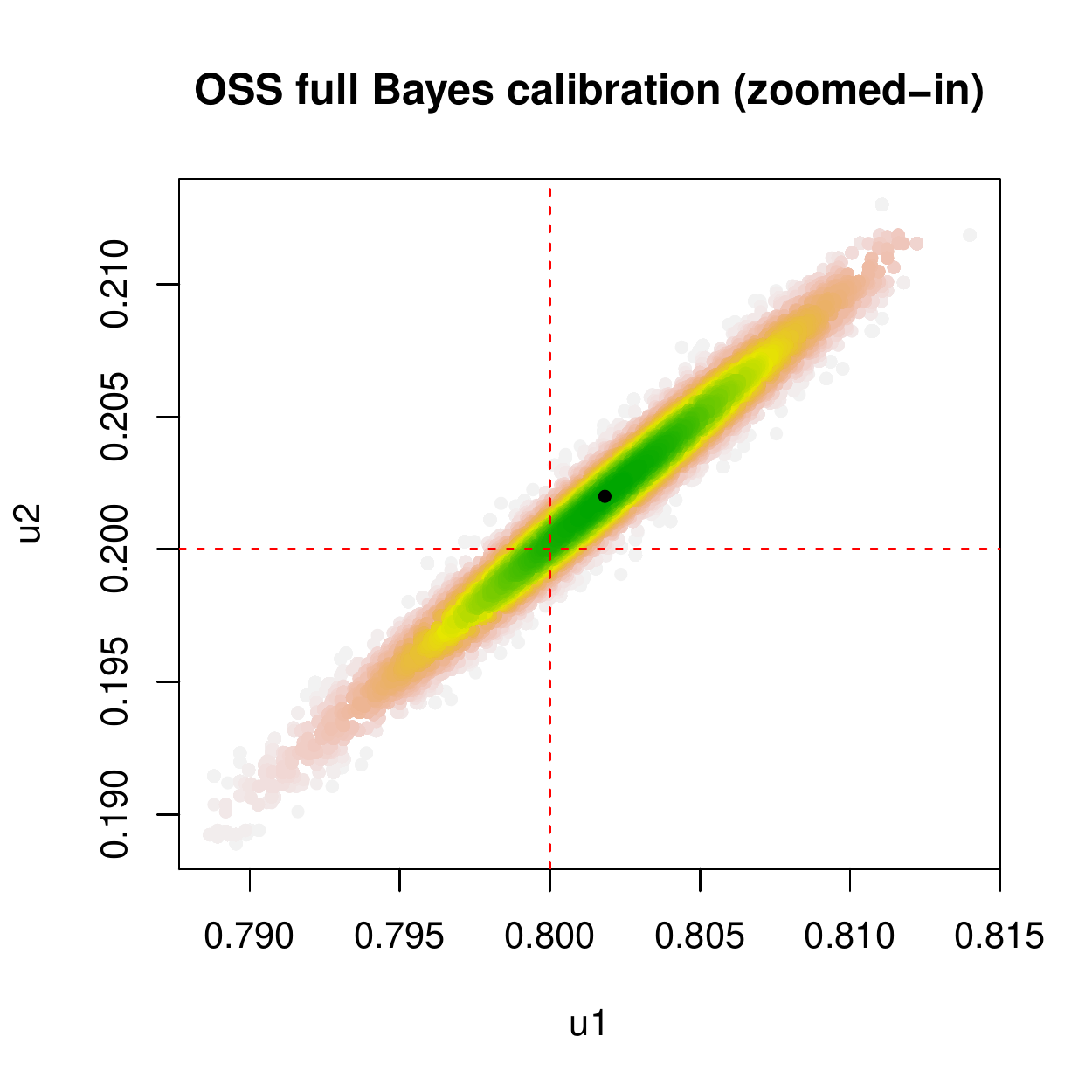}
\caption{Calibration results from optimization and full Bayes for the
toy example. Terrain colors are derived from ranks of log-scaled
posteriors as a visual aid; black dots indicate the MAP setting; red dashed
lines are the true values of calibration parameters, $\hat{\mathbf{u}}$.}%
\label{fig:toyresult}%
\end{figure}

Next, we compare calibration results from global surrogate optimization, OSSs
via modularization/optimization [Section \ref{sec:optim}, and OSSs via full
Bayes [Section \ref{sec:fullBayes}]. In this simple toy example, uniform
priors $u_i \stackrel{\mathrm{iid}}{\sim}\mathrm{Unif}(0,1)$ are sufficient
for good performance. The first row of Figure \ref{fig:toyresult} shows the
distributions of converged $\hat{\mathbf{u}}$ via Eq.~(\ref{eq:opt}) from the
optimization approach described in Section \ref{sec:optim}. The left panel
corresponds to the lower-fidelity global surrogate and the right panel to the
higher-fidelity OSSs.  Converged solutions from 500 random initializations
are shown. Terrain colors on the ranked log posteriors are provided to aid in
visualization. The best single coordinate $\hat{\mathbf{u}}$ is indicated by
the black dot. For comparison, the true $\mathbf{u}^*$ value is shown as
red-dashed crosshairs.  Although the best $\hat{\mathbf{u}}$ values found
cluster near the truth, both are sometimes fooled by a posterior ridge in
another quadrant of the space.
The second row of Figure \ref{fig:toyresult} shows the  posterior distribution
of $\mathbf{u}$ in full (left) and zoomed-in ranges (right). Compared with the
OSS-based optimization approach, the KOH analog found $\mathbf{u}$'s
tightly coupled around the truth.

In this simple example, posterior uncertainty is low, in part because a
relatively large computer experiment could be entertained in a
small input dimension. In fact all three methods worked reasonably well.
However, as we entertain more realistic settings, such as the honeycomb in 17
dimensions, only the methods based on OSSs are viable computationally
(assuming a relatively dense sampling of the computer model is viable).

\subsection{KOH versus modularized optimization: on honeycomb}
\label{sec:calibresults}
Here we return to our motivating honeycomb seal example, first providing a
qualitative comparison between our two approaches based on OSSs, via
modularized optimization [Section \ref{sec:optim}] and KOH [Section
\ref{sec:fullBayes}].  We then turn to an out-of-sample comparison, pitting
the KOH framework against the initial NLS analysis.  Throughout, we use a
regularizing independent $\mathrm{Beta}(2,2)$ prior on the components of
$\mathbf{u}$.  Appendix \ref{ap:unif} provides an analog presentation under a
uniform prior, accompanied by a brief discussion.

\begin{figure}[ht!]
\centering
\includegraphics[width=0.425\linewidth, trim=0 60 15 50, clip]{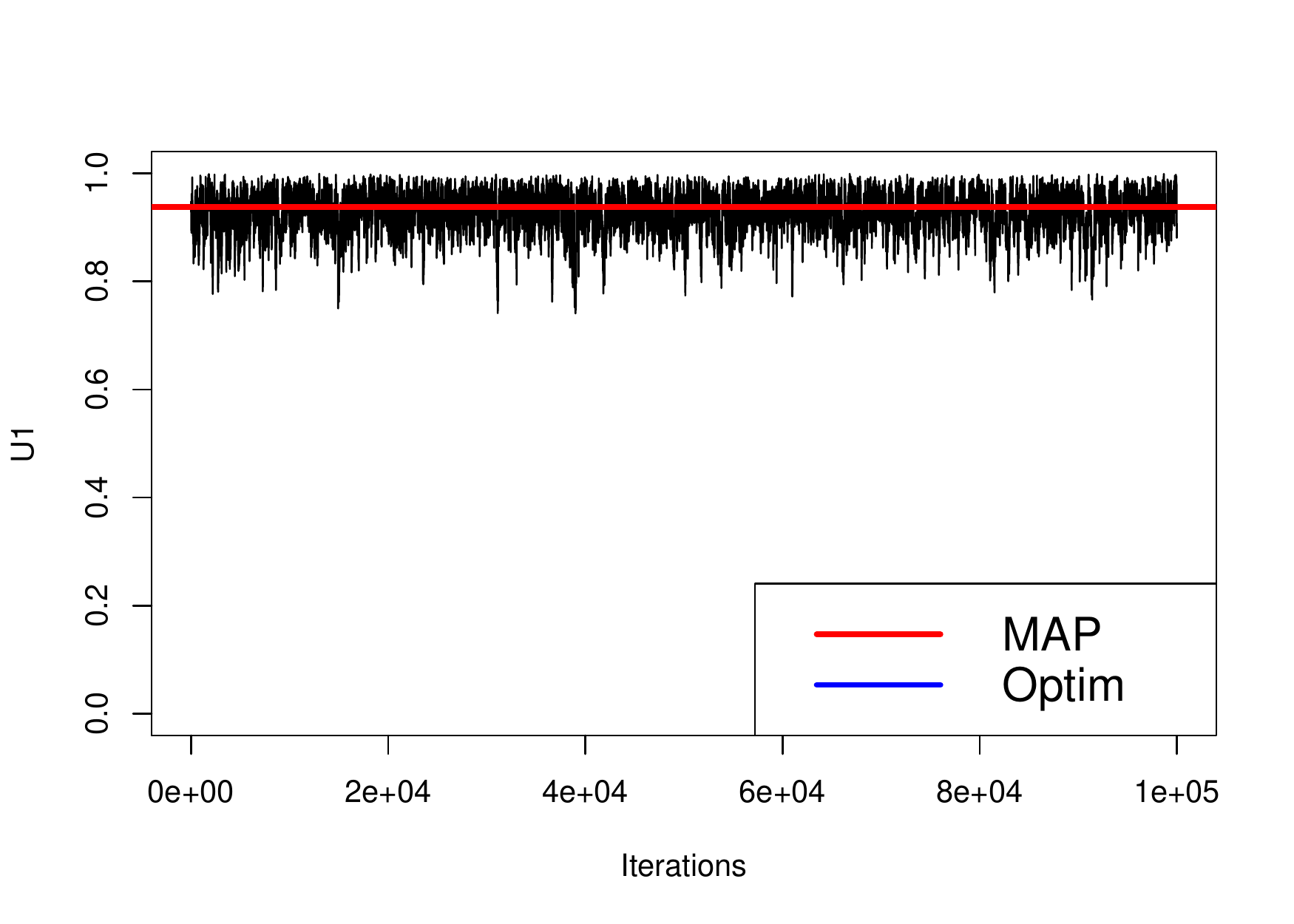}
\includegraphics[width=0.425\linewidth, trim=0 60 15 50, clip]{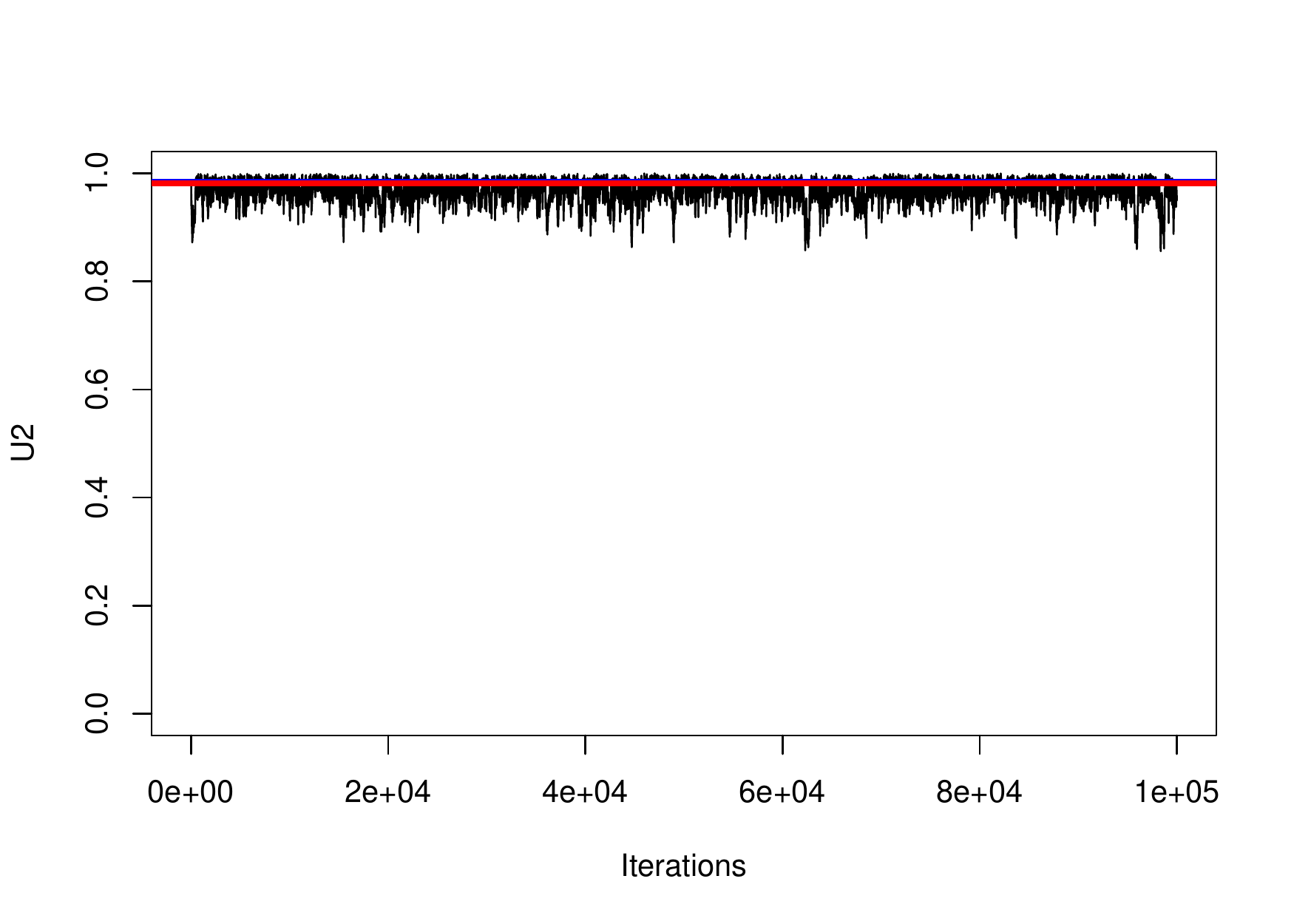}\\
\includegraphics[width=0.425\linewidth, trim=0 15 15 50, clip]{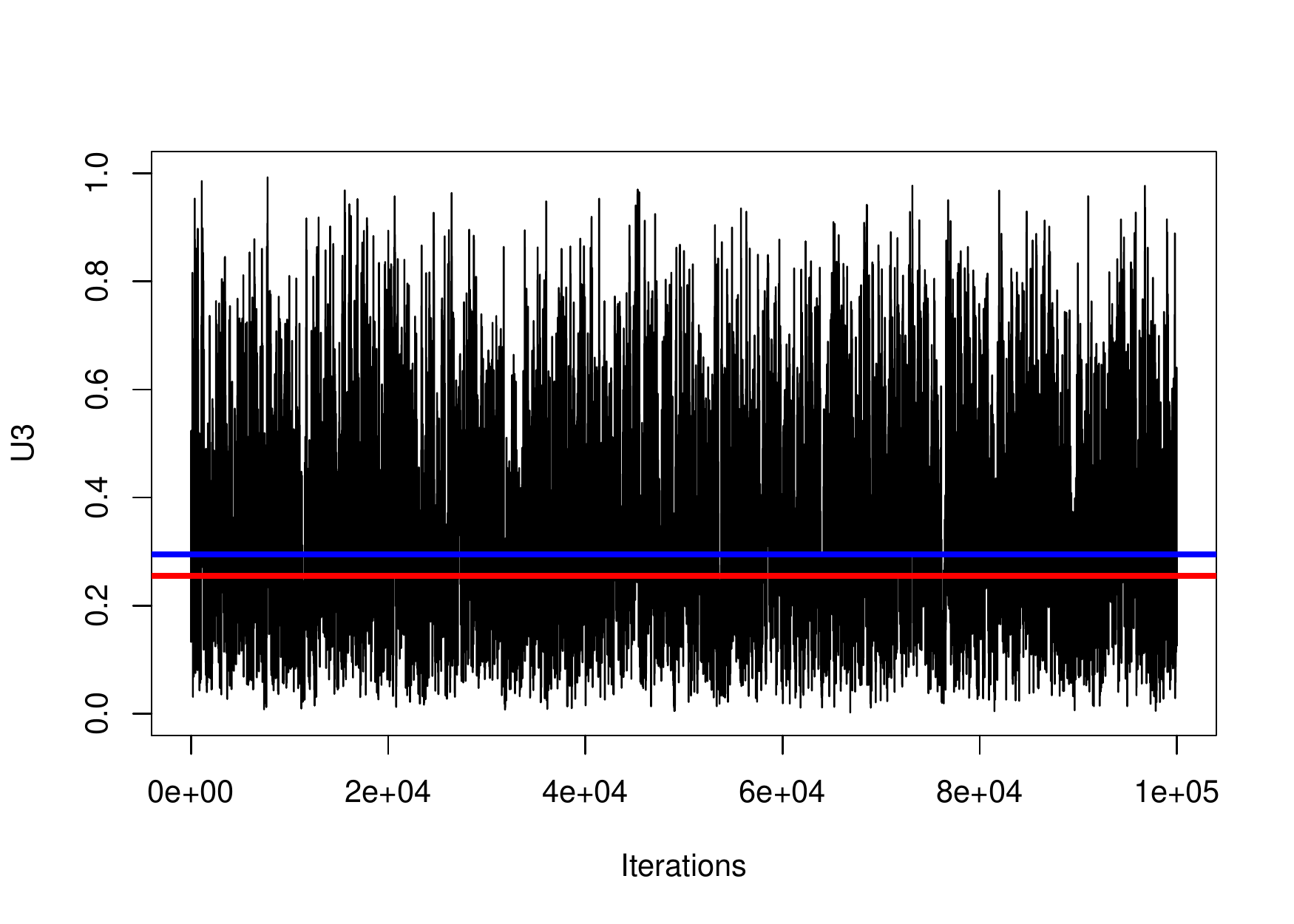}
\includegraphics[width=0.425\linewidth, trim=0 15 15 50, clip]{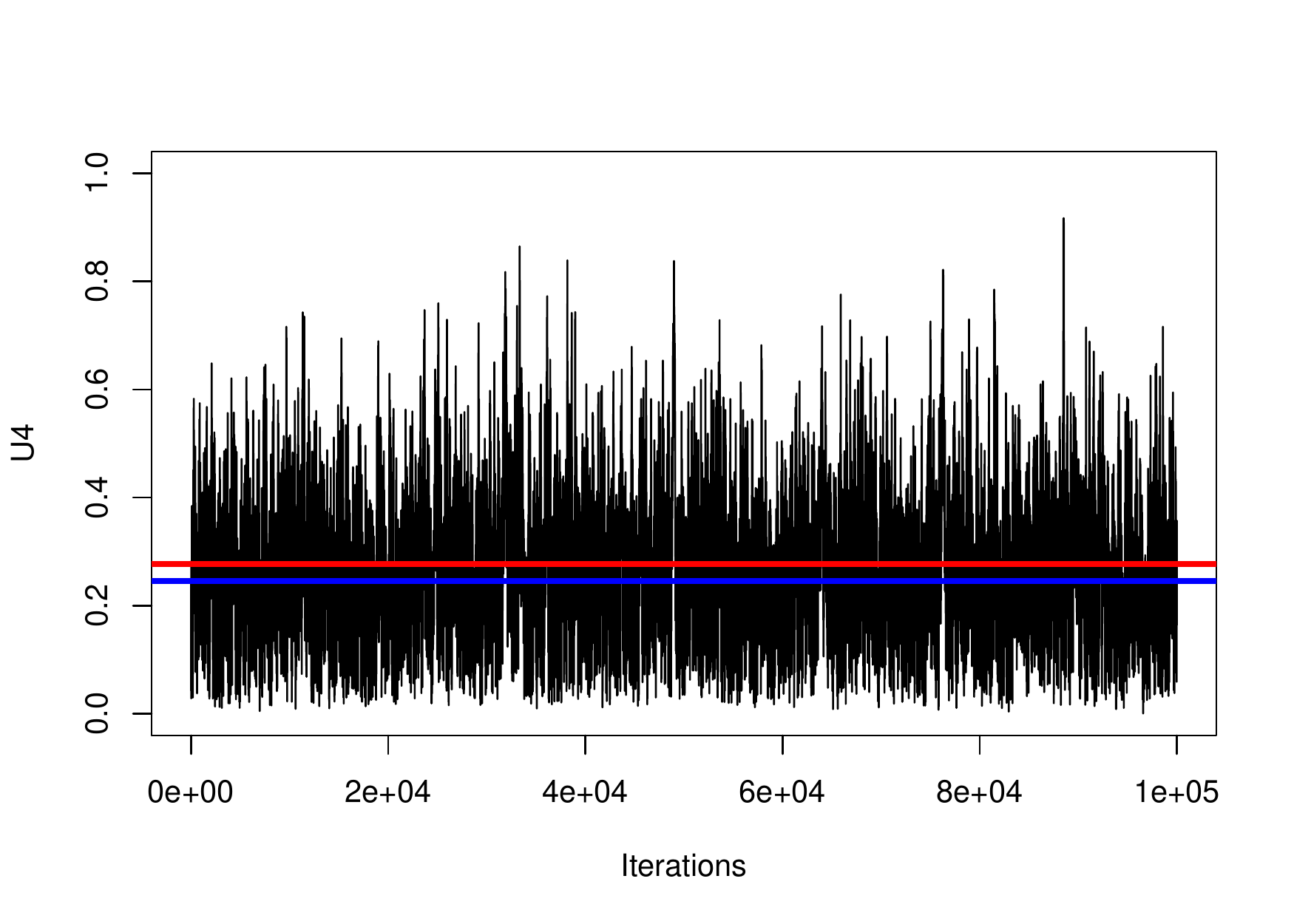}
\caption{Trace plots of MCMC samples for all calibration parameters $\mathbf{u}$ after
burn-in. Blue line indicates the best setting for $\mathbf{u}$ from
optimization. Red line indicates the MAP
$\mathbf{u}$ extracted from the samples; modular/opt added for
comparison.}%
\label{fig:trace}%
\end{figure}

Figure \ref{fig:trace} shows traces of the samples obtained
via our Metropolis-within-Gibbs scheme, described in Section \ref{sec:mcmc}.  
The figure indicates clear convergence to the stationary distribution with
mixing that is qualitatively quite good. The effective sample sizes (ESS)
\citep{kass:1998}, marginally for all four friction factors, are sufficiently high at
$\mathrm{ESS}_{u_1} = 1026$, $\mathrm{ESS}_{u_2}= 684$,
$\mathrm{ESS}_{u_3}=2062$, $\mathrm{ESS}_{u_4}=1462$, respectively.

\begin{figure}[ht!]
\centering
\includegraphics[width=1\linewidth, trim=25 25 25 40,clip=TRUE]{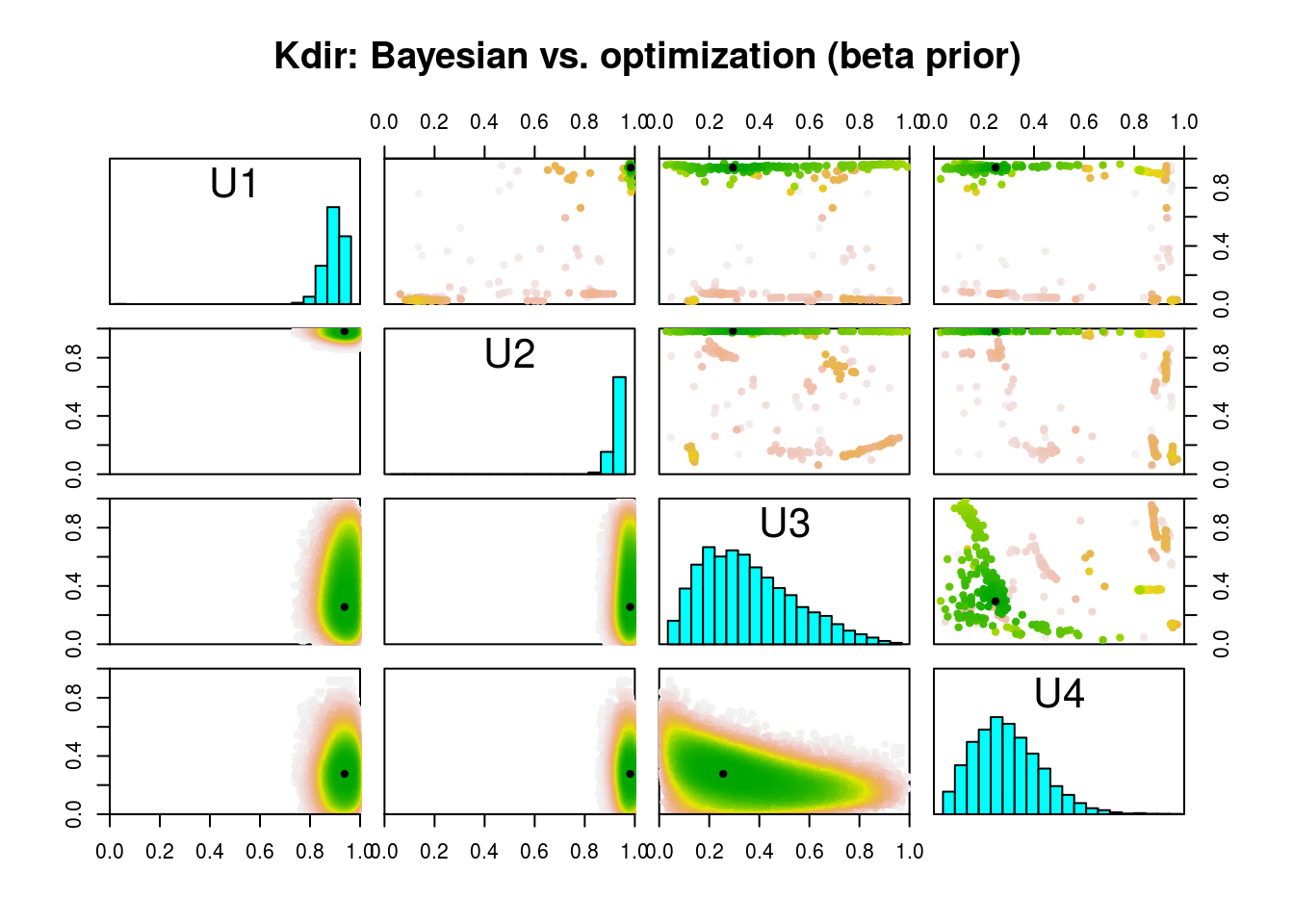}
\includegraphics[width=1\linewidth, trim=50 40 16 55,clip=TRUE]{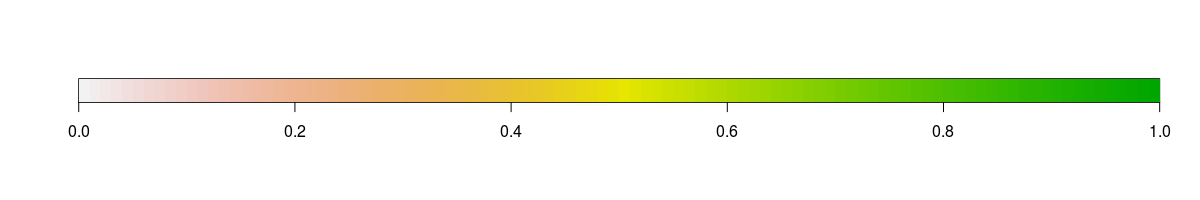}
\caption{Bayesian KOH (lower and diagonal) posterior for $\mathbf{u}$ vs.~modularlized
optimization (upper) analog, both under an independent $\mathrm{Beta}(2,2)$
prior for $\mathbf{u}$. Colors are derived from ranks of posterior
probabilities to aid in visualization. Modularized results are from 500
converged optimization under random initialization. Black dots indicate MAP
values.}
\label{fig:res}%
\end{figure}

Figure \ref{fig:trace} clearly shows that the posterior is, at least
marginally, far more concentrated for the first two friction factors (first
row) than for the last two.  For a better joint glimpse at the
four-dimensional posterior distribution of $\mathbf{u}$, the
bottom-left panels of Figure \ref{fig:res} show these samples via pairs of
coordinates.  The points are colored by a rank-transformed log-scaled
posterior evaluation as a means of better visualizing the high concentrations
in a cramped space.  Histograms along the diagonal panels show individual
margins; panels on the top-right mirror those on the bottom-left but instead
show solutions found by the modular/optimal approach [Section \ref{sec:optim}]
in 500 random restarts.

Several notable observations can be drawn from the plots in that figure.  For
one, consistency is high between the two approaches: KOH and
modular/opt.   Although the values of log posteriors evaluations are not
directly comparable across the models, both agree on most probable values
(black dots in the off-diagonal panels).  A diversity of solutions from the
optimization-based approach indicates that the solver struggles to navigate
the log posterior surface but usually finds estimates that are in the right
ballpark.  The full posterior distribution via KOH indicates that the first
two friction factors are well pinned-down by the posterior.  However, posterior
concentration is more diffuse for the latter two.  A complicated correlation
structure is evident in $(u_3, u_4)$.

A similar suite of results under an independent uniform prior is provided in
Appendix \ref{ap:unif}.  The story there is similar, except that the
posterior sampling concentrates more heavily on the boundary of
$\mathbf{u}$-space for all four parameters.  Considering that we know our {\tt
ISOTSEAL} simulator is less reliable in those regimes, leading to far more
missing values and thus requiring greater degree of extrapolation from our
OSSs, we prefer the more stable regime (better emulation and MCMC mixing)
offered by light penalization under a $\mathrm{Beta}(2,2)$ prior.

\subsection{Out-of-sample prediction}
\label{sec:oosresults}

To close the loop on our NLS comparison from Section \ref{sec:nls},
particularly Figure \ref{fig:nls} highlighting in-sample prediction, we
conclude our empirical work on the honeycomb with an exercise \blu{measuring}
out-of-sample \blu{predictive accuracy}. \blu{Pointwise comparators based on
several variations are entertained, e.g., with/without OSSs, with/without
estimated discrepancies compensating for bias. Finally, we complete the
Bayesian KOH OSS setup (Section \ref{sec:fullBayes}) with a predictor that
tractably propagates uncertainty through the sparse covariance structure.}

\begin{figure}[ht!]
\centering
\includegraphics[width=0.49\linewidth, trim=0 0 0 0, clip]{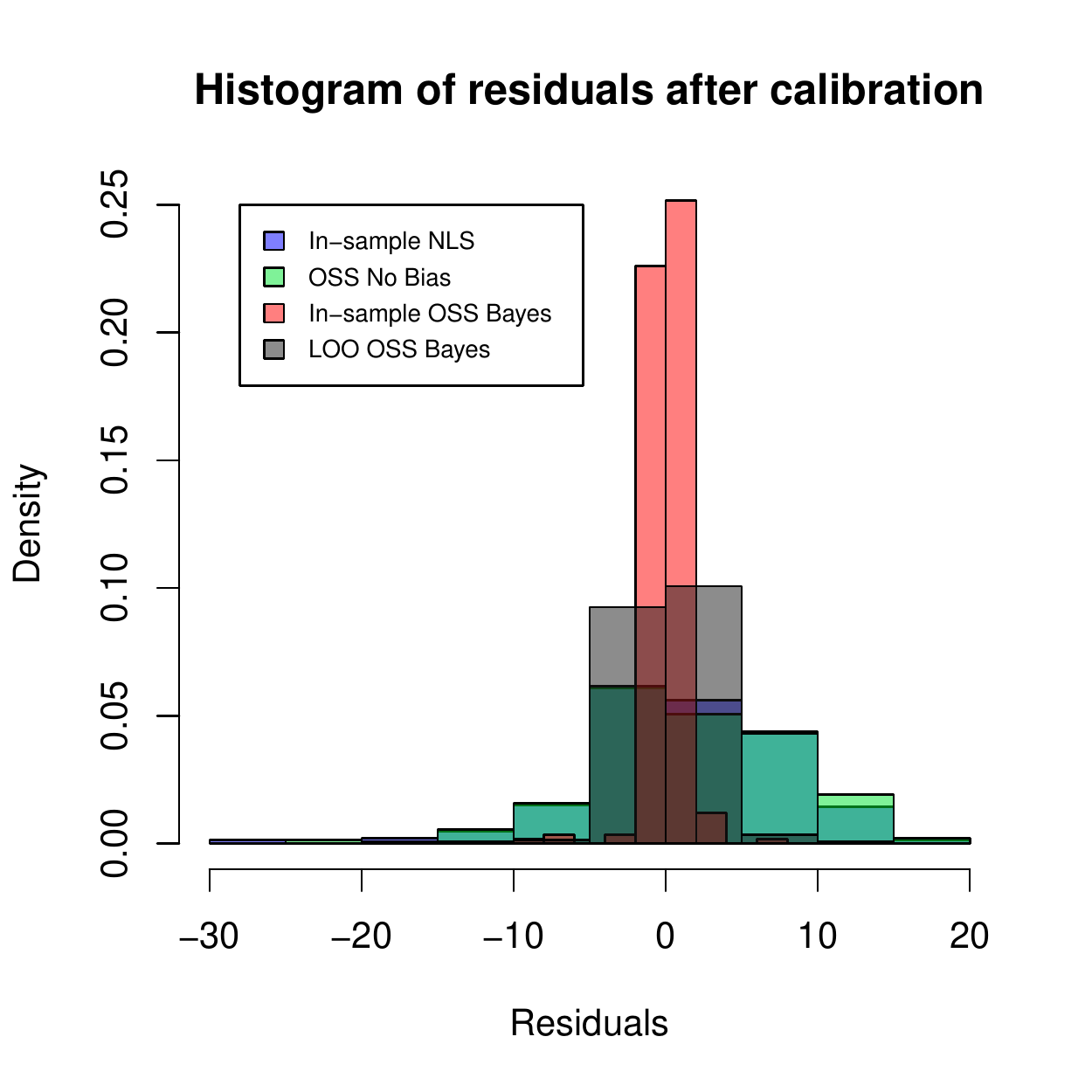}
\includegraphics[width=0.49\linewidth, trim=0 0 0 0, clip]{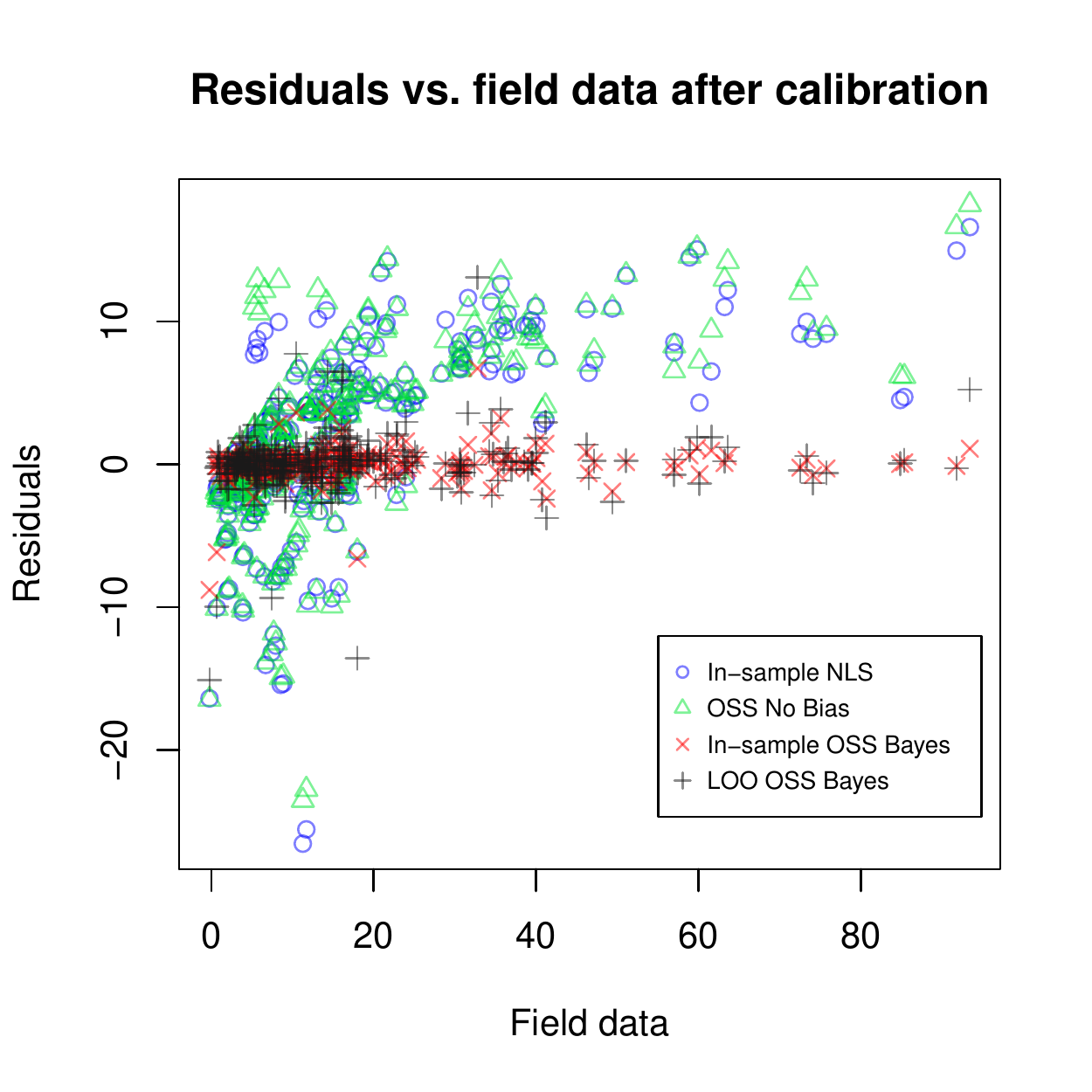}
\caption{Residuals to honeycomb field data. The left panel shows 
histograms comparing three approaches; the right panel plots them versus the
true response.}
\label{fig:loo}
\end{figure}

\blu{The NLS baseline from Section \ref{sec:nls} involves direct application
of {\tt ISOTSEAL} for new physical (testing) site
$\mathbf{x}_{\mathrm{\mathrm{new}}}$, paired with plug-in $\hat{\mathbf{u}}$
furnished by our BHGE colleagues: }
\blu{
\begin{align}
 \hat{y}^F(\mathbf{x}_{\mathrm{\mathrm{new}}}) = 
 {y}^M(\mathbf{x}_{\mathrm{\mathrm{new}}},\hat{\mathbf{u}}^\text{NLS}) 
\end{align}
}
\blu{No bias correction is applied.  Figure \ref{fig:loo}, augmenting
Figure \ref{fig:nls}, provides a view into residuals under this comparator,
and others explained momentarily. We clarify that these NLS results are
"in-sample" as they use the same data our BHGE colleagues trained on. Precise
root mean-squared errors (RMSEs) and $\hat{\mathbf{u}}$-values are summarized
in Table \ref{tab:pred}.}

\blu{Feeding $\hat{\mathbf{u}}$ directly into {\tt ISOTSEAL} is problematic
because simulation dynamics are nonstationary, unstable, and unreliable.} We
had trouble getting an implementation of this variation to behave reliably
enough in order to report meaningful out-of-sample results. \blu{As
demonstrated in Section \ref{sec:isotseal}, and the second row in Figure
\ref{fig:loc2}, {\tt ISOTSEAL} can fail to converge and instead return
${y}^M(\mathbf{x}_{\mathrm{\mathrm{new}}},\hat{\mathbf{u}})= $ {\tt NA}
especially at $\mathbf{u}$ around the upper limit of their range(s). Our BHGE
colleagues carefully engineered their NLS search to avoid problematic
$\mathbf{u}$-settings.  OSSs were proposed in order to more gracefully cope
with {\tt NA}s and to correct for other idiosyncrasies. When predicting out of
sample, a new OSS $\hat{y}^M(\mathbf{x}_{\mathrm{\mathrm{new}}},
\cdot)$ must be fit via new on-site design
$\mathbf{U}_{\mathrm{\mathrm{new}}}$ paired with
$\mathbf{x}_{\mathrm{\mathrm{new}}}$. As with OSS training described in
Section \ref{sec:oss}, we shall utilize a size $n=1000$ maximin LHS. }

\blu{Surrogate $\hat{y}^M(., .)$ enables a full search of the entire
$\mathbf{u}$-space, offering the potential of finding a better
$\hat{\mathbf{u}}$ especially nearby regions where direct {\tt ISOTSEAL} runs
may fail. Acting on OSSs without discrepancy correction, we find
$\hat{\mathbf{u}}^{\text{OSS}}_\text{nobias}$ slightly different from the
$\hat{\mathbf{u}}^\text{NLS}$ using direct {\tt ISOTSEAL} runs. See Table
\ref{tab:pred}.  To compare the predictive performance directly to the
in-sample NLS, we plug-in $\hat{\mathbf{u}}^{\text{OSS}}_\text{nobias}$ for
new site $\mathbf{x}_{\mathrm{\mathrm{new}}}$ though the new OSS: }
\blu{
\begin{align}
 \hat{y}^F(\mathbf{x}_{\mathrm{\mathrm{new}}}) = 
 \hat{y}^M(\mathbf{x}_{\mathrm{\mathrm{new}}},\hat{\mathbf{u}}^{\text{OSS}}_\text{nobias}).
\end{align}
}
\blu{Figure \ref{fig:loo} indicates similar residual behavior for these two
comparators.  OSSs without bias correction fares slightly worse than the NLS
analog, however note that the latter is truly out-of-sample and the former was
technically in-sample.  The OSS version
$\hat{y}^M(\mathbf{x}_{\mathrm{\mathrm{new}}},\hat{\mathbf{u}}^{\text{OSS}}_\text{nobias})$
offers fuller uncertainty quantification in predictions, via local GP
predictive variances.}

\blu{Now consider variations which correct for potential bias between OSS
and field data measurements.} Feed $\hat{\mathbf{u}}$ through
the OSS and obtain
\begin{align}
 \hat{y}^F(\mathbf{x}_{\mathrm{\mathrm{new}}}) = 
 \hat{y}^M(\mathbf{x}_{\mathrm{\mathrm{new}}},\hat{\mathbf{u}}) +
  \hat{b}(\mathbf{x}_{\mathrm{\mathrm{new}}})
\end{align}
To benchmark \blu{these} predictions \blu{out of sample} we designed the
following leave-one-out (LOO) cross-validation (CV) experiment. Alternately
excluding each field data location $i=1,\dots,N_F=292$, we fit 292 LOO
discrepancy terms $\hat{b}^{(-i)}(\cdot)$ via residuals
$\mathbf{y}^F_{(-i)}-\hat{\mathbf{y}}^M_{(-i)}$ and $\mathbf{X}^F_{(-i)}$. We
could build a new OSS for $\mathbf{x}_i$, treating it as a
$\mathbf{x}_{\mathrm{\mathrm{new}}}$ as described above, but instead it is
equivalent (and computationally more thrifty) to use the $\mathbf{U}_i$ we
already have. Based on those calculations, point predictions are composed of
\begin{align}
\hat{y}^F({\mathbf{x}_i}) =
\hat{y}^M(\mathbf{x}_i,
\hat{\mathbf{u}})+\hat{b}^{(-i)}(\mathbf{x}_i), \quad \mbox{for} \quad
i=1, 2, \dots, N_F.
\label{eq:LOO_bias}
\end{align}
Predictions thus obtained are compared with true outputs $\mathbf{y}^F$ and
residuals for RMSE calculations. We note that this experiment focuses
primarily on bias correction.  New $\hat{\mathbf{u}}_{(-i)}$ are not
calculated for each of $i=1,\dots,N_F$ \blu{due to the prohibitive
computational cost.}

Figure \ref{fig:loo} shows those LOO residuals graphically alongside our other
comparators. Only results for $\hat{\mathbf{u}}$ via modular/opt framework are
shown here since $\hat{\mathbf{u}}$ from the fully Bayes KOH setup are
similar. The panels in the figure indicate that bias correction
offers substantial improvement over NLS: in-sample NLS residuals are worse
than LOO OSS Bayes results. Summarizing those residuals,
\blu{modular/opt calibration with discrepancy has an leave-one-out RMSE of
$2.126$, being even smaller than both the in-sample NLS value of $6.605$ reported
in Section \ref{sec:nls} and the in-sample OSS no-bias value of $6.818$.
Furthermore,  LOO OSS modular/opt RMSE is comparable to its in-sample analog of
$1.125$.}

\begin{table}[ht!]
\blu{
\centering
\begin{tabular}{r | r | r | r | r | r}
Method  &$\hat{u}_1$ &$\hat{u}_2$ & $\hat{u}_3$& $\hat{u}_4$ & RMSE \\
\hline
In-sample NLS & 0.00000 &	0.00000	& 0.82123 & 0.99615 & 6.605\\
OSS No Bias & 0.00877  &  0.17352  &  0.94893 &   0.94474& 6.818 \\
In-sample OSS Bayes &  0.93659 &  0.98348 &  0.28441  &  0.25975 & 1.125 \\
LOO OSS modular/opt &  0.93659  & 0.98348 & 0.28441 & 0.25975  & 2.126\\
LOO OSS KOH full Bayes &  0.93659  & 0.98348 & 0.28441 & 0.25975  &  1.957\\
\end{tabular}
\caption{Estimated  $\hat{\mathbf{u}}$ and RMSEs from in-sample and LOO
comparisons.}
\label{tab:pred}
}
\end{table}

\blu{
Next we develop fully Bayesian prediction for
$y^F(\mathbf{X}^F_{\mathrm{new}})$ at $N'_F$ new physical locations
$\mathbf{X}^F_{\mathrm{new}}=(\mathbf{x}^{\mathrm{new}}_1,
\mathbf{x}^{\mathrm{new}}_2, \dots, \mathbf{x}^{\mathrm{new}}_{N'_F})^\top$.
As in the pointwise case, $N'_F$ new OSSs must be built on $N'_M$ new on-site
simulations $\mathbf{y}^M_{\mathrm{new}} = (\mathbf{y}_{N_F+1}, \dots,
\mathbf{y}_{N_F+N'_F})^\top$. Following from Eq.~(\ref{eq:lik}), 
\begin{align}
\begin{bmatrix} \mathbf{y}^M \\ \mathbf{y}^F \\
 \mathbf{y}^M_{\mathrm{new}} \\ \mathbf{y}^F_{\mathrm{new}} \end{bmatrix}
=  \begin{bmatrix} \mathbf{y}_1 \\ \vdots \\ \mathbf{y}_{N_F} \\
 \mathbf{y}^F  \\ \mathbf{y}_{N_F+1}  \\ \vdots \\\mathbf{y}_{N_F + N'_F} \\ \mathbf{y}^F_{\mathrm{new}}
\end{bmatrix}
= \begin{bmatrix} 
y_1(\mathbf{U}_1)\\ \vdots \\ y_{N_F}(\mathbf{U}_{N_F}) \\ y^M(\mathbf{U}) + b(\mathbf{X}^F) \\
y_{N_F+1}(\mathbf{U}_{N_F+1})\\ \vdots \\ y_{N_F + N'_F}(\mathbf{U}_{N_F + N'_F}) \\
y^M_{\mathrm{new}}(\mathbf{U}_{N'_F}) + b(\mathbf{X}_{\mathrm{new}}^F) 
\end{bmatrix} \sim \mathcal{N}
(\mathbf{0}, \mathbb{V}^P(\mathbf{u})) 
\label{eq:d_new}
\end{align}
where $\mathbf{U}_{N'_F}=[\mathbf{u}^\top; \cdots; \mathbf{u}^\top]^\top$
stacks $N'_F$ identical $p_u$-dimensional row vectors $\mathbf{u}^\top$. The
$(N_M + N_F + N'_M + N'_F) \times (N_M + N_F + N'_M + N'_F)$ covariance matrix
$\mathbb{V}^P(\mathbf{u})$, combining OSS training data and out-of-sample data
elements, may be built as follows
 \begin{align}
\mathbb{V}^P(\mathbf{u}) =\begin{bmatrix} \mathbb{V}_{o} & \mathbb{V}^\top_{ob}(\mathbf{u}) &  \mathbf{0}  &  \mathbf{0} \\
  \mathbb{V}_{ob}(\mathbf{u}) &   \mathbb{V}_{b} & \mathbf{0} & \mathbb{V}_{b}^\top(\mathbf{X}^F_{\mathrm{new}}, \mathbf{X}^F) \\ 
  \mathbf{0} & \mathbf{0} &  \mathbb{V}^{\mathrm{new}}_{o}  
  & \mathbb{V}^{\mathrm{new}}_{ob}(\mathbf{u})^\top\\
  \mathbf{0} & \mathbb{V}_{b}( \mathbf{X}^F_{\mathrm{new}}, \mathbf{X}^F)  
  &  \mathbb{V}^{\mathrm{new}}_{ob}(\mathbf{u}) 
  &  \mathbb{V}^{\mathrm{new}}_{b} \end{bmatrix},
\label{eq:Vdnew} 
\end{align} 
borrowing notation for $\mathbb{V}(\mathbf{u})$ from Eq.~(\ref{eq:Vd}). }

\blu{Like $\mathbb{V}(\mathbf{u})$, $\mathbb{V}^P(\mathbf{u})$ emits sparse
block-wise structure due to the OSSs.  Extending from Eq.~(\ref{eq:Vd}), we
have $\mathbb{V}^{\mathrm{new}}_{o}=\Diag[\mathbf{V}_i(\mathbf{U}_i,
\mathbf{U}_i)]$, for $i = N_F+1, \dots, N_F+ N'_F$, an upper-left block
diagonal submatrix.  Similarly
$\mathbb{V}^{\mathrm{new}}_{b}=\mathbf{v}_{\mathrm{new}} \mathbb{I}_{N'_F}
+ V_b(\mathbf{X}^F_{\mathrm{new}})$, where $\mathbf{v}_{\mathrm{new}} \mathbb{I}_{N'_F}$ 
is a diagonal of nugget effects from the new OSSs, and $ V_b(\mathbf{X}^F_{\mathrm{new}})$ 
is the covariance matrix on $\mathbf{X}^F_{\mathrm{new}}$ from the bias correction. 
$\mathbb{V}^{\mathrm{new}}_{ob}(\mathbf{u})$ and 
$\mathbb{V}^{\mathrm{new}}_{ob}(\mathbf{u})^\top$ are similar 
to $\mathbb{V}_{ob}(\mathbf{u})$ and $\mathbb{V}^\top_{ob}(\mathbf{u})$, 
composed of  $V_i(\mathbf{U}_{N'_F})$ with
$i = N_F + 1, \dots, N_F + N'_F$.  Each $V_i(\mathbf{U}_{N'_F})$ is 
sparse with single row of non-zero entries. 
In Eq.~(\ref{eq:Vdnew}), the new OSS on $\mathbf{X}^F_{\mathrm{new}}$ is sparse between training 
data $(\mathbf{y}^M, \mathbf{y}^F)$ and new data $(\mathbf{y}^M_{\mathrm{new}}, 
\mathbf{y}^F_{\mathrm{new}})$, involving only the small $N'_F \times N_F$
 bias covariance $\mathbb{V}_{b}(\mathbf{X}^F_{\mathrm{new}}, \mathbf{X}^F) $.
}

\blu{
Using those definitions, the predictive distribution of $\mathbf{y}^F_{\mathrm{new}}$ 
conditioning on both data sources, 
$(\mathbf{y}^M,  \mathbf{y}^M_{\mathrm{new}})$ from simulation 
and $\mathbf{y}^F$ from physical experiments, 
hyperparameters $\bm{\Phi}$ and calibration parameter
 $\mathbf{u}$, is MVN with mean $\mathbf{m}_{\mathrm{new}}$ and covariance $\mathbf{V}_{\mathrm{new}}$ following
\begin{align}
\mathbf{m}_{\mathrm{new}} &= \mathbb{V}_{b}(\mathbf{X}^F_{\mathrm{new}}, \mathbf{X}^F) 
 \mathbb{C}^{-1}(\mathbf{u})[ \mathbf{y}^F -\mathbb{V}_{ob}(\mathbf{u})
  \mathbb{V}_{o}^{-1} \mathbf{y}^M] +  \mathbb{V}^{\mathrm{new}}_{ob}(\mathbf{u}) 
 (\mathbb{V}^{\mathrm{new}}_{o})^{-1}\mathbf{y}^M_{\mathrm{new}} 
 \label{eq:mnew}  \\
\mathbf{V}_{\mathrm{new}} &= \mathbb{V}^{\mathrm{new}}_{b}- \mathbb{V}_{b}(\mathbf{X}^F_{\mathrm{new}}, 
 \mathbf{X}^F) \mathbb{C}^{-1}(\mathbf{u})
  \mathbb{V}_{b}(\mathbf{X}^F_{\mathrm{new}}, \mathbf{X}^F)^\top
  -\mathbb{V}^{\mathrm{new}}_{ob}(\mathbf{u}) (\mathbb{V}^{\mathrm{new}}_{o})^{-1}   \mathbb{V}^{\mathrm{new}}_{ob}(\mathbf{u})^\top.
\label{eq:vnew}
\end{align}
Fully Bayesian uncertainty quantification using
Eqs.~(\ref{eq:mnew}--\ref{eq:vnew}) is tractable. Sparse-matrix decompositions
can be applied in a manner similar to likelihood evaluation Section
\ref{sec:kohoss}. }

\begin{figure}[ht!]
\centering
\includegraphics[width=0.65\linewidth, trim=0 0 0 0, clip]{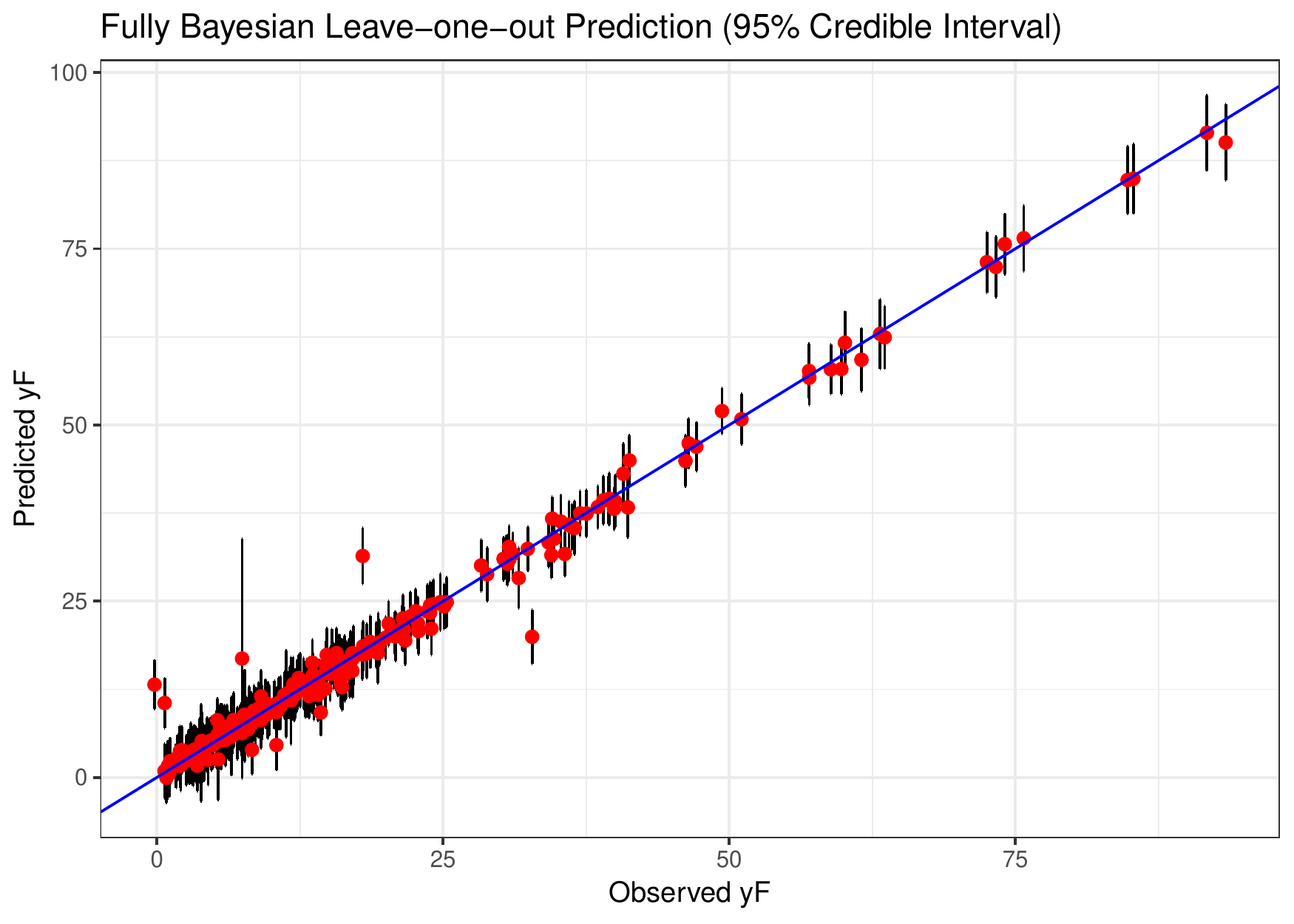}
\caption{\blu{Fully Bayesian out-of-sample predicted $\mathbf{y}^F$ with 95\% credible interval over observed honeycomb field data $\mathbf{y}^F$.}}
\label{fig:predBayes}
\end{figure}

\blu{Consider deploying these equations in our out-of-sample setup, re-using
the new OSSs trained for the pointwise comparisons.  Following a similar
LOO setup, we derive  $(\mathbf{y}^F_{i}  \mid \mathbf{y}^M_{-i},  \mathbf{y}^F_{-i}, 
 \mathbf{y}^M_{i},  \bm{\Phi}, \mathbf{u}^{(t)}) 
 \sim \mathcal{N}_{i}(\mathbf{m}_{i}, \mathbf{V}_{i} )$ 
via Eqs.~(\ref{eq:mnew}--\ref{eq:vnew}) integrating over $\mathbf{u}$ by
aggregating over Monte Carlo samples for $\{\mathbf{u}^{(t)}\}_{t=1}^{T}$ shown in
bottom-left panels of Figure \ref{fig:res}.  When aggregating covariances,
covariances of sample means are incorporated respecting the law of
total variance. Figure \ref{fig:predBayes} shows this fully Bayesian predicted
mean with 95\% credible interval over each observed $\mathbf{y}^F$. In contrast to
to the previous leave-one-out experiments described in Eq. \ref{eq:LOO_bias},
which involved 292 LOO discrepancy terms $\hat{b}^{(-i)}(\cdot)$ via residuals
$\mathbf{y}^F_{(-i)}-\hat{\mathbf{y}}^M_{(-i)}$ and $\mathbf{X}^F_{(-i)}$,
results in Figure \ref{fig:predBayes} provide full out-of-sample posterior
predictive uncertainty for both the simulation and the discrepancy correction.}


\section{Discussion}
\label{sec:discussion}

Motivated by a computer model calibration problem the design of a seal used in
turbines, we developed a thrifty new method to address several challenging
features.  Those challenges include a high-dimensional input space, local
instability in computer model simulations, nonstationary simulator dynamics,
and modeling for large computer experiments. Taken alone, each of these
challenges has solutions that are, at least in some cases, well established in
the literature.  Taken together, a more deliberate and custom development was
warranted.  To meet those challenges, we developed the method of on-site
surrogates. The construction of OSSs is motivated by the unique structure of
the posterior distribution under study in the canonical
\citeauthor{Kennedy:O'Hagan:2001} calibration framework, where predictions are
needed only at a limited number of field data sites, no matter how big the
computer experiment is.  This unique structure allowed us to map a single,
potentially high-dimensional problem, into a multitude of low-dimensional ones
where computation can be performed in parallel.  Two OSS-based calibration
settings were entertained, one based on simple bias-corrected maximization and
the other akin to the original KOH framework.  Both were shown to empirically
outperform simpler, yet high-powered, alternatives.

Despite its many attractive features, there is clearly much potential to
refine this approach, in particular the design and modeling behind the OSSs.
While simple Latin hypercube samples and GPs with exponential kernels and
nuggets work well, several simple extensions could be quite powerful. The
need for such extensions, along at least one avenue, is perhaps revealed by
the final row of Figure \ref{fig:loc2}.  Those plots show bifurcating {\tt
ISOTSEAL} runs due to numerical instabilities.  Although inflated nuggets
enable smoothing over those regimes, the result is uniformly high uncertainty
for all inputs rather than just near the trouble spot.  The reason is that the
GP formulation being used is still (locally) stationary.  Specifically, the
error structure is homoskedastic.  Using a heteroskedastic GP instead
\citep{Binois2017}, say via {\tt hetGP} on CRAN \citep{hetGP}, could offer a
potential remedy.  In a follow-in paper \cite{Binois2018} showed how designs
for effective {\tt hetGP} modeling could be built up sequentially, balancing
an appropriate amount of exploration and replication in order to effectively
learn signal-to-noise relationships in the data.  Such an approach could
represent an attractive alternative to simple LHSs in $\mathbf{u}$-space.

Here we only entertained a single output $k_{\mathrm{dir}}$,
at a single frequency, among a multitude of others and at other frequencies.
In future work we plan to investigate a multiple output approach to
calibration.  Much work remains to assess the potential for such an approach,
say via simple {\em co-kriging} \citep{verhoef:barry:1997} or a linear model
of co-regionalization \citep[e.g.,][]{wack:1998}.  Our BHGE collaborators'
pilot study also indicated that there could potentially be input-dependent
variations in the best setting of the friction factors.  That is, we could be
looking at a $\hat{\mathbf{u}}(\mathbf{x})$, perhaps for a subset of the
coordinates of the 13-dimensional $\mathbf{x}$ input.  Whether a simple
partition-based or linear scheme might be appropriate, or if something more
nonparametric like \citep{brown:atam:2016} is required, remains an open
question.

\blu{We'd like to close with a thought on confounding and
identifiability, an ever-present concern in the KOH setting.  OSSs are no help
here, essentially chopping up the design space, limiting information sharing
and reducing the (Bayesian) learning that could transpire about calibration
parameters compared to the usual (global) setup.  Although we have seen no
evidence of concern, it is possible that OSSs would exacerbate the problem.
However, we note that the underlying framework -- linking a latent
$\mathbf{u}$-variable to a nonparametric discrepancy -- is identical whether
or not OSSs are deployed.  Accordingly, simplifications \citep{tuo2015} or
extensions \citep{plumlee2017bayesian} are similarly viable as a means of
limiting sources of confounding that challenges identifiability.}

\blu{There are many reasons to calibrate, with KOH or otherwise.  One is simply
predictive; another is to get a sense of how the apparatus could be tuned, or
to quantify how much information is in the data (and prior) about promising
$\mathbf{u}$ settings.  Both are very doable, and worth doing, even in the
face of confounding.  Our posterior summaries for $\mathbf{u}$ are a testament
in this regard.  In our toy example, which has many features in common with
the motivating honeycomb seal, the posterior is quite peaked.  Does this mean
our $\hat{\mathbf{u}}$ or Bayesian samples $\mathbf{u}^{(t)}$ have identified
the right $\mathbf{u}^*$? Possibly not in general, except that we know the
truth in this case and identification can be confirmed.  Our posterior for
$\mathbf{u}$ in the honeycomb example shows sharp concentration for some
inputs, less for others, and interpretable correlation in one pair $(u_3,
u_4)$.  Our colleagues at BHGE were not surprised by these results, and found
them to be helpful in designing new field experiments.  Although we cannot
be confident about identification in this example, KOH has been a
useful exercise.}

\if0\blind{
\subsubsection*{Acknowledgments}

Authors JH, RBG, and MB are grateful for support from National Science
Foundation grants DMS-1521702 and DMS-1821258.  JH and RBG also gratefully
acknowledge funding from a DOE LAB 17-1697 via subaward from Argonne National
Laboratory for SciDAC/DOE Office of Science ASCR and High Energy Physics.  The
work of MB is partially supported by the U.S. Department of Energy, Office of
Science, Office of Advanced Scientific Computing Research, under Contract No.
DE-AC02-06CH11357. \blu{We thank Andrea Panizza for early NLS work on this
project, and for initiating the line of research.  We are grateful to two
referees for thoughtful suggestions which led to many improvements. }
Finally, we wish to thank our collaborators at Baker
Hughes/GE for being generous with their data, their time, and their expertise.

\blu{} 
}\fi

\section{Appendix: Calibration under uniform prior}
\label{ap:unif}

\begin{figure}[ht!]
\centering
\includegraphics[width=1\linewidth, trim=25 25 25 40,clip=TRUE]{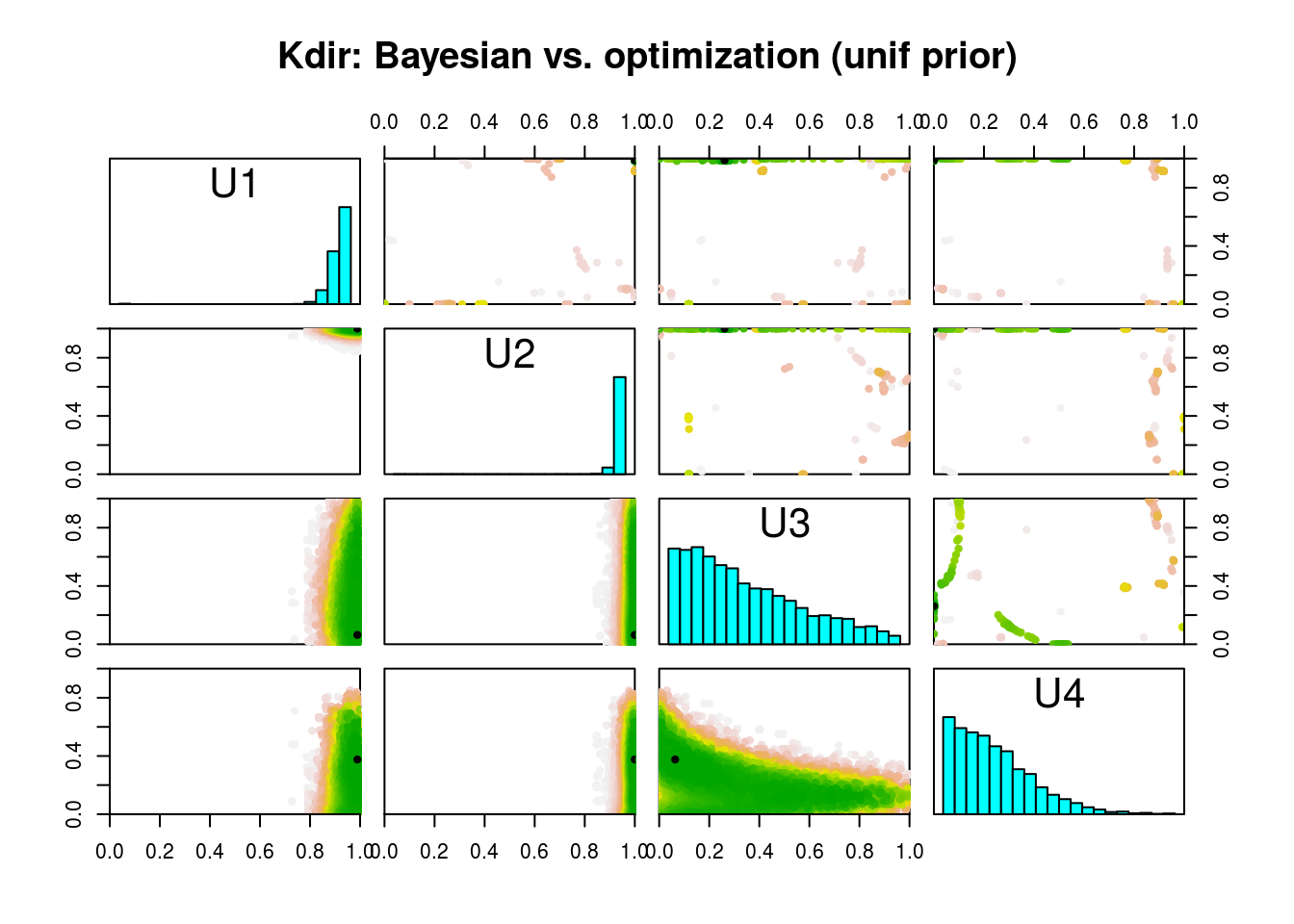}
\includegraphics[width=1\linewidth, trim=50 40 16 55,clip=TRUE]{colorscale.png}
\caption{Bayesian KOH (lower and diagonal) posterior for $\mathbf{u}$ vs.~modularized
optimization (upper) analog, both under an independent uniform
prior for $\mathbf{u}$. Colors are derived from ranks of posterior
probabilities to aid in visualization. Modularized results are from 500
converged optimization under random initialization. Black dots indicate MAP
values. }
\label{fig:res2}
\end{figure}

For completeness, we provide calibration results under a uniform prior in
Figure \ref{fig:res2}, complementing those from Figure \ref{fig:res} under
$\mathrm{Beta}(2,2)$. 
Compared with those results, the ones shown here more
heavily concentrate on the boundaries of the study region. Also, somewhat more inconsistency exists between the modular/opt results and the fully
Bayesian analog.  The regularization effect of the $\mathrm{Beta}(2,2)$ 
leads to better numerics.

\bibliography{honeycomb}

\begin{thebibliography}{40}
\newcommand{\enquote}[1]{``#1''}
\expandafter\ifx\csname natexlab\endcsname\relax\def\natexlab#1{#1}\fi

\bibitem[\protect\citename{Abramson et~al., }2013]{AuCo04a}
Abramson, M., Audet, C., Couture, G., {Dennis, Jr.}, J., {Le~Digabel}, S., and
  Tribes, C. (2013).
\newblock \enquote{The {NOMAD} project.}
\newblock Software available at \url{http://www.gerad.ca/nomad}.

\bibitem[\protect\citename{Ba and Joseph, }2012]{ba:joseph:2012}
Ba, S. and Joseph, V. (2012).
\newblock \enquote{Composite {G}aussian process models for emulating expensive
  functions.}
\newblock {\em Annals of Applied Statistics\/}, 6, 4, 1838--1860.

\bibitem[\protect\citename{Bastos and O'Hagan, }2009]{bastos:ohagan:2009}
Bastos, L. and O'Hagan, A. (2009).
\newblock \enquote{Diagnostics for {G}aussian Process Emulators.}
\newblock {\em Technometrics\/}, 51, 4, 425--438.

\bibitem[\protect\citename{Binois and Gramacy, }2018]{hetGP}
Binois, M. and Gramacy, R.~B. (2018).
\newblock {\em {\tt hetGP}: {H}eteroskedastic {G}aussian Process Modeling and
  Design under Replication\/}.
\newblock R package version 1.0.3.

\bibitem[\protect\citename{Binois et~al., }2018]{Binois2017}
Binois, M., Gramacy, R.~B., and Ludkovski, M. (2018).
\newblock \enquote{Practical heteroskedastic {G}aussian process modeling for
  large simulation experiments.}
\newblock {\em Journal of Computational and Graphical Statistics\/}, 27, 4,
  808--821.

\bibitem[\protect\citename{Binois et~al., }2019]{Binois2018}
Binois, M., Huang, J., Gramacy, R.~B., and Ludkovski, M. (2019).
\newblock \enquote{Replication or exploration? Sequential design for stochastic
  simulation experiments.}
\newblock {\em Technometrics\/}, 61, 1, 7--23.

\bibitem[\protect\citename{Bornn et~al., }2012]{bornn:shaddick:zidek:2012}
Bornn, L., Shaddick, G., and Zidek, J. (2012).
\newblock \enquote{Modelling Nonstationary Processes Through Dimension
  Expansion.}
\newblock {\em J.~of the American Statistical Association\/}, 107, 497,
  281--289.

\bibitem[\protect\citename{{Brown} and {Atamturktur}, }2018]{brown:atam:2016}
{Brown}, D.~A. and {Atamturktur}, S. (2018).
\newblock \enquote{{Nonparametric Functional Calibration of Computer Models}.}
\newblock {\em Statistica Sinica\/}, 28, 721--742.

\bibitem[\protect\citename{Byrd et~al., }1995]{byrd:etal:1995}
Byrd, R., Qiu, P., Nocedal, J., , and Zhu, C. (1995).
\newblock \enquote{A Limited Memory Algorithm for Bound Constrained
  Optimization.}
\newblock {\em Journal on Scientific Computing\/}, 16, 5, 1190--1208.

\bibitem[\protect\citename{Carnell, }2018]{lhs}
Carnell, R. (2018).
\newblock {\em {\tt lhs}: {L}atin Hypercube Samples\/}.
\newblock {\sf R} package version 0.16.

\bibitem[\protect\citename{{D'Souza} and {Childs}, }2002]{D'Souza:Childs:2002}
{D'Souza}, R.~J. and {Childs}, D.~W. (2002).
\newblock \enquote{{A Comparison of Rotordynamic-Coefficient Predictions for
  Annular Honeycomb Gas Seals Using Three Different Friction-Factor Models}.}
\newblock {\em Journal of Tribology\/}, 124, 3, 524--529.

\bibitem[\protect\citename{Gelfand and Smith, }1990]{Gelfand:1990}
Gelfand, A.~E. and Smith, A. F.~M. (1990).
\newblock \enquote{Sampling-Based Approaches to Calculating Marginal
  Densities.}
\newblock {\em Journal of the American Statistical Association\/}, 85, 410,
  398--409.

\bibitem[\protect\citename{Gramacy, }2016]{gramacy:jss:2016}
Gramacy, R. (2016).
\newblock \enquote{{laGP}: Large-Scale Spatial Modeling via Local Approximate
  {G}aussian Processes in {R}.}
\newblock {\em Journal of Statistical Software, Articles\/}, 72, 1, 1--46.

\bibitem[\protect\citename{Gramacy and Lee, }2012]{gra:lee:2012}
Gramacy, R. and Lee, H. (2012).
\newblock \enquote{Cases for the nugget in modeling computer experiments.}
\newblock {\em Statistics and Computing\/}, 22, 3, 713--722.

\bibitem[\protect\citename{{Gramacy} and {Apley}, }2015]{gramacy:apley:2015}
{Gramacy}, R.~B. and {Apley}, D.~W. (2015).
\newblock \enquote{Local Gaussian process approximation for large computer
  experiments.}
\newblock {\em Journal of Computational and Graphical Statistics\/}, 24, 2,
  561--578.

\bibitem[\protect\citename{Gramacy et~al., }2015]{gra:etal:2015}
Gramacy, R.~B., Bingham, D., Holloway, J.~P., Grosskopf, M.~J., Kuranz, C.~C.,
  Rutter, E., Trantham, M., and Drake, R.~P. (2015).
\newblock \enquote{Calibrating a large computer experiment simulating radiative
  shock hydrodynamics.}
\newblock {\em Annals of Applied Statistics\/}, 9, 3, 1141--1168.

\bibitem[\protect\citename{Gramacy and Sun, }2018]{laGP}
Gramacy, R.~B. and Sun, F. (2018).
\newblock {\em {\tt laGP}: {L}ocal approximate {G}aussian process
  regression\/}.
\newblock R package version 1.5-2.

\bibitem[\protect\citename{Gul et~al., }2018]{gul:2018}
Gul, E., Joseph, V.~R., Yan, H., and Melkote, S.~N. (2018).
\newblock \enquote{Uncertainty quantification of machining simulations using an
  in situ emulator.}
\newblock {\em Journal of Quality Technology\/}, 50, 3, 253--261.

\bibitem[\protect\citename{Hastings, }1970]{Hastings:1970}
Hastings, W.~K. (1970).
\newblock \enquote{{Monte Carlo} Sampling Methods Using {M}arkov Chains and
  Their Applications.}
\newblock {\em Biometrika\/}, 57, 1, 97--109.

\bibitem[\protect\citename{Higdon et~al., }2004]{Higdon:2004}
Higdon, D., Kennedy, M., Cavendish, J.~C., Cafeo, J.~A., and Ryne, R.~D.
  (2004).
\newblock \enquote{Combining field data and computer simulations for
  calibration and prediction.}
\newblock {\em SIAM Journal on Scientific Computing\/}, 26, 2, 448--466.

\bibitem[\protect\citename{Hirs, }1973]{Hirs:1973}
Hirs, G.~G. (1973).
\newblock \enquote{{A Bulk-Flow Theory for Turbulence in Lubricant Films}.}
\newblock {\em Journal of Lubrication Technology\/}, 95, 2, 137--145.

\bibitem[\protect\citename{Hoff, }2009]{hoff2009first}
Hoff, P.~D. (2009).
\newblock {\em A first course in {B}ayesian statistical methods\/}.
\newblock Springer Science \& Business Media.

\bibitem[\protect\citename{Kass et~al., }1998]{kass:1998}
Kass, R.~E., Carlin, B.~P., Gelman, A., and Neal, R.~M. (1998).
\newblock \enquote{{Markov Chain Monte Carlo} in Practice: A Roundtable
  Discussion.}
\newblock {\em The American Statistician\/}, 52, 2, 93--100.

\bibitem[\protect\citename{Kennedy and O'Hagan, }2001]{Kennedy:O'Hagan:2001}
Kennedy, M.~C. and O'Hagan, A. (2001).
\newblock \enquote{Bayesian calibration of computer models.}
\newblock {\em Journal of the Royal Statistical Society, Series B\/}, 63, 3,
  425--464.

\bibitem[\protect\citename{Kleynhans and Childs, }1997]{isotseal}
Kleynhans, G. and Childs, D. (1997).
\newblock \enquote{The Acoustic Influence of Cell Depth on the Rotordynamic
  Characteristics of Smooth-Rotor/Honeycomb-Stator Annular Gas Seals.}
\newblock {\em ASME Journal of Engineering for Gas Turbines and Power\/},
  949--957.

\bibitem[\protect\citename{{Le~Digabel}, }2011]{Le09b}
{Le~Digabel}, S. (2011).
\newblock \enquote{Algorithm 909: {NOMAD}: Nonlinear Optimization with the
  {MADS} algorithm.}
\newblock {\em {ACM} Transactions on Mathematical Software\/}, 37, 4,
  44:1--44:15.

\bibitem[\protect\citename{Liu et~al., }2009]{Liu:2009}
Liu, F., Bayarri, M., and Berger, J. (2009).
\newblock \enquote{Modularization in {B}ayesian analysis, with emphasis on
  analysis of computer models.}
\newblock {\em Bayesian Analysis\/}, 4, 1, 119--150.

\bibitem[\protect\citename{Morris and Mitchell, }1995]{morris:1995}
Morris, M.~D. and Mitchell, T.~J. (1995).
\newblock \enquote{Exploratory designs for computational experiments.}
\newblock {\em Journal of Statistical Planning and Inference\/}, 43, 381--402.

\bibitem[\protect\citename{Nash, }2016]{nlmrt}
Nash, J.~C. (2016).
\newblock {\em {\tt nlmrt}: Functions for Nonlinear Least Squares Solutions\/}.
\newblock {\sf R} package version 2016.3.2.

\bibitem[\protect\citename{Petersen et~al., }2008]{Petersen:2008}
Petersen, K.~B., Pedersen, M.~S., et~al. (2008).
\newblock \enquote{The matrix cookbook.}
\newblock {\em Technical University of Denmark\/}, 7, 15.

\bibitem[\protect\citename{Plumlee, }2017]{plumlee2017bayesian}
Plumlee, M. (2017).
\newblock \enquote{Bayesian calibration of inexact computer models.}
\newblock {\em Journal of the American Statistical Association\/}, 112, 519,
  1274--1285.

\bibitem[\protect\citename{{R Core Team}, }2018]{lbfgsb}
{R Core Team} (2018).
\newblock {\em {\tt optim}: General-purpose Optimization\/}.
\newblock {\sf R} package version 3.6.0.

\bibitem[\protect\citename{Rasmussen and Williams, }2006]{Rasmussen2006}
Rasmussen, C.~E. and Williams, C. (2006).
\newblock {\em {Gaussian Processes for Machine Learning}\/}.
\newblock MIT Press.

\bibitem[\protect\citename{Sacks et~al., }1989]{Sacks1989}
Sacks, J., Welch, W.~J., Mitchell, T.~J., and Wynn, H.~P. (1989).
\newblock \enquote{Design and analysis of computer experiments.}
\newblock {\em Statistical science\/}, 4, 409--423.

\bibitem[\protect\citename{Santner et~al., }2003]{Santner2003}
Santner, T.~J., Williams, B.~J., and Notz, W.~I. (2003).
\newblock {\em The design and analysis of computer experiments\/}.
\newblock Springer Science \& Business Media.

\bibitem[\protect\citename{Tuo and Wu, }2015]{tuo2015}
Tuo, R. and Wu, C. F.~J. (2015).
\newblock \enquote{Efficient calibration for imperfect computer models.}
\newblock {\em Ann. Statist.\/}, 43, 6, 2331--2352.

\bibitem[\protect\citename{Vannarsdall, }2011]{vann:2011}
Vannarsdall, M.~L. (2011).
\newblock \enquote{Measured Results for a New Hole-Pattern Annular Gas Seal
  Incorporating Larger Diameter Holes, Comparisons to Results for a Traditional
  Hole-Pattern Seal and Predictions.}
\newblock Available electronically from
  http://hdl.handle.net/1969.1/ETD-TAMU-2011-08-9759.

\bibitem[\protect\citename{Ver~Hoef and Barry, }1998]{verhoef:barry:1997}
Ver~Hoef, J. and Barry, R.~P. (1998).
\newblock \enquote{Constructing and Fitting Models for Cokriging and
  Multivariate Spatial Prediction.}
\newblock {\em Journal of Statistical Planning and Inference\/}, 69, 275--294.

\bibitem[\protect\citename{Wackernagel, }1998]{wack:1998}
Wackernagel, H. (1998).
\newblock {\em Multivariate Geostatistics\/}.
\newblock New York: Springer.

\bibitem[\protect\citename{Ypma et~al., }2017]{nloptr}
Ypma, J., Borchers, H.~W., and Eddelbuettel, D. (2017).
\newblock {\em {\tt nloptr}: R interface to NLopt\/}.
\newblock {\sf R} package version 1.0.4.

\end{thebibliography}

\bibliographystyle{jasa}
\appendix


\end{document}